\documentclass[aps,prl,nofootinbib,twocolumn,superscriptaddress,preprintnumbers,balancelastpage,longbibliography]{revtex4-1}
\usepackage{aas_macros}

\usepackage{placeins}
\usepackage{amsmath,amssymb,mathtools,bm}
\usepackage{graphicx, color, hepunits}
\usepackage[dvipsnames]{xcolor}
\usepackage{cancel}
\usepackage[normalem]{ulem}
\usepackage{float}
\usepackage{filecontents}
\usepackage{multirow}
 \usepackage{hyperref} 
\hypersetup{
    colorlinks=true,       
    linkcolor=blue,        
    citecolor=blue,        
    filecolor=magenta,     
    urlcolor=blue          
}
\usepackage[utf8]{inputenc}
\usepackage[english]{babel}

\begin{document}

\title{Leading Axion-Photon Sensitivity with NuSTAR Observations of M82 and M87}

\author{Orion Ning}
\affiliation{Berkeley Center for Theoretical Physics, University of California, Berkeley, CA 94720, U.S.A.}
\affiliation{Theoretical Physics Group, Lawrence Berkeley National Laboratory, Berkeley, CA 94720, U.S.A.}

\author{Benjamin R. Safdi}
\affiliation{Berkeley Center for Theoretical Physics, University of California, Berkeley, CA 94720, U.S.A.}
\affiliation{Theoretical Physics Group, Lawrence Berkeley National Laboratory, Berkeley, CA 94720, U.S.A.}

\date{\today}

\begin{abstract}
We perform the most sensitive search to-date for the existence of ultralight axions using data from the NuSTAR telescope.   We search for stellar axion production in the M82 starburst galaxy and the M87 central galaxy of the Virgo cluster and then the subsequent conversion into hard X-rays in the surrounding magnetic fields. 
We sum over the full stellar populations in these galaxies when computing the axion luminosity, and we account for the conversion of axions to photons by using magnetic field profiles in simulated IllustrisTNG analogue galaxies.  
We show that analyzing NuSTAR data towards these targets between roughly 30 to 70 keV shows no evidence for axions and leads to robust constraints on the axion-photon coupling at the level of $|g_{a\gamma\gamma}| \lesssim 6.4 \times 10^{-13}$ GeV$^{-1}$ for $m_a \lesssim 10^{-10}$ eV at 95\% confidence.  
\end{abstract}
\maketitle

Low-mass axions with weak couplings to the Standard Model are strongly-motivated particle candidates. They are predicted to solve observational conundrums such as the strong CP problem of the neutron electric dipole moment and the dark matter  of our Universe~\cite{Peccei:1977hh,Peccei:1977ur,Weinberg:1977ma,Wilczek:1977pj,Preskill:1982cy,Abbott:1982af,Dine:1982ah}.  Axions also emerge generically in string theory from the compactification of the higher-dimensional theory to four dimensions~\cite{Witten:1984dg,Choi:1985je,Barr:1985hk,Svrcek:2006yi,Arvanitaki:2009fg}.  Low-mass axions may leave observable signatures in astrophysical data sets through their feeble interactions, which may be magnified in extreme environments. In this work we produce the strongest constraints to-date on the axion-photon coupling at low axion masses, finding no evidence for new physics, from searches for hard X-rays from the M82 and M87 galaxies.  

\begin{figure}[!htb]
\centering
\includegraphics[width=0.48\textwidth]{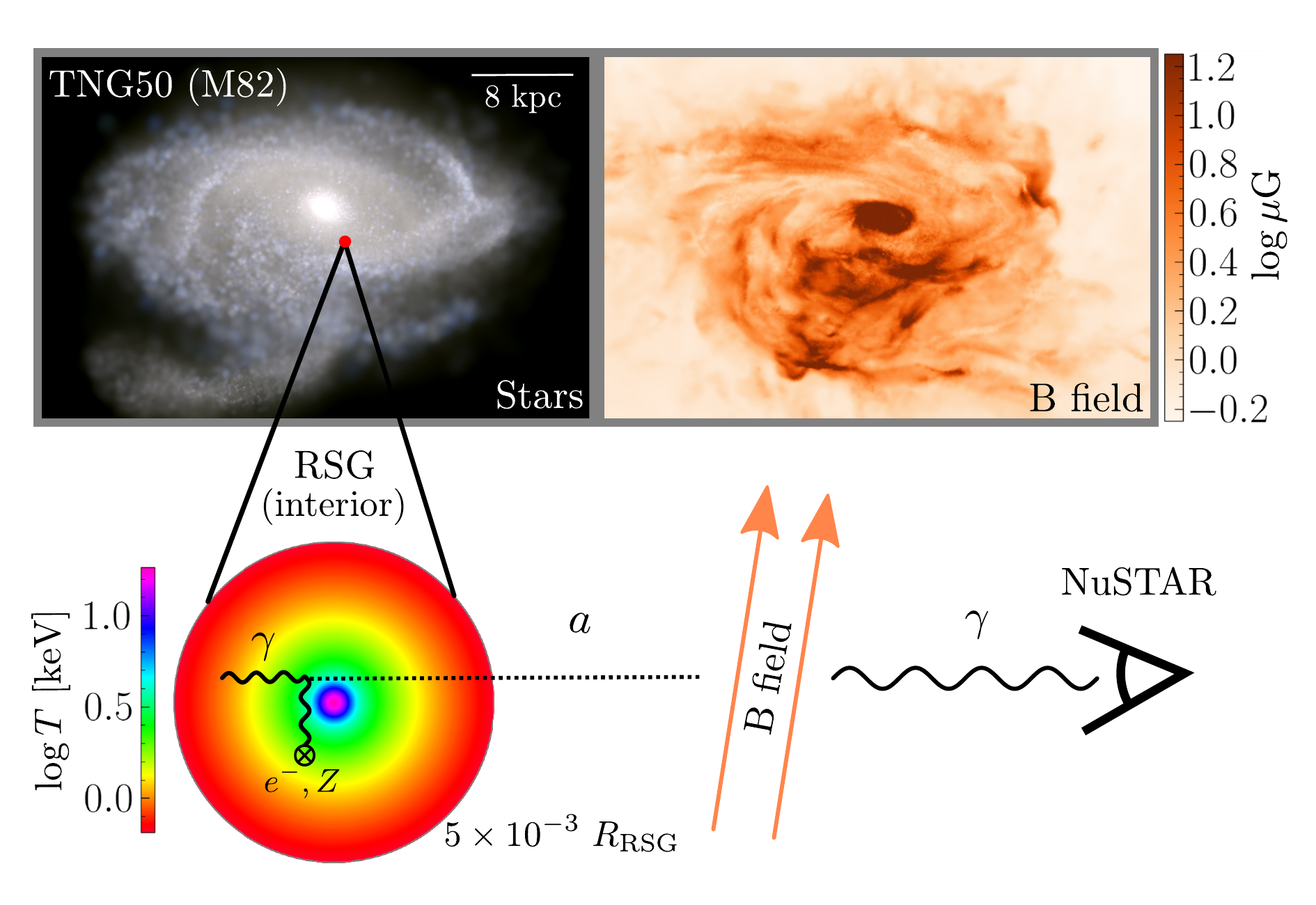}
\vspace{-0.4cm}
\caption{ 
Axions are produced through the Primakoff process within the ensemble of stars in the M82 and M87 galaxies and then convert to hard X-rays in the strong magnetic fields permeating these systems.  We search for the axion-induced hard X-rays with archival NuSTAR data. We illustrate the Primakoff process occurring in the core of a typical red supergiant (RSG) star in M82; RSGs are a class of stars contributing a dominant component of the total axion luminosity predicted from M82. The conversion probabilities are computed using an analogue galaxy in the TNG50 simulation within the IllustrisTNG cosmological hydrodynamic simulation project; we show here a stellar composite image as well as the corresponding magnetic field from one such analogue galaxy.}
\label{fig:ill}
\end{figure}

The basic idea behind the proposed search is illustrated in Fig.~\ref{fig:ill}.  Axions are produced through the Primakoff process within the hot stellar cores of stars in the galaxy and then the axions convert to photons through the inverse Primakoff process in the strong galactic magnetic fields, leading to hard X-ray signals on Earth that we search for using data from the NuSTAR telescope. The hot photons within the stellar cores are trapped by their electromagnetic interactions and unable to directly escape the stars.
In this sense, the searches we perform towards M82 and M87 are analogous to terrestrial light-shining-through-walls experiments~\cite{Ehret:2010mh,DIAZORTIZ2022100968}.

\begin{figure}[!t]
\centering
\includegraphics[width=0.49\textwidth]{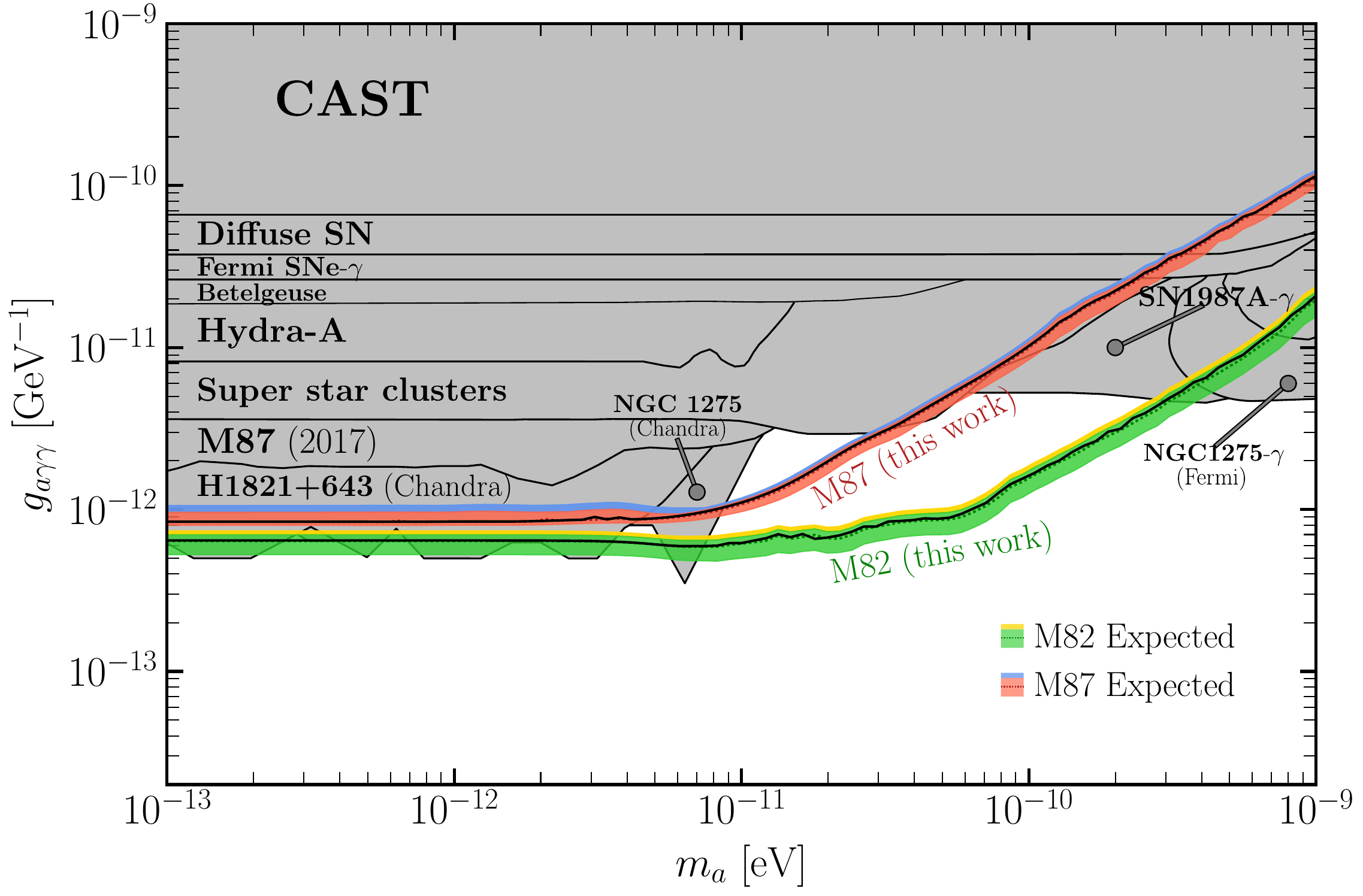}
\vspace{-0.4cm}
\caption{The parameter space of axion-photon couplings versus axion mass $m_a$ for ultra-light axions~\cite{AxionLimits}. The 95\% upper limits on $|g_{a\gamma\gamma}|$ from this work for our analyses of NuSTAR data towards M82 and M87 are indicated, along with the expectations for the upper limits at 1$\sigma$/2$\sigma$ confidence under the null hypothesis. Our upper limits at low masses are competitive with those from the Chandra spectral modulation searches ({\it e.g.},~\cite{Marsh:2017yvc,Conlon:2017qcw,Reynolds:2019uqt,Reynes:2021bpe}) and surpass those from searches for axion-induced X-ray signatures from SSCs~\cite{Dessert:2020lil}, which also used NuSTAR data.     }
\label{fig:axion}
\end{figure}

The region of axion parameter space we probe in this Letter is illustrated in Fig.~\ref{fig:axion}.  We are primarily sensitive to axions with any mass $m_a \lesssim 10^{-9}$ eV. Such ultra-light axions are motivated by string theory compactifications, which predict a large number $N \sim {\mathcal O}(10 - 10^3)$ of axion-like particles that arise as the zero modes of compactified gauge fields (see, {\it e.g.},~\cite{Demirtas:2018akl,Halverson:2019cmy,Mehta:2021pwf,Gendler:2023kjt}). One linear combination of these axions couples to quantum chromodynamics (QCD) and is responsible for solving the strong-CP problem. The other $N - 1$ axions, however, primarily receive their masses from non-QCD instantons, such as stringy instantons, and can be ultralight ($m_a \ll 10^{-9}$ eV).  Both the QCD axion and axion-like particles (which we refer to simply as axions) are expected to couple to Standard Model matter and gauge fields through dimension-5 operators that are suppressed by the axion decay constant $f_a$.
In field theory axion constructions the axion arises as the pseudo-Goldstone mode of $U(1)$ symmetry breaking. In this case, $f_a$ is related to the vacuum expectation value of the complex scalar charged under the $U(1)$ symmetry. In string theory constructions instead $f_a$ is inversely proportional to the size of the extra-dimensional cycle that gives rise to the axion in the low-energy effective field theory (EFT). (See~\cite{Hook:2018dlk,DiLuzio:2020wdo,Safdi:2022xkm,Caputo:2024oqc} for axion reviews.)

The axion EFT contains the interaction \mbox{${\mathcal L} \supset  g_{a\gamma\gamma} a {\bm E} \cdot {\bm B}$}, with $a$ the axion field and ${\bm E}$ (${\bm B}$) the electric (magnetic) field.  Here, $g_{a \gamma \gamma} \propto \alpha_{\rm EM} / f_a$, with $\alpha_{\rm EM}$ the fine-structure constant, is the dimension-full axion-photon coupling, which we constrain in this work.  Modern string theory constructions  
motivate the strongest coupled axion having $g_{a\gamma\gamma} \sim 10^{-13}$ GeV$^{-1}$ for $N \sim 100$ total axions~\cite{Halverson:2019cmy,Gendler:2023kjt}.  This is directly the level of coupling strength probed in this work.

There are two notable existing axion search strategies that are especially relevant to this work. The X-ray spectral modulation searches use data mostly from the Chandra telescope below $\sim$$10$ keV to search for spectral distortions of otherwise smooth X-ray sources in galaxies and galaxy clusters hosting large and extended magnetic fields~\cite{Marsh:2017yvc,Conlon:2017qcw,Reynolds:2019uqt,Reynes:2021bpe}. In the presence of extended magnetic fields and non-trivial photon plasma frequencies, the photons have a non-trivial, energy-dependent survival probability $P_{\gamma \to \gamma}(\omega)$, due to energy loss to axions, that distorts the otherwise smooth spectra. 
The absence of such spectral distortions in Chandra data from the quasar H1821+643 leads to the currently-leading upper limit (before this work) on $g_{a\gamma\gamma}$ at low axion masses~\cite{Reynes:2021bpe}, though the magnetic field of the host cluster is uncertain (see, {\it e.g.},~\cite{Libanov:2019fzq,Matthews:2022gqi}).  Searches for spectral modulations with Chandra data from the Perseus cluster produced similar though slightly weaker limits~\cite{Reynolds:2019uqt}.

Less sensitive upper limits at low axion masses also come from NuSTAR observations of Betelgeuse~\cite{Xiao:2021} as well as Galactic super star clusters (SSCs)~\cite{Dessert:2020lil}. Our search is especially similar to that performed in~\cite{Dessert:2020lil}. In~\cite{Dessert:2020lil}, the axions would be produced in the hot and massive Wolf-Rayet (WR) stars through Primakoff emission and then convert to hard X-rays in the Milky Way's magnetic fields. The absence of hard X-rays towards the Quintuplet SSC near the Galactic Center leads to the upper limit shown in Fig.~\ref{fig:axion}. 

It is useful to roughly compare the SSC NuSTAR search in~\cite{Dessert:2020lil} with the M82 search performed in this work. While the Quintuplet cluster has around $10^4$ $M_\odot$ in stellar mass, the M82 galaxy, which is a starburst galaxy with a high star-formation rate, has a stellar mass $\sim$$10^{10}$ $M_\odot$~\cite{Oehm:2017}. M82 is around 3.6 Mpc from Earth~\cite{1994ApJ...427..628F, Gerke:2011} and is viewed almost edge on. Making the very rough assumptions that the axion-luminosity-per-stellar mass in the hard X-ray band (between around 30 to 70 keV) is constant between Quintuplet and M82, that the axion-to-photon conversion probability $P_{a \to \gamma}$ for M82 is the same as for Quintuplet, that the NuSTAR observations of M82 and Quintuplet have the same exposure time, we estimate just based off of stellar mass and distance alone that the axion-induced X-ray emission in M82 should be around 10 times brighter than that from Quintuplet.  The majority of the M82 stars are within a few kpc of its center, which means that most of the stars fall within the NuSTAR 90\% containment radius of the point spread function (PSF) or around $1.7'$ and, like for Quintuplet, can be treated as a point source. This strongly suggests that M82 will be more sensitive than Quintuplet.

\noindent
{\bf Axion production calculation.---}We model the production of axions within the population of stars in the M82 and M87 galaxies by accounting for the Primakoff production rate in populations of simulated stars.  We treat the stellar interiors as thermal systems, and we focus on stars with plasma frequencies $\omega_{\rm pl}$ much less than the temperature $T$, so that there exists a thermal distribution of photons. These thermal photons may convert to axions via processes of the form $\gamma + (e^-,Z) \to a + (e^-,Z)$, with $(e^-,Z)$ standing for either free electrons or ions, using the axion-two-photon vertex.  The differential axion luminosity per unit volume from this process is computed in~\cite{Raffelt:1990yz,Dessert:2020lil} as a function of temperature, density, and composition, and is described in detail in the Supplementary Material (SM).

The axion luminosity from an individual star is computed from radial profiles of $T$ and $\{n_i\}$, where $\{n_i\}$ stands for the set of ion and electron number densities. We compute these profiles using the Modules for Experiments in Stellar Astrophysics (MESA) code package~\cite{2011ApJS..192....3P,2013ApJS..208....4P}.  The inputs for the MESA simulations are the initial stellar mass and metallicity. MESA then evolves the star over time until its end-point; we may thus extract the stellar profiles at the desired stellar age. Note that since we are only interested in stars with $T > \omega_{\rm pl}$, we do not track compact objects such as neutron stars and white dwarfs. As we show below, we are primarily sensitive to hot massive stars.  Additionally, we verify that for the values of $g_{a\gamma\gamma}$ probed in this work we may, to good approximation, neglect the back-reaction of the axion emission on stellar evolution.  

We fix the initial metallicity to $Z = 0.02$ for all stars in M82~\cite{Origlia:2004} and M87~\cite{Liu:2005, Kuntschner:2010}, though as we show in the SM our results are robust to variations in $Z$. Then, we construct an ensemble of time-resolved MESA simulations over a grid consisting of mass points spaced between $0.5$ $M_\odot$ and $100$ $M_\odot$.  We sample from this ensemble of MESA simulations in both mass and time according to the stellar distributions, which we describe below.  

For the initial mass function (IMF) for M82 we use that determined in~\cite{ForsterSchreiber:2003ft} by comparing population synthesis models with infrared spectroscopy data.  The IMF behaves as $dn/dM \propto 1/M^{2.35}$ at high masses and is flatter below $\sim$3$M_\odot$, with a high-mass cut-off of $100$ $M_\odot$.  
For M87, the determination of the IMF is more difficult due to its history of continuous merger assembly~\cite{Montes:2014}; following~\cite{Cook:2020}, we adopt the canonical Salpeter IMF ($dn/dM \propto 1/M^{2.35}$). 

For M82 we follow~\cite{ForsterSchreiber:2003ft} and describe the star formation rate as taking place in two bursts, one at an age of $t_{\rm burst} = 4.1$ Myr and the other at $t_{\rm burst} = 9.0$ Myr.  The star formation rate (SFR) for $t < t_{\rm burst}$ is modeled as $R_0 e^{-(t_{\rm burst} - t) / t_{\rm sc}}$, where $t_{\rm sc}$ is the characteristic decay time scale. For the older burst we take $R_0 = 31$ $M_\odot$/yr, while for the younger burst $R_0 = 18$ $M_\odot$/yr. For both bursts we follow~\cite{ForsterSchreiber:2003ft} and assume $t_{\rm sc} = 1.0$ Myr. For M87 we model the SFR by following~\cite{Cook:2020}, adopting an exponential $\tau$ model with $\tau = 5$ Gyr for ages $t \geq 1$ Gyr, and another exponential $\tau$ model with $\tau = 3.5$ Gyr for ages $t < 1$ Gyr, although~\cite{Cook:2020} notes that below $1$ Gyr the SFR is relatively poorly constrained. As we show in the SM, for a fixed number of stars today, our results are not strongly dependent on the details of the time dependence of the SFR. With that in mind, we also assume that the SFR is the same throughout all of the galaxy.

We estimate the number of stars in M82 as $N_{\rm tot} = (1.8 \pm 0.4) \times 10^{10}$ (consistent with~\cite{Oehm:2017}) using color-mass-to-light ratio relations~\cite{McGaugh:2014} and the $B - V$ color of M82 from the NASA/IPAC Extragalactic Database (NED)\footnote{\url{http://ned.ipac.caltech.edu/}}.  We determine the number of stars in M87 to be $N_{\rm tot} =(2.1 \pm 0.5) \times 10^{12}$ using the total stellar mass inference in~\cite{DeLaurentis:2022} combined with our choice of IMF. (See the SM.) 
For both systems the uncertainties on the number of stars are sub-leading sources of uncertainty in our final results, and so we adopt the central values in our analyses.

\begin{figure}[!t]
\centering
\includegraphics[width=0.49\textwidth]{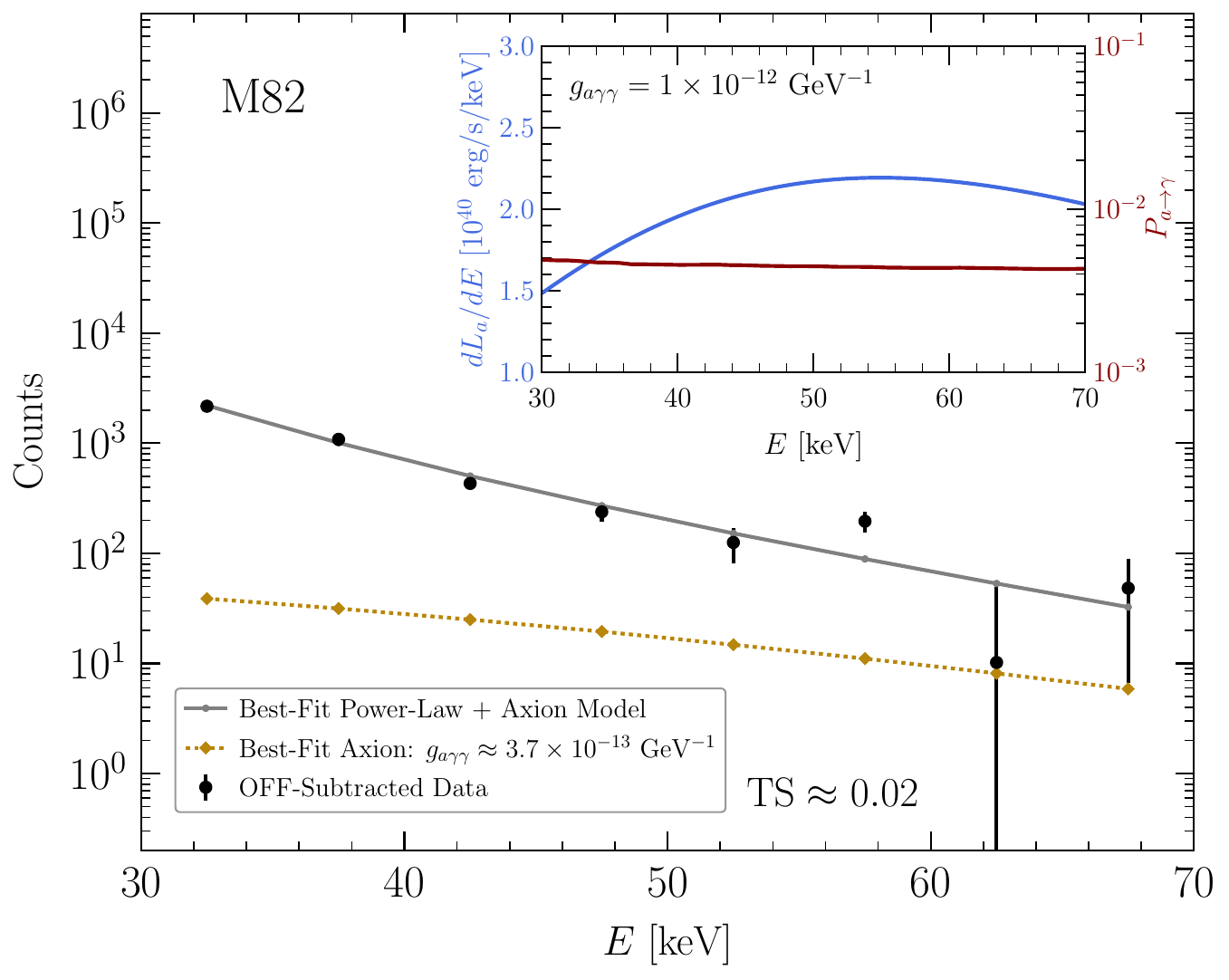}
\vspace{-0.4cm}
\caption{The stacked and binned OFF-subtracted NuSTAR data towards M82 in our analysis energy range. Also shown are our best-fit axion-induced signal model (gold) and our combined best-fit power-law and axion signal model (gray), all for our fiducial model in the low $m_a$ limit. In the inset we illustrate, for an example $g_{a\gamma \gamma}$ as indicated, the total differential axion luminosity (blue) as well as the median conversion probability (red) from our ensemble of conversion probabilities calculated using the IllustrisTNG TNG50 simulations.}
\label{fig:M82_spectra}
\vspace{-0.4cm}
\end{figure}

We draw from our distribution of MESA simulations $N_{\rm tot}$ times according to the appropriate IMF and age distributions, as described above. The axion luminosity may be computed for each stellar type. We classify stellar types from the MESA output following the procedures outlined in~\cite{Weidner:2010, Smith:2004}. For M82, we find that the axion luminosity in the energy range of interest is dominated by red supergiant (RSG) stars, which make up $\sim$$49\%$ of the emission (in terms of erg/s) integrated between 30 and 70 keV. O-type stars are the next most important contribution, representing $\sim$$34\%$ of the integrated emission between 30 and 70 keV. 
For M87, a combination of O-types ($\sim$$25\%$), blue supergiant (BSG) ($\sim$$27\%$), and RSG ($\sim$$20\%$) stars dominate the emission integrated between 30 and 70 keV.  
Fixing $g_{a\gamma\gamma} = 10^{-12}$ GeV$^{-1}$, we predict on average a total axion luminosity from M82 (M87) stars of around $\sim$$8 \times 10^{41}$ ($\sim$$1 \times 10^{42}$) erg/s. The differential axion luminosity spectrum for M82 is illustrated in Fig.~\ref{fig:M82_spectra} (inset).  In comparison, we note that the Quintuplet SSC total expected axion luminosity integrated between 30 and 70 keV, as computed in~\cite{Dessert:2020lil}, is $\sim$$8 \times 10^{34}$ erg/s.

 \noindent

{\bf Axion-photon conversion calculation.---} 
The axion-photon interaction causes an incoming axion state to rotate into a polarized electromagnetic wave in the presence of an external magnetic field (see, \textit{e.g.},~\cite{Raffelt:1987im,Safdi:2022xkm,Caputo:2024oqc}). The conversion probability $P_{a\to \gamma}$ depends on the magnetic field, the axion mass $m_a$, and the plasma frequency $\omega_{pl}$, which is determined by the free-electron density $n_e$ (see the SM).
In order to compute $P_{a\to \gamma}$, we need magnetic field profiles and $n_e$ distributions along the lines of sight. There are indications that the central kpc of M82 hosts strong, turbulent and ordered magnetic fields with magnitudes on the order of 0.1 - 1 mG~\cite{2006ApJ...645..186T,Lacki:2013ry,2021ApJ...914...24L} 
as a result of the starburst activity.  On the other hand, for our purposes we need a full-galaxy magnetic field model, along with a full-galaxy model of $n_e$. While such models are available for the Milky Way ({\it e.g.}, the Jansson and Farrar (JF) magnetic field model~\cite{2012ApJ...757...14J} (and more recently~\cite{Unger:2023lob}) and the NE2001 free-electron model~\cite{Cordes:2002wz}), there are no such detailed models for M82.  Instead, we leverage recent developments in magnetohydrodynamic cosmological simulations to obtain simulated analogue M82 galaxies that we can use both to draw the stellar locations and to calculate the axion-to-photon conversion probabilities. 

For M82, we use the simulation data releases from the IllustrisTNG TNG50 cosmological magnetohydrodynamical simulations~\cite{Pillepich:2019bmb,Nelson:2019jkf}. 
The {IllustrisTNG} simulations, of which TNG50 is the highest resolution version at present, have been shown to self-consistently produce magnetic field and free-electron profiles, along with other stellar and baryonic distributions, broadly consistent with those found in nature~\cite{Marinacci:2017wew}. 
We identify M82 analogue galaxies in the TNG50 simulation data by searching for candidates which have a stellar mass close to $\sim$$10^{10}$ $M_{\odot}$~\cite{Oehm:2017} and a star formation rate close to $\sim$$10$ $M_{\odot}$/yr~\cite{deGrijs:2001}. We identify three close candidates to this set of criteria, which are described further in the SM. Then, we orient the selected TNG50 galaxy to match the observed orientation of M82. This is important because M82 is nearly edge-on as seen from the Milky Way, with the normal vector to the disk at an angle of around 80$^\circ$ relative to our line-of-sight~\cite{McKeith:1995}. The magnetic fields are largest within the M82 disk, so the edge-on nature of M82 enhances the average conversion probability.  

We draw stellar positions for each star in M82 using the baryon density in the TNG50 simulation output as a probability distribution for stellar locations.  For each stellar location we compute the conversion probability for axions that propagate out of the Galaxy starting from that point. Note that we must identify an orientation for the galaxy in the azimuthal direction, as seen from the normal of the M82 galactic plane. We consider all possible orientations in all three simulation candidates, and for each orientation and candidate we compute the upper limit on $g_{a\gamma\gamma}$ at asymptotically low $m_a$. We select as our fiducial model the orientation and simulated galaxy that leads to the 84\% percentile ({\it i.e.}, $1 \sigma$) weakest upper limit at low $m_a$.  
Our fiducial galaxy is illustrated in Fig.~\ref{fig:ill}, where we show both the baryon distribution and the magnitude of the magnetic field.  The median conversion probability $P_{a\to\gamma}$ for stars in M82 for our fiducial analogue galaxy is shown in the inset of Fig.~\ref{fig:M82_spectra}. 

The magnetic field uncertainty
is the dominant source of uncertainty in the prediction of the axion signal.
On the other hand, since the predicted signal scales as $g_{a\gamma\gamma}^4$, the magnetic field uncertainty translates to a minor uncertainty on the upper limit. For example, the average upper limit across our ensemble of simulations and orientations is around 10\% stronger than that for our fiducial choice. On the other hand, we note that none of our simulated galaxies exhibit magnetic fields with fields strengths above $\sim$40 $\mu$G; the field strengths in the simulated galaxies are much smaller than the $\sim$mG level field strengths claimed for M82 in~\cite{2006ApJ...645..186T,Lacki:2013ry,2021ApJ...914...24L}.  As we show in the SM, if we instead use the magnetic field model for M82 in~\cite{2021ApJ...914...24L}, which only covers the inner $\sim$2 kpc, the upper limit at low $m_a$ would be $|g_{a\gamma\gamma}| \lesssim 3.8 \times 10^{-13}$ GeV$^{-1}$, which is much stronger than our fiducial upper limit at low $m_a$,  $|g_{a\gamma\gamma}| \lesssim 6.4 \times 10^{-13}$ GeV$^{-1}$.

In our fiducial approach to M87 we follow an analogous procedure to M82, whereby we find analogue Virgo clusters in the TNG300 simulation output. The TNG300 simulations, which are produced over much larger spatial scales than TNG50, are needed in order to identify the Virgo cluster in which M87 is nestled, as well as the cluster-level magnetic fields which are crucial for this analysis.  In our fiducial M87 analogue galaxy we find $P_{a\to\gamma} \approx 3 \times 10^{-2}$ for $g_{a\gamma\gamma} =10^{-12}$ GeV$^{-1}$ in the energy range of interest; this conversion probability is nearly an order of magnitude larger than that in our fiducial M82 model.  As an independent check of the conversion probability, we use the stochastic, domain-based Virgo magnetic field model developed in~\cite{Marsh:2017yvc}, which yields upper limits consistent within $\sim$10\% of those from our fiducial simulated-based analysis (see the SM).  We note that for both M82 and M87 the free-electron density profiles play minor roles, given the high axion energies; setting $n_e=0$ in the conversion probability calculations changes the upper limits by less than 1\%.  

\noindent
{\bf NuSTAR data analysis and results.---}
We reduce and analyze all archival NuSTAR data for M82 (M87), amounting to $\sim$2.07 Ms ($\sim$0.51 Ms) of exposure time for each of two Focal Plane Modules (FPM). 
The NuSTAR data reduction is performed with HEASoft software version 6.28~\cite{2014ascl.soft08004N}, giving us sets of counts for source and background regions for each energy bin and for each exposure. The background spectra accounts for a variety of sources, including the cosmic X-ray background, reflected solar X-rays, and instrumental backgrounds such as Compton-scattered gamma rays and detector and fluorescence emission lines~\cite{Wik:2014}. We choose as our source region (ON region) the 68\% containment region of the NuSTAR PSF centered on our galaxy target. The background region (OFF region) is chosen to be spatially separated from the source as an annulus with inner and outer radii corresponding to the 80\% and 90\% containment radii of the NuSTAR PSF. These localized regions around our sources minimize possible systematic biases from background mismodeling and are illustrated in the SM. We use data from both FPMs, and we stack and re-bin the data in 5-keV-wide energy bins, using as our fiducial energy range 30-70 keV. We begin our analysis range at 30 keV to minimize the mismodeling of low-energy astrophysical X-ray emission from the galaxies, although we show in the SM that our limits would improve if we extended the analysis to lower energies.  

For a given $g_{a\gamma\gamma}$ we determine the predicted signal counts NuSTAR would observe by forward-modeling the axion signal model through the instrument response. To compare with the background-subtracted source spectra (which we call the OFF-subtracted data) 
from NuSTAR, we model the OFF-subtracted data as the forward-modeled axion signal plus a floating power-law background model, with the power-law parameters treated as nuisance parameters.  The power-law model describes the astrophysical hard X-ray emission from the galaxies.  
We construct a Gaussian spectral likelihood, joint over all energy bins, and profile over the nuisance parameters associated with the power-law background. 
In Fig.~\ref{fig:M82_spectra} we illustrate an example of the OFF-subtracted NuSTAR data along with our best-fit total signal model. We then use standard frequentist statistics in the context of the profile likelihood to constrain $g_{a\gamma\gamma}$ (see~\cite{Safdi:2022xkm} and the SM).

In Fig.~\ref{fig:axion} we illustrate the 95\% power-constrained~\cite{Cowan:2011an} upper limits on $g_{a\gamma \gamma}$ as a function of the axion mass found from our analyses of M82 and M87. For a fixed axion mass $m_a$, the one-sided upper limits on $g_{a\gamma \gamma}$ and the discovery significances are calculated via Wilk's theorem~\cite{Cowan:2010js}, and the $1\sigma$/$2\sigma$ expectations for the 95\% upper limits under the null hypothesis are constructed from the Asimov procedure~\cite{Cowan:2010js}. In the massless limit, our 95\% upper limit from M82 (M87) is $|g_{a\gamma \gamma}| \lesssim 6.4 \times 10^{-13}$ GeV$^{-1}$ ($ \lesssim 8.4 \times 10^{-13}$ GeV$^{-1}$), with the evidence in favor of the axion model being $\sim$$0.15\sigma$ ($\sim$$0.83\sigma$).  The best fit couplings from the two analyses are $|g_{a\gamma\gamma}| \approx 3.7 \times 10^{-13}$ GeV$^{-1}$ and $|g_{a\gamma\gamma}| \approx 7.9 \times 10^{-13}$ GeV$^{-1}$, respectively.

\noindent
{\bf Discussion.---}
In this work we produce leading constraints on the axion-photon coupling for low-mass axions by searching for hard X-ray signatures from all stars in the M82 and M87 galaxies. The high-energy axions would be produced in the cores of hot, massive stars, primarily RSG and O-type stars, and then convert to X-rays in the strong magnetic fields in the M82 galaxy or in the Virgo cluster, in the case of M87. Our results directly impact ultraviolet models
that predict the existence of these particles~\cite{Halverson:2019cmy,Gendler:2023kjt}.
As explored further in the SM, our results are robust to systematic tests that probe uncertainties in the stellar populations and magnetic field structures of these systems. 

While we focus on M82 and M87 in this work, we have not performed an extensive survey of other possible targets. It is possible, for example, that NGC 1275 in the center of the Perseus cluster is also a promising target, and one could even imagine doing a stacked analysis over many clusters. Other nearby galaxies, such as M31 or even Milky Way satellites, may also produce competitive sensitivity.  We also note that strong sensitivity to the axion-electron and axion-nucleon couplings, multiplied by the axion-photon coupling, can also likely be obtained by considering production through bremsstrahlung processes in stellar populations in these systems. 

\section{Acknowledgements}

{\it
We thank Chris Dessert for collaboration during the early stages of this project, and we thank Joshua Benabou, Andrea Caputo, Andrew Long, Claudio Manzari, Yujin Park, and Georg Raffelt for helpful conversations. 
B.R.S was supported in part by the DOE Early Career Grant DESC0019225.  The work of O.N. was supported in part by the NSF Graduate Research Fellowship Program under Grant DGE2146752.
This research used resources of the National Energy Research Scientific Computing Center (NERSC), a U.S. Department of Energy Office of Science User Facility located at Lawrence Berkeley National Laboratory, operated under Contract No. DE-AC02-05CH11231 using NERSC award HEP-ERCAP0023978.  Supplementary data to reproduce the main figures may be found \href{https://github.com/orionning676/M82_M87_NuSTAR_Axion}{here}.  
}

\bibliography{refs}

\clearpage

\onecolumngrid
\begin{center}
  \textbf{\large Supplementary Material for Leading Axion-Photon Sensitivity with NuSTAR Observations of M82 and M87}\\[.2cm]
  \vspace{0.05in}
  {Orion Ning and Benjamin R. Safdi}
\end{center}

\twocolumngrid

\setcounter{equation}{0}
\setcounter{figure}{0}
\setcounter{table}{0}
\setcounter{section}{0}
\setcounter{page}{1}
\makeatletter
\renewcommand{\theequation}{S\arabic{equation}}
\renewcommand{\thefigure}{S\arabic{figure}}
\renewcommand{\thetable}{S\arabic{table}}

\onecolumngrid

This Supplementary Material (SM) contains further explanations and illustrations of the methods and results presented in the main Letter. First, we present additional details regarding the data reduction, analysis, simulations, and calculations used in this work. Then, we extend the results of our main Letter by performing systematic analyses across all major sources of uncertainties in our work, comparing them against our fiducial results at the very end.

\section{Data Reduction and Analysis}

For NuSTAR data reduction, we use the NuSTARDAS software included with HEASoft 6.28~\cite{2014ascl.soft08004N}. We first reprocess the data with the NuSTARDAS task \texttt{nupipeline}. This outputs calibrated and screened events files. Then, we extract the energy spectra in counts for both focal plane modules (FPMs) within the NuSTAR field of view (FOV) with \texttt{nuproducts}, and we rebin the counts in energy bins of width $5$ keV from $5 - 80 $ keV. Additionally, we generate the ancillary response files (ARFs) and the redistribution matrix files (RMFs) for each FPM, which are used in the forward modeling of our axion-induced signal.

The data used in this analysis are summarized in Tab. \ref{tab:obs}, which are downloaded from the HEASARC NuSTAR archive. We use all archival data dedicated to M82 and M87 with exposure time greater than 100 seconds.

\begin{table}[hb]

\begin{tabular}[t]{ccc}
ObsID & $t_{\rm exp}$ [s] & Target \\ \hline \hline
30101045002       & $189231$  & M82        
\\
50002019004       & $160740$  & M82         
\\
30702012002       & $127605$  & M82     
\\
30901038002       & $123396$ & M82           
\\
30702012004      & $117506$  & M82           
\\
30502022004       & $95922$ & M82            
\\
30502020002       & $88651$  & M82         
\\
30502022002       & $88348$ & M82          
\\
30502020004       & $88040$ & M82           
\\
30502021002       & $85157$ & M82  
\\
90201037002       & $80194$  & M82        
\\
30502021004       & $78591$  & M82         
\\
30602027002       & $71868$  & M82     
\\
30602027004       & $69992$ & M82           
\\
30602028004      & $68045$  & M82           
\\
30602028002       & $66381$ & M82            
\\
30202022004       & $47035$  & M82         
\\
90202038002       & $45475$ & M82 
\\
30202022010       & $44368$ & M82
\\
\vdots & &         
\end{tabular}
\begin{tabular}[t]{ccc}
ObsID & $t_{\rm exp}$ [s] & Target \\ \hline \hline
\vdots & & 
\\
90202038004       & $43319$ & M82  
\\
30202022008       & $42844$  & M82        
\\
80202020008       & $40355$  & M82
\\
30202022002       & $39021$  & M82     
\\
90101005002       & $37407$ & M82           
\\
80202020002      & $36133$  & M82           
\\
80202020004       & $31667$ & M82            
\\
50002019002       & $31243$  & M82         
\\
80202020006       & $30501$ & M82          
\\
30202022006       & $2108$ & M82           
\\
90201037001       & $405$ & M82  
\\
30502022003       & $299$  & M82    
\\
30202022003       & $271$ & M82            
\\
80202020007       & $239$  & M82         
\\
30502021001       & $150$ & M82          
\\
30202022005       & $129$ & M82           
\\
90101005001       & $116$ & M82  
\\
30901038001       & $112$  & M82 
\\
\hline
Total & 2072864 & M82
\\
\hline 
\end{tabular}
\begin{tabular}[t]{ccc}
ObsID & $t_{\rm exp}$ [s] & Target \\ \hline \hline
60201016002       & $50342$ & M87  
\\
60801007002       & $50238$ & M87  
\\
60602021002       & $48770$ & M87 
\\
60702051001       & $48565$ & M87  
\\
60602021004       & $47965$ & M87  
\\
60702051002       & $45538$ & M87 
\\
50502002006       & $31636$ & M87  
\\
50502002004       & $31576$ & M87 
\\
50502002002       & $28160$ & M87  
\\
90202052002       & $24414$ & M87  
\\
60502010002       & $23549$ & M87 
\\
90202052004       & $22480$ & M87  
\\
60466002002       & $21082$ & M87  
\\
60502010004       & $20899$ & M87 
\\
60801001002       & $19421$ & M87  
\\
\hline
Total & 514635 & M87
\\
\hline
\end{tabular}

\caption{\label{tab:obs} Information on the observations used in the NuSTAR analysis. The first column is the observation ID, while the second is the exposure time in [s], and the third is the astrophysical target.}
\end{table}

With the reduced NuSTAR data, we apply the analysis procedure described in the main Letter to obtain upper limits on $g_{a \gamma \gamma}$ using our signal model. However, to compare our predicted axion signal model to the NuSTAR source spectrum, we must first forward-model the predicted axion-induced flux through the NuSTAR instrument response. Our prescription for this forward-modeling procedure returns a set of expected signal counts that we can compare to the data:
\begin{equation}
    \mu^e_{S, i}(\boldsymbol{\theta}_S) = t^e \int dE' \text{RMF}_i^e(E') \text{ARF}^e(E') S(E' | \boldsymbol{\theta}_S).
\label{eq:mu_S}
\end{equation}
Here, $t^e$ is the exposure time corresponding to the exposure $e$ in [s], while the signal is the expected intensity spectrum in [erg/cm$^2$/s/keV] with parameters $\boldsymbol{\theta}_S = \{ m_a, g_{a \gamma \gamma} \}$. The expected total stacked signal counts $\sum_e \mu^e_{S, i} = \mu_{S, i}(\boldsymbol{\theta}_S)$ then represent the axion-induced spectra across energy bins $i$ and all exposures $e$ that can then be incorporated into our likelihood, described later. 

To compare to the OFF-subtracted source spectra, $d_i$, we additionally model our total signal $\mu_i$ as containing an astrophysical power-law background model $B(E|\boldsymbol{\theta}_B) =  \alpha E^{\beta}$, where $\boldsymbol{\theta}_B = \{ \alpha, \beta \}$ are nuisance parameters to be profiled over in our analysis. This power-law background is folded through the detector response, as in~\eqref{eq:mu_S}, to obtain the total expected astrophysical power-law background counts over all observations, $\mu_{B,i}(\boldsymbol{\theta}_B)$. 

Our total predicted counts from the signal model plus power-law background is thus $\mu_i = \mu_{S,i}(\boldsymbol{\theta}_S) + \mu_{B,i}(\boldsymbol{\theta}_B)$. We then construct the Gaussian likelihood 
\begin{equation}
p(\boldsymbol{d} | \{ \boldsymbol{\theta}_S, \boldsymbol{\theta}_B \} ) = \prod_i \frac{1}{\sqrt{2 \pi} \sigma_i} \exp \left[ -\frac{(d_i - \mu_i)^2}{2\sigma_i^2}  \right] \,,
\end{equation}
and we profile the likelihood over the nuisance parameters
by maximizing the log-likelihood at each fixed $\boldsymbol{\theta}_S$ over the $\boldsymbol{\theta}_B$. Note that the $\sigma_i$ are taken to be the appropriate Poisson errors of the OFF-subtracted source spectra $d_i$, while $\boldsymbol{d} = \{ d_i \}$ is the data set consisting of all stacked counts $d_i$ within our energy analysis range.  The Gaussian likelihood is justified, as an approximation to the Poisson likelihood, because in each energy bin the total number of counts in the source and background data is much larger than 10. 

Given our likelihood and a fixed $m_a$, the upper limits on $g_{a \gamma \gamma}$ are constructed by analyzing the test statistic 
\begin{equation}
    q(g_{a \gamma \gamma} | m_a) \equiv 2 \ln p(\boldsymbol{d} | \{m_a, g_{a\gamma \gamma}\}) - 2 \ln p(\boldsymbol{d} | \{ m_a, \Bar{g}_{a \gamma \gamma}\})\,,
\end{equation}
where $\Bar{g}_{a\gamma \gamma}$ is the signal strength that maximizes the likelihood, allowing for the possibility of negative signal strengths as well (see, \textit{e.g.}~\cite{Cowan:2010js}). Through Wilk's theorem, the 95\% upper limit is given by the value $g_{a\gamma \gamma} > \Bar{g}_{a\gamma \gamma}$ such that $q(g_{a\gamma\gamma} | m_a) \approx 2.71$, with the $1\sigma$ and $2\sigma$ expectations for the 95\% upper limits under the null hypothesis constructed from the Asimov procedure~\cite{Cowan:2010js}. (See~\cite{Safdi:2022xkm} for a review.)

\begin{figure}[!htb]
\centering
\includegraphics[width=0.5\textwidth]{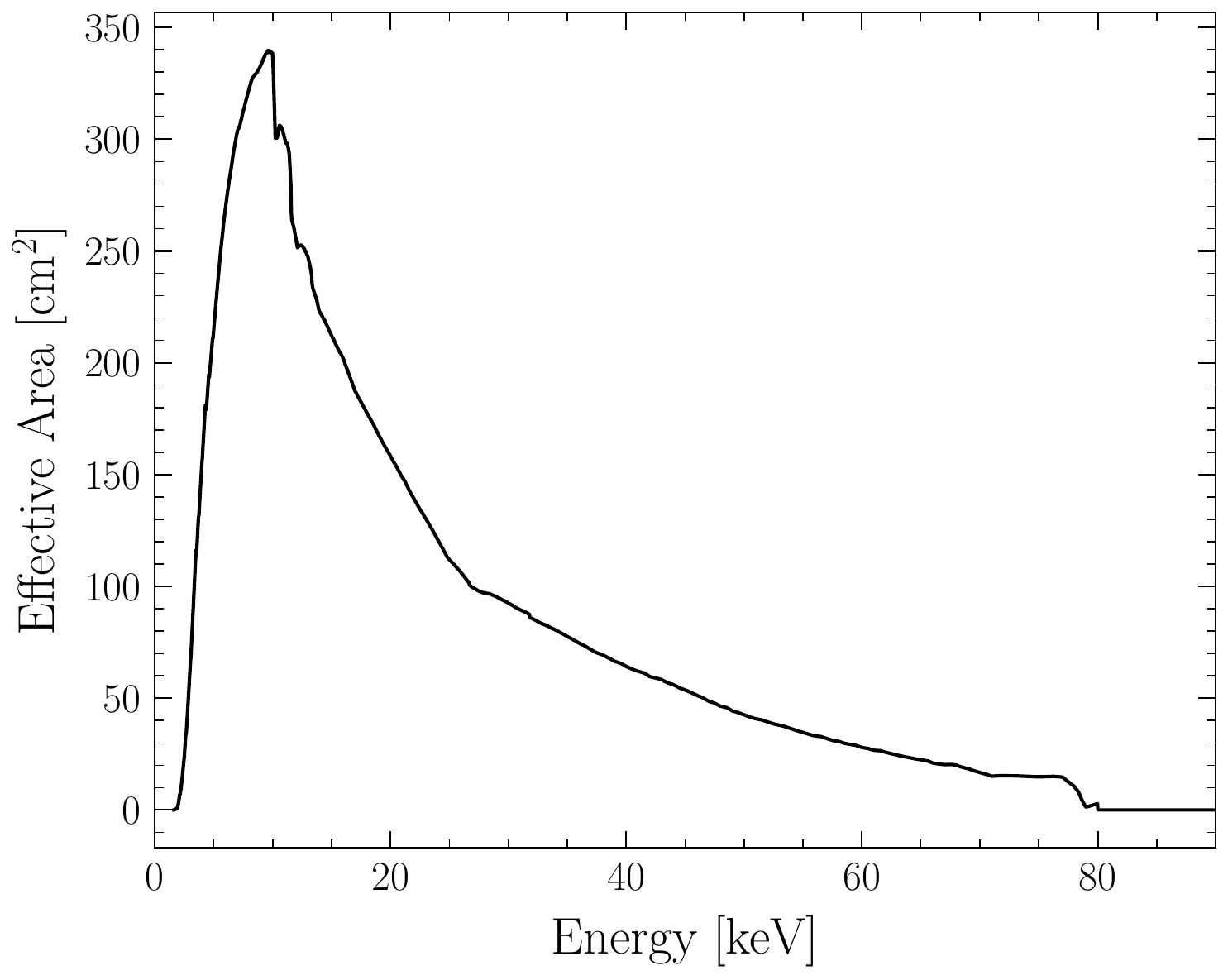}
\caption{The NuSTAR effective area. This particular effective area is extracted from one of our observations, Obs. 30202022004, and is typical across the observations used in our work.}
\label{fig:NuSTAR_effarea}
\end{figure}

\begin{figure}[!htb]
\centering
\includegraphics[width=0.7\textwidth]{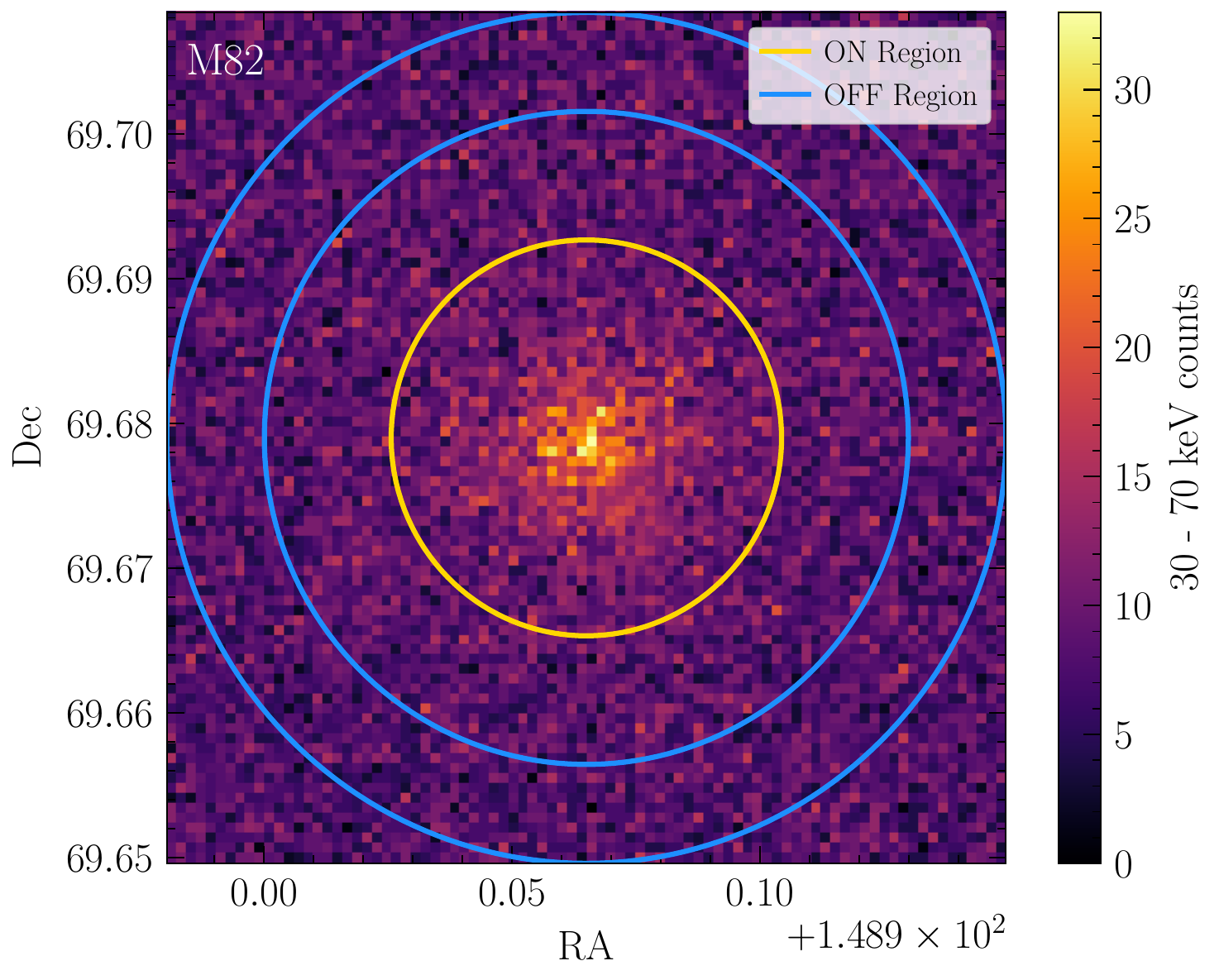}
\caption{Spatial map of total stacked counts between 30-70 keV from all archival NuSTAR observations of M82.  The hard X-ray emission from the galaxy is clearly visible, though the spectral morphology of these excess counts is not consistent with an axion origin.}
\label{fig:M82_map}
\end{figure}

\begin{figure}[!htb]
\centering
\includegraphics[width=0.7\textwidth]{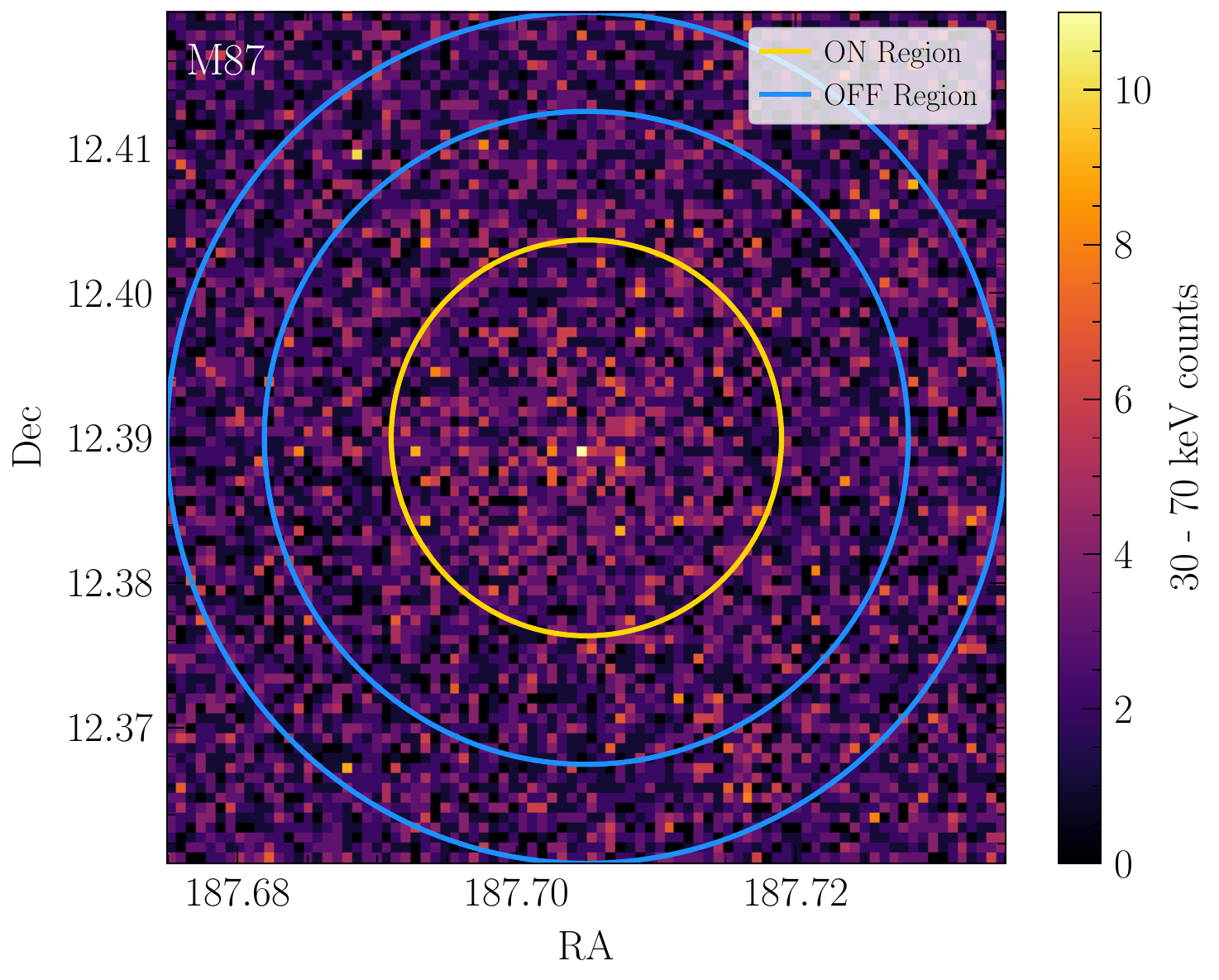}
\caption{As in Fig.~\ref{fig:M82_map}, but for M87.  }
\label{fig:M87_map}
\end{figure}

\begin{figure}[!htb]
\centering
\includegraphics[width=0.6\textwidth]{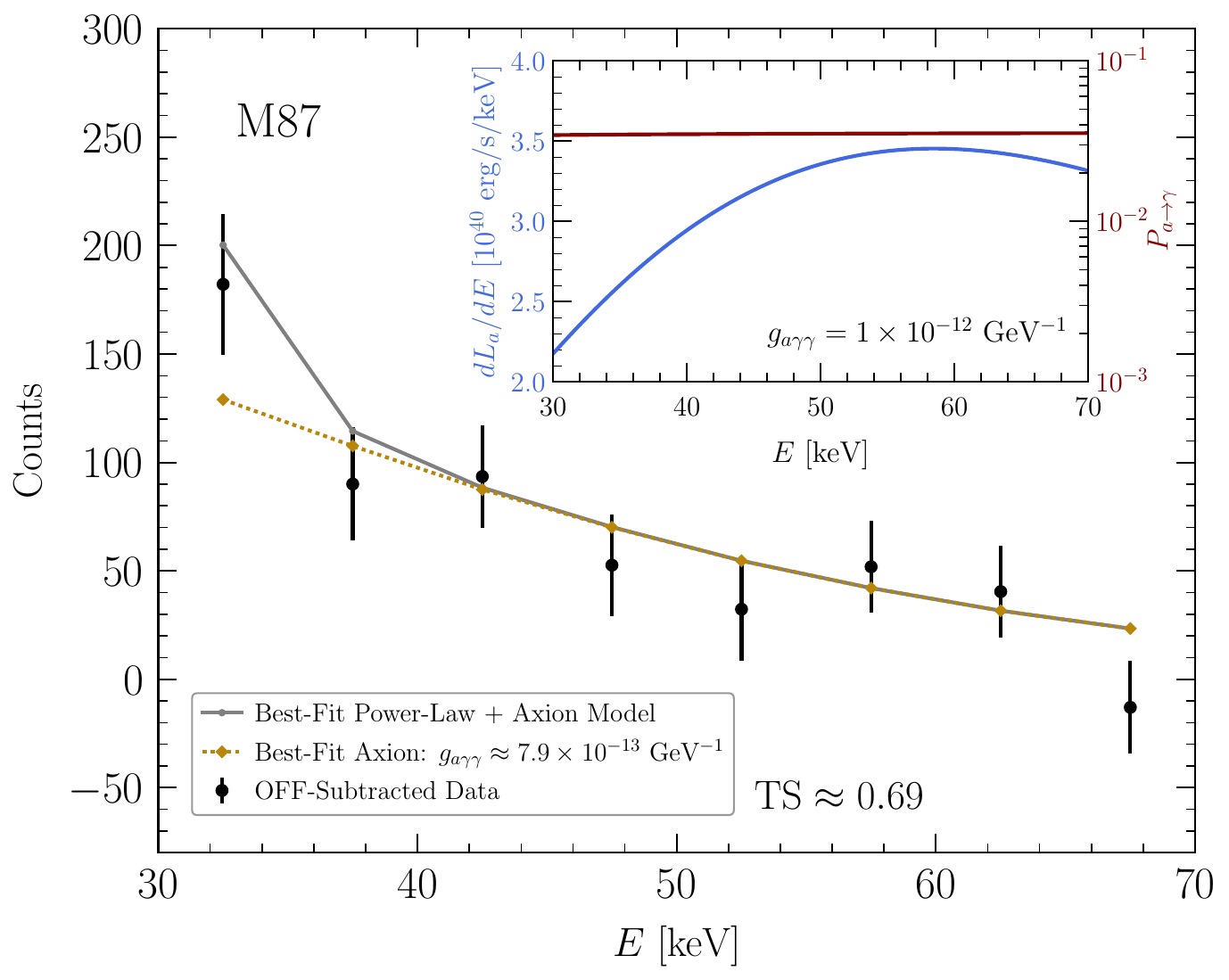}
\caption{As in Fig.~\ref{fig:M82_spectra}, but for M87. 
}
\label{fig:M87_spectra}
\end{figure}

Here we illustrate several figures relevant to the data portion of this work. In Fig.~\ref{fig:NuSTAR_effarea} we show the NuSTAR effective area, taken from one of our observations and typical across others. In Figs.~\ref{fig:M82_map} and \ref{fig:M87_map} we display spatial maps of the total stacked counts in our fiducial analysis energy range of 30-70 keV from all archival NuSTAR observations of M82 and M87, illustrated with our chosen ON and OFF regions used in our analysis, as discussed in the main Letter. 
 The hard X-ray emission from these galaxies is clearly visible in these images, so we find that the spectral morphology of the excess counts is not consistent with an axion-induced origin in both cases. In Fig.~\ref{fig:M87_spectra} we illustrate the analogous data and fits as in Fig.~\ref{fig:M82_spectra} but for M87, which ends up producing weaker upper limits on $g_{a \gamma \gamma}$ than M82.

\section{Axion Production in Stars}

In this section we overview how we derive the total axion luminosity and spectra emitted from a given star.  As mentioned in the main Letter, we treat the stellar interiors as thermal systems, focusing on stars with plasma frequencies $\omega_{\rm pl}$ much less than the temperature $T$, so that there exists a thermal distribution of photons. Through the Primakoff process, the differential axion luminosity per unit volume is given by~\cite{Raffelt:1990yz,Dessert:2020lil} 
\begin{equation}
{d L_a(E) \over dE} = {g_{a\gamma\gamma}^2 \over 8 \pi^2} {\xi^2 T^3 E \over e^{E/T} - 1} \sqrt{1 - {\omega_{\rm pl}^2 \over E^2}} 
\left[ (E^2 + \xi^2 T^2)
 \log\left(1 + {E^2 \over \xi^2 T^2} \right) 
- E^2 \right] \,,
\label{eq:dLadE}
\end{equation}
with $E > \omega_{\rm pl}$ the axion energy. 
The parameter $\xi$ is defined by $\xi \equiv \kappa / (2 T)$, where $\kappa^2 \equiv (4 \pi \alpha_{\rm EM} / T) \sum_i Z_i^2 n_i$ is the Debye screening scale, $Z_i$ is the charge of the scattering target (electron or ion), and $n_i$ is the target number density.  The quantity $dL_a / dE$ has units of erg$/$s$/$keV$/$cm$^3$, and so it should be integrated over the star in order to obtain a differential luminosity. Note that the photon plasma frequency is given by $\omega_{\rm pl}^2 = 4 \pi \alpha_{\rm EM} \sum_i Z_i^2 n_i / m_i$, which tends to be dominated by the electron contribution since it has the smallest mass.

For an individual star we are able to compute the axion luminosity through the Primakoff process given radial profiles for $T$ and $\{n_i\}$, where $\{n_i\}$ stands for the set of ion and electron number densities. These profiles are given by MESA simulations, described later.  
Finally, the axion-induced photon spectrum at Earth, for a given star, is 
\begin{equation}
    \frac{dF}{dE}(E) = P_{a\to \gamma} (E) \frac{1}{4 \pi d^2}\frac{dL_a(E)}{dE} \,,
\label{eq:dFdE}
\end{equation}
where $d$ is the distance, and $P_{a\to \gamma}$ is the conversion probability, computed later.

\section{MESA Simulations}

To simulate the individual stellar components used in this work, we use MESA (release \texttt{r23.05.1}), a one-dimensional stellar evolution code which solves the equations of stellar structure and returns detailed stellar profiles at any point in the evolution of the star. In our fiducial analysis for both M82 and M87 we construct and evolve stellar models at a metallicity of $Z = 0.02$ over a grid of stellar masses spaced betweeen $0.5$ $M_{\odot}$ and $100$ $M_{\odot}$, as indicated in the main Letter. We note that the determination of metallicity for both M82 and M87 is difficult; in both galaxies, the abundances of different metal species can be quite different, and there is additionally some degree of spatial dependence. This makes the overall metallicity a challenging quantity to estimate. Still, there is evidence to support the use of near-solar metallicity as a very rough estimate for an overall $Z$. We adopt this choice for M82, based on IR and X-ray spectral data that come with overall uncertainties of $\sim$$\pm 0.004$~\cite{Origlia:2004}, as well as for M87, based on optical photometry and spectroscopic measurements with overall uncertainties of $\sim$$\pm 0.006$~\cite{Liu:2005, Kuntschner:2010}. We show in Figs.~\ref{fig:M82_systematics} and~\ref{fig:M87_systematics} how the extent of these metallicity uncertainties affect our final upper limits.

For stars of mass $\sim$8 $M_{\odot}$ and greater we use the default suite and in-list for high-mass stars, \texttt{20M\_pre\_ms\_to\_core\_collapse}, evolving the stars all the way from pre-main sequence (pre-MS) to the onset of core-collapse. This suite sets a number of parameters required for high-mass evolution, namely the use of Type 2 opacities. For lower mass stars, we use the default suite and in-list for low-mass stars, \texttt{1M\_pre\_ms\_to\_wd}, which fully evolves the stars from the pre-MS to the white dwarf stage. We note that, as mentioned in the main Letter, since we are only interested in stars with $T > \omega_{pl}$, the compact objects found at the end of some of these simulations are unimportant to our analysis. Additionally, the lifetimes of especially low mass stars can exceed the SFH of M82 and, to a lesser extent, M87, making the end points of low-mass simulations even less important in our case.

The output of these MESA simulations are sets of radial profiles at many time steps along the stellar evolution. These profiles contain information about, for example, the temperature, density, and composition of the star, from which other quantities such as plasma frequency, $\omega_{pl}$, can be derived. These profiles allow us to compute the axion spectrum for a given time step by integrating the axion volume emissivity ({\it i.e.},~\eqref{eq:dLadE}) over the stellar interior.
Here, we show detailed results for a selection of representative stars illustrating the salient aspects of our MESA simulations. 

\begin{figure}[!htb]
\centering
\includegraphics[width=0.46\textwidth]{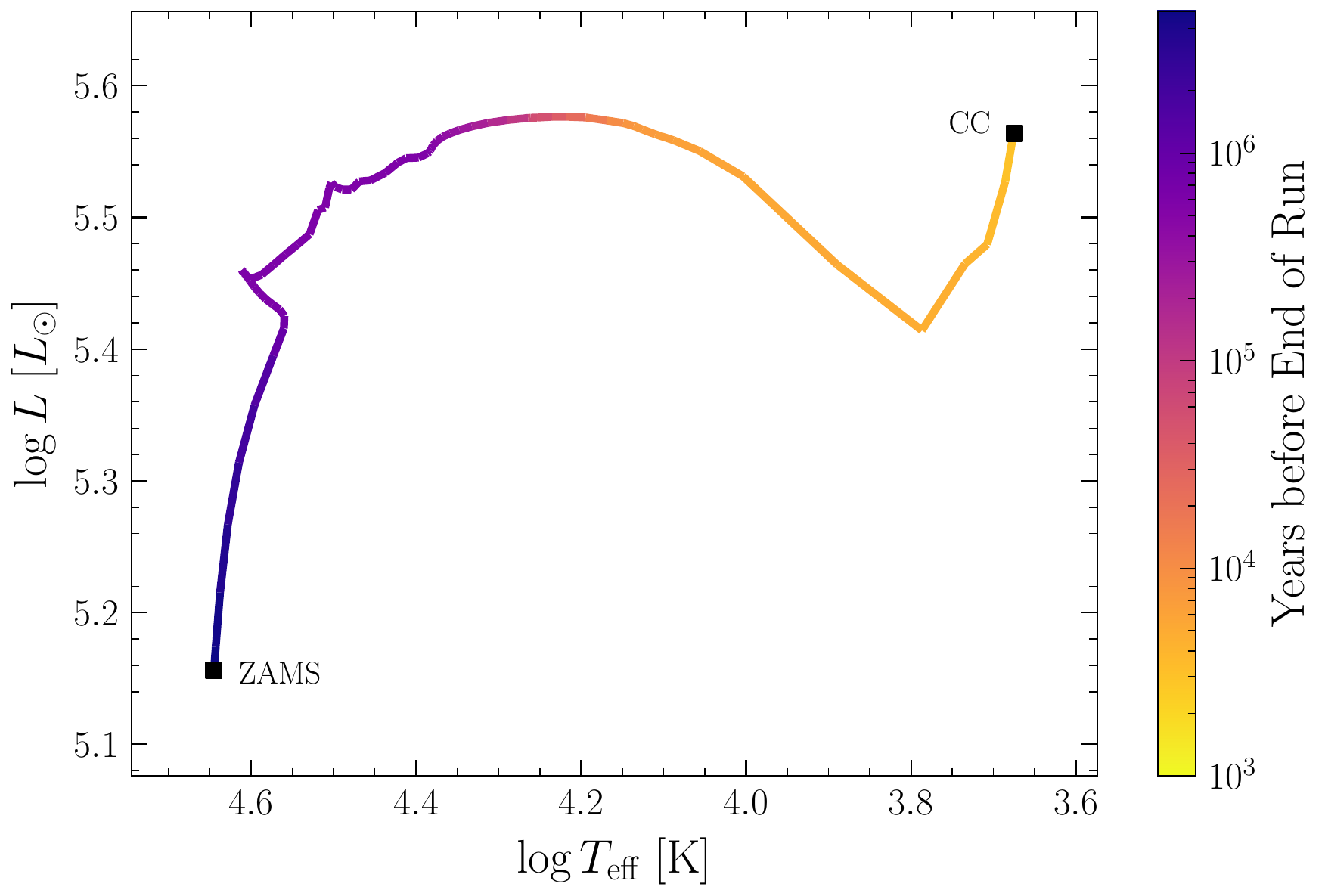}
\hspace{0.5cm}
\includegraphics[width=0.46\textwidth]{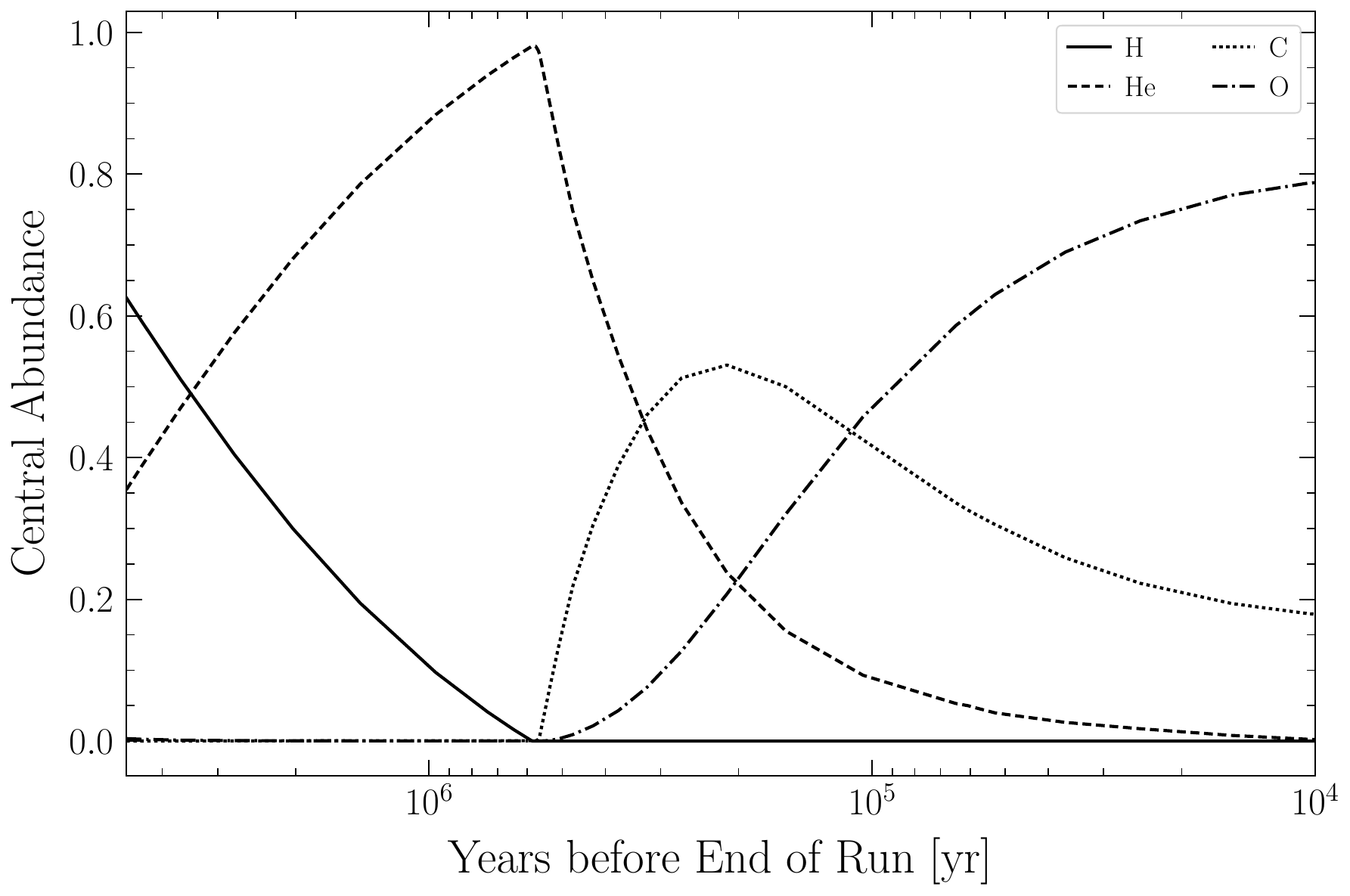}
\caption{(Left) The HR diagram for our representative high-mass ($30$ $M_{\odot}$) star from the beginning of its MS hydrogen-burning phase (zero-age MS, or ZAMS) all the way to the onset of core-collapse (CC) where the run ends. The coloring indicates the year before the end of the run (CC). (Right) The central abundances of hydrogen, helium, carbon, and oxygen for our representative high-mass star as a function of time. As described in the text, these elements illustrate important steps in the stellar evolution, such as the onset of helium and carbon burning in the stellar core. In our full simulations, heavier elements after oxygen also eventually fuse in the core as the star rapidly approaches core-collapse.}
\label{fig:MESA_HR_abund}
\end{figure}

\begin{figure}[!htb]
\centering
\includegraphics[width=0.45\textwidth]{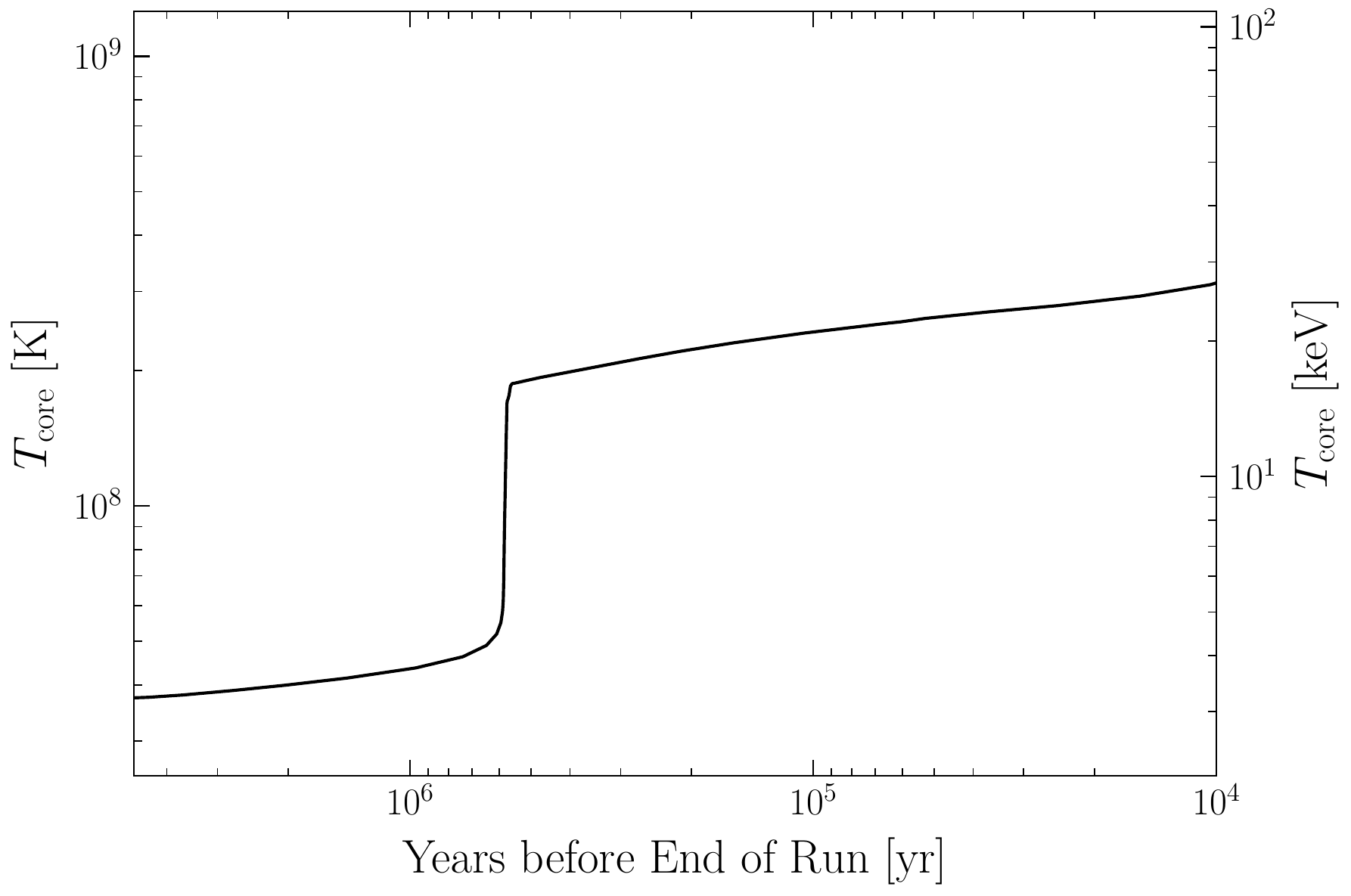}
\hspace{0.5cm}
\includegraphics[width=0.45\textwidth]{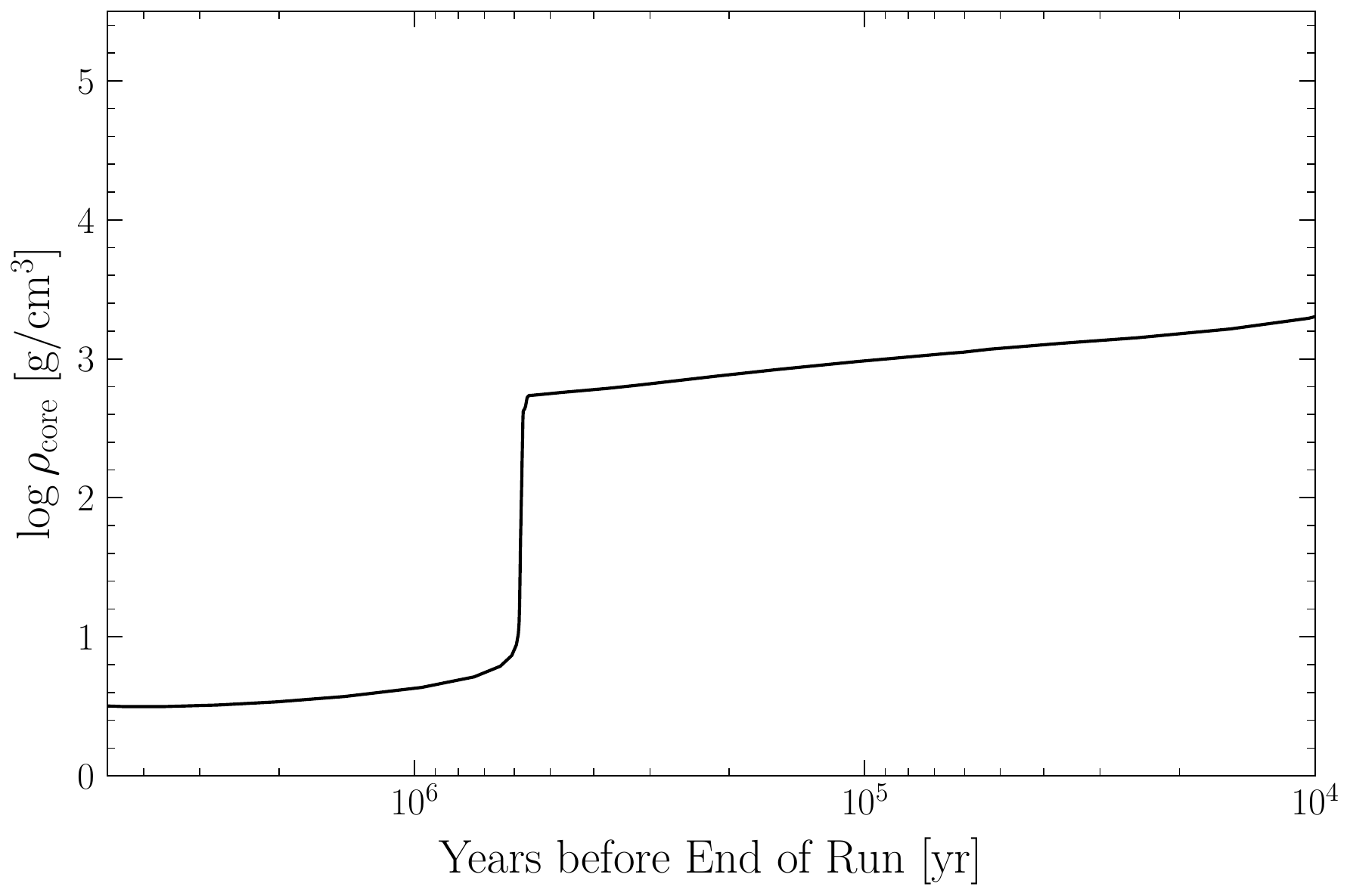}
\caption{(Left) The stellar core temperature as a function of time for the simulation described in Fig.~\ref{fig:MESA_HR_abund}. (Right) As in the left panel but for the stellar core density.}
\label{fig:MESA_T_rho}
\end{figure}

\begin{figure}[!htb]
\centering
\includegraphics[width=0.5\textwidth]{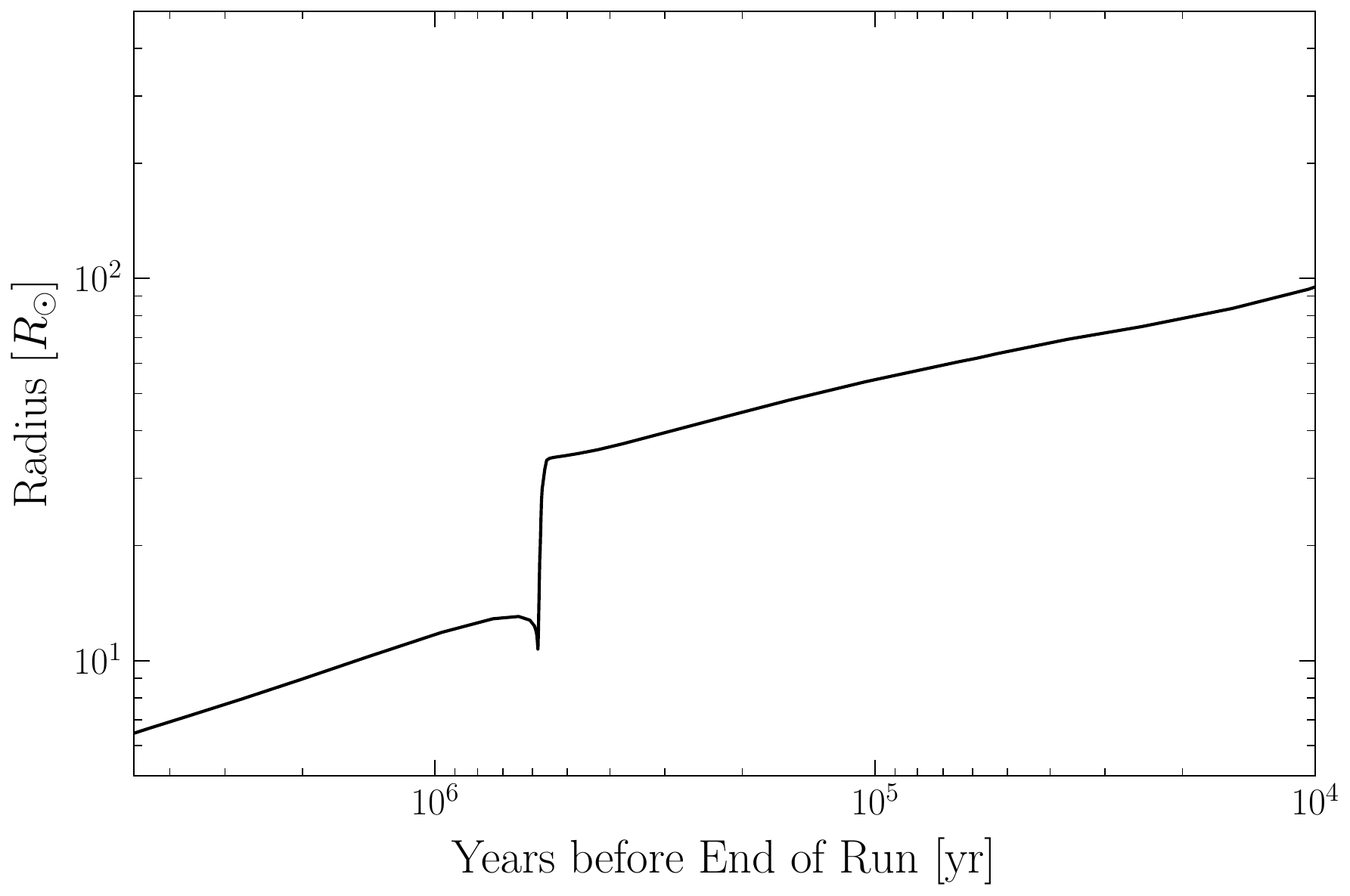}
\caption{The stellar radius as a function of time for the simulation described in Fig.~\ref{fig:MESA_HR_abund}.}
\label{fig:MESA_rad}
\end{figure}

\begin{figure}[!htb]
\centering
\includegraphics[width=0.45\textwidth]{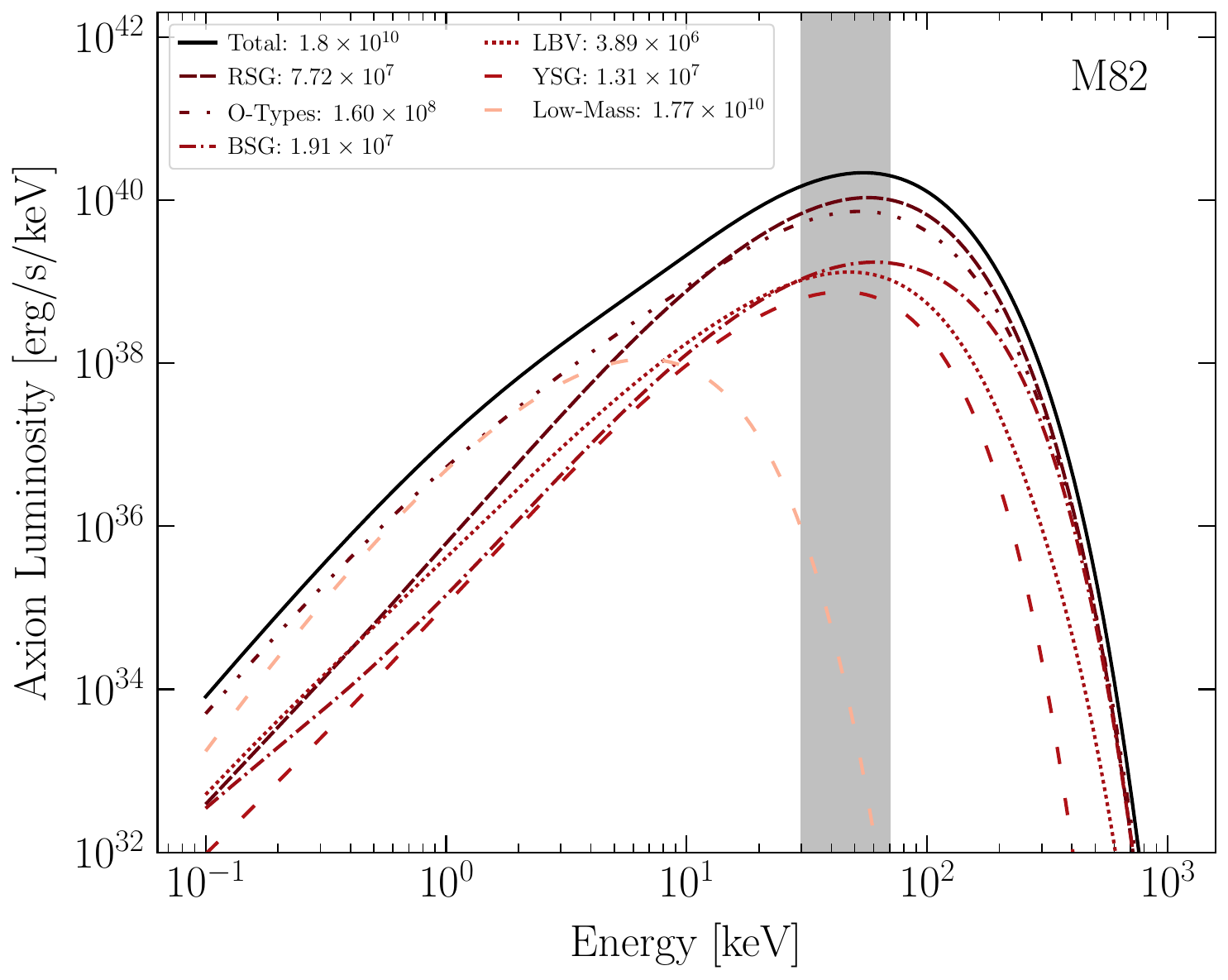}
\hspace{0.5cm}
\includegraphics[width=0.45\textwidth]{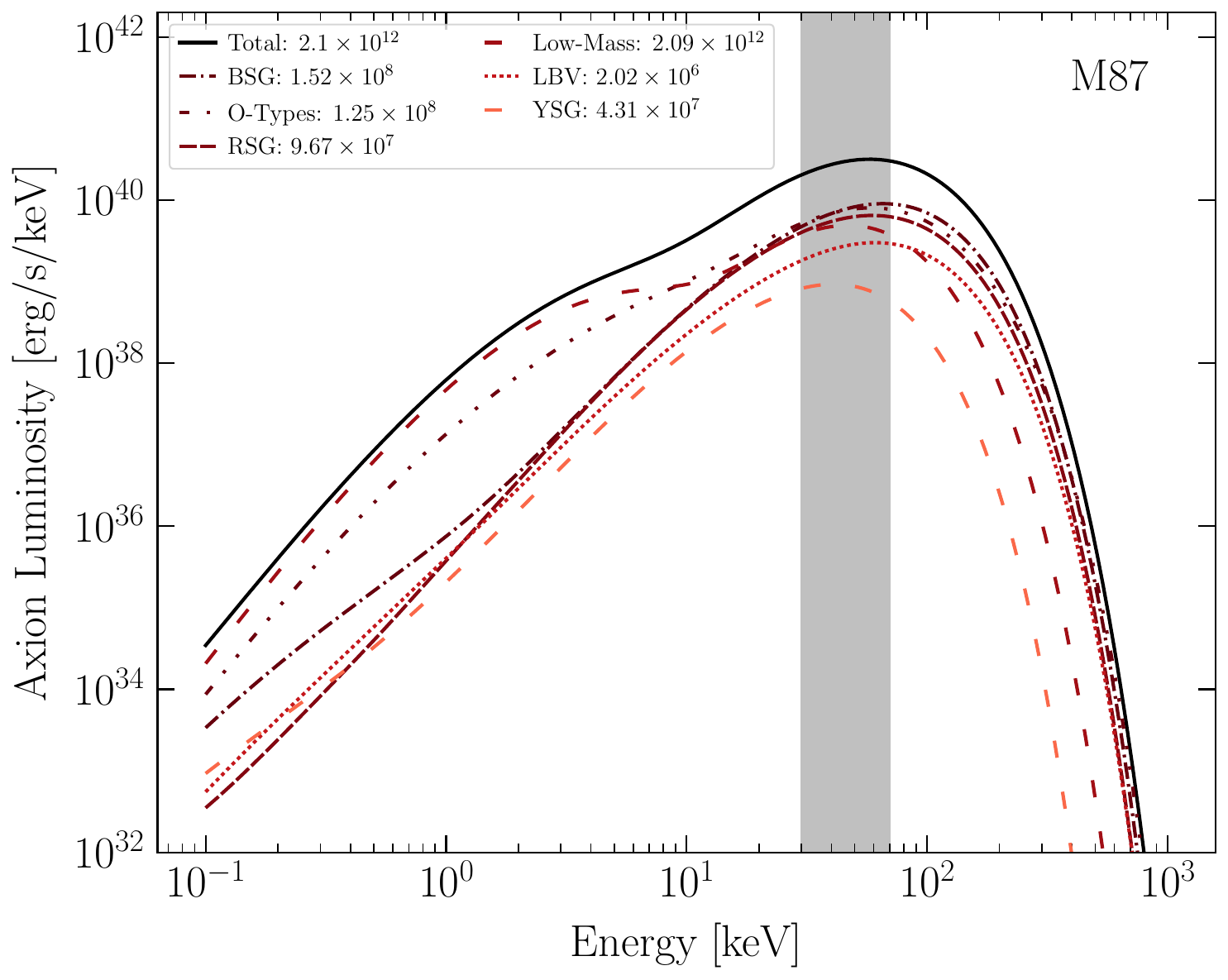}
\caption{(Left) The axion luminosity spectrum for a fixed $g_{a \gamma \gamma} = 1 \times 10^{-12}$ GeV$^{-1}$, over all the stars in M82, with breakdown by stellar classification. Our analysis region is shaded in gray, with approximate numbers of each stellar type in the legend. (Right) The same but for M87.}
\label{fig:signal_model_total}
\end{figure}

In Figs.~\ref{fig:MESA_HR_abund}-\ref{fig:MESA_rad} we first illustrate the time evolution of several quantities for a representative $30$ $M_{\odot}$ high-mass star. High-mass stars (which include supergiants, O-Types, and LBVs) are the most important type of stars in our overall analyses for both M82 and M87; their high temperatures and dense cores admit high axion production, and they clearly dominate the total axion luminosity spectrum, as seen in Fig.~\ref{fig:signal_model_total}. This is especially true in the case of M82, which, as a starburst galaxy, is host to large numbers of young, hot, massive stars. The core temperature, core density, radius, central abundances of key elements, and a Hertzsprung-Russell (HR) diagram for our representative high-mass star are shown in Figs.~\ref{fig:MESA_HR_abund}-\ref{fig:MESA_rad} as a function of stellar age. These figures tell the typical story of our high-mass stars; the star's life begins on the MS, where it initiates core hydrogen burning. When the core hydrogen fuel runs out, the star begins helium ignition to prevent core-collapse (see Fig.~\ref{fig:MESA_HR_abund}), and the core contracts to obtain the thermal energy needed to reach helium-burning temperatures (see Fig.~\ref{fig:MESA_T_rho}). When core helium is depleted, the star again contracts and begins burning carbon in its core. The star repeats this process, igniting and fusing heavier and heavier elements in its core, all the while increasing its radius as radiation pressure pushes the stellar atmospheric envelope outwards (see Fig.~\ref{fig:MESA_rad}). Eventually, the massive star ends its life with an iron core, at which point no more fusion can occur, and begins its collapse into a compact remnant. 

\begin{figure}
\begin{minipage}{.5\linewidth}
\centering
\includegraphics[scale=0.28]{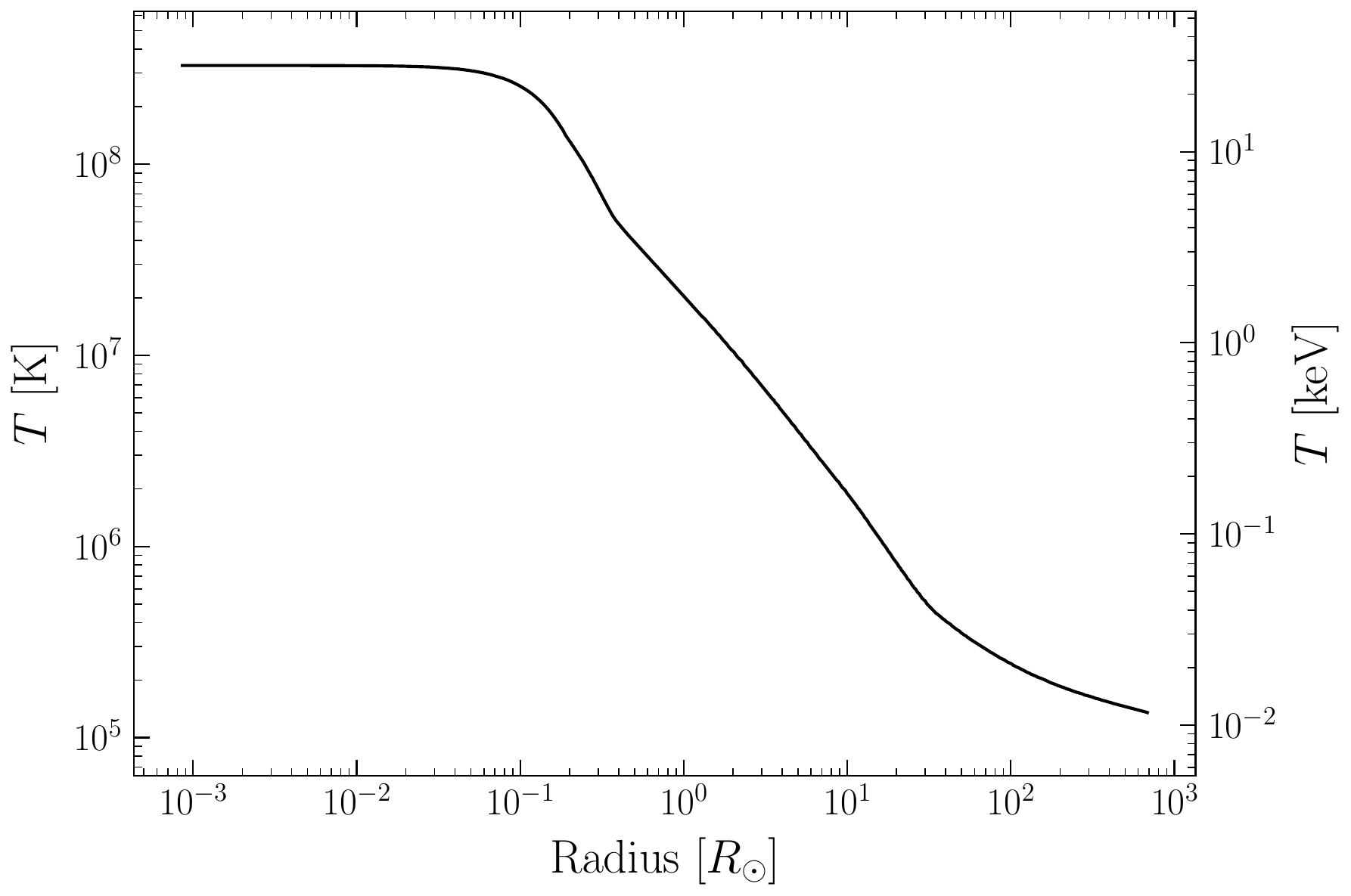}
\end{minipage}%
\begin{minipage}{.5\linewidth}
\centering
\includegraphics[scale=0.28]{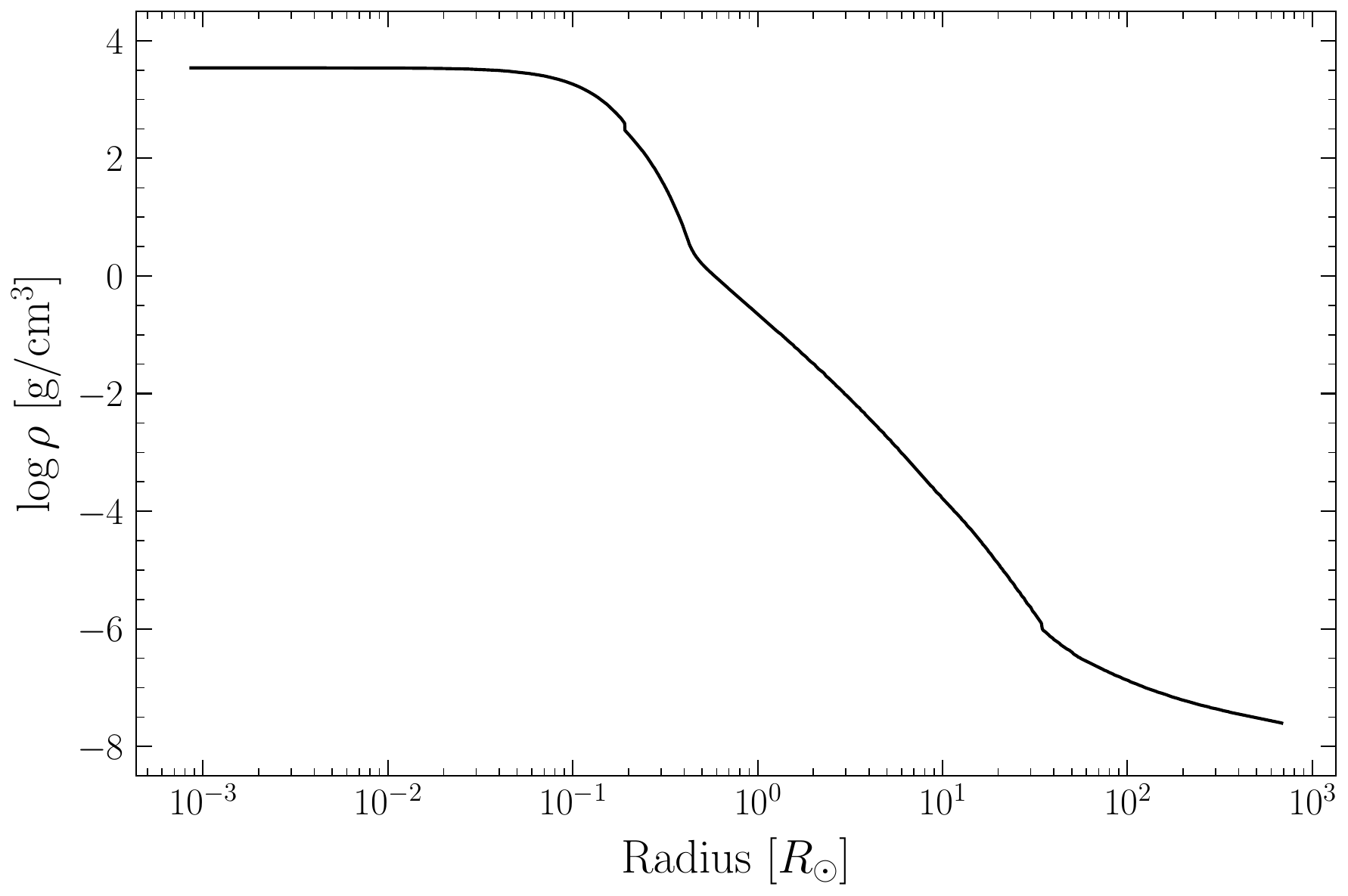}
\end{minipage}\par\medskip
\centering
\includegraphics[scale=0.28]{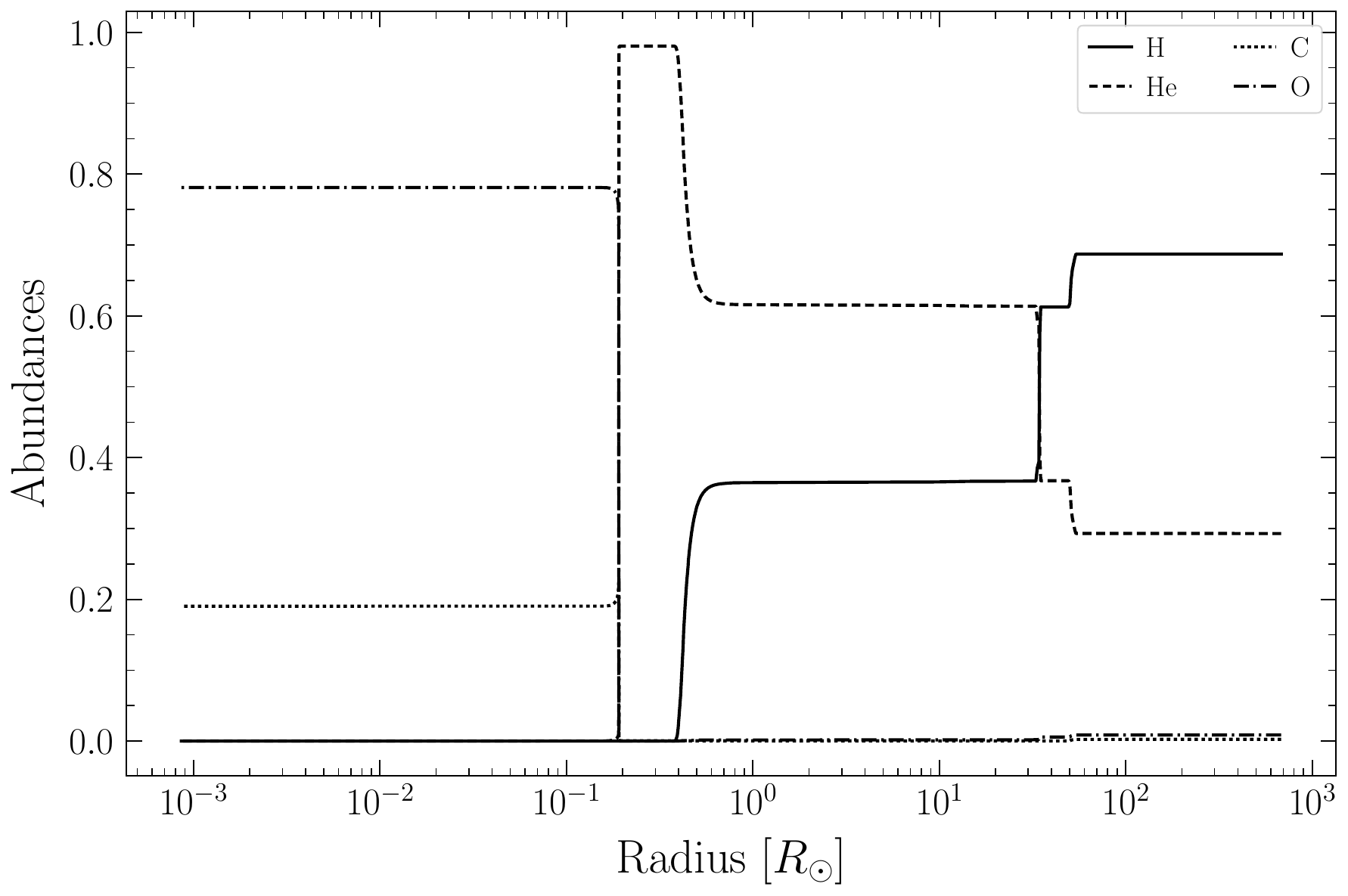}
\caption{Radial profiles over the interior of an initial $20$ $M_{\odot}$ star undergoing an RSG phase. The profiles are taken at an age of $\sim$$8.3$ Myr, which falls under the star's core carbon-burning phase. (Top Left) The temperature profile. (Top Right) The density profile. (Bottom) The abundance profile of hydrogen, helium, carbon, and oxygen. In this phase of the star's life, carbon is being burned in the core at high temperatures while the surrounding cooler helium and hydrogen envelopes continue to expand outwards.}
\label{fig:MESA_RSG}
\end{figure}

\begin{figure}
\begin{minipage}{.5\linewidth}
\centering
\includegraphics[scale=0.28]{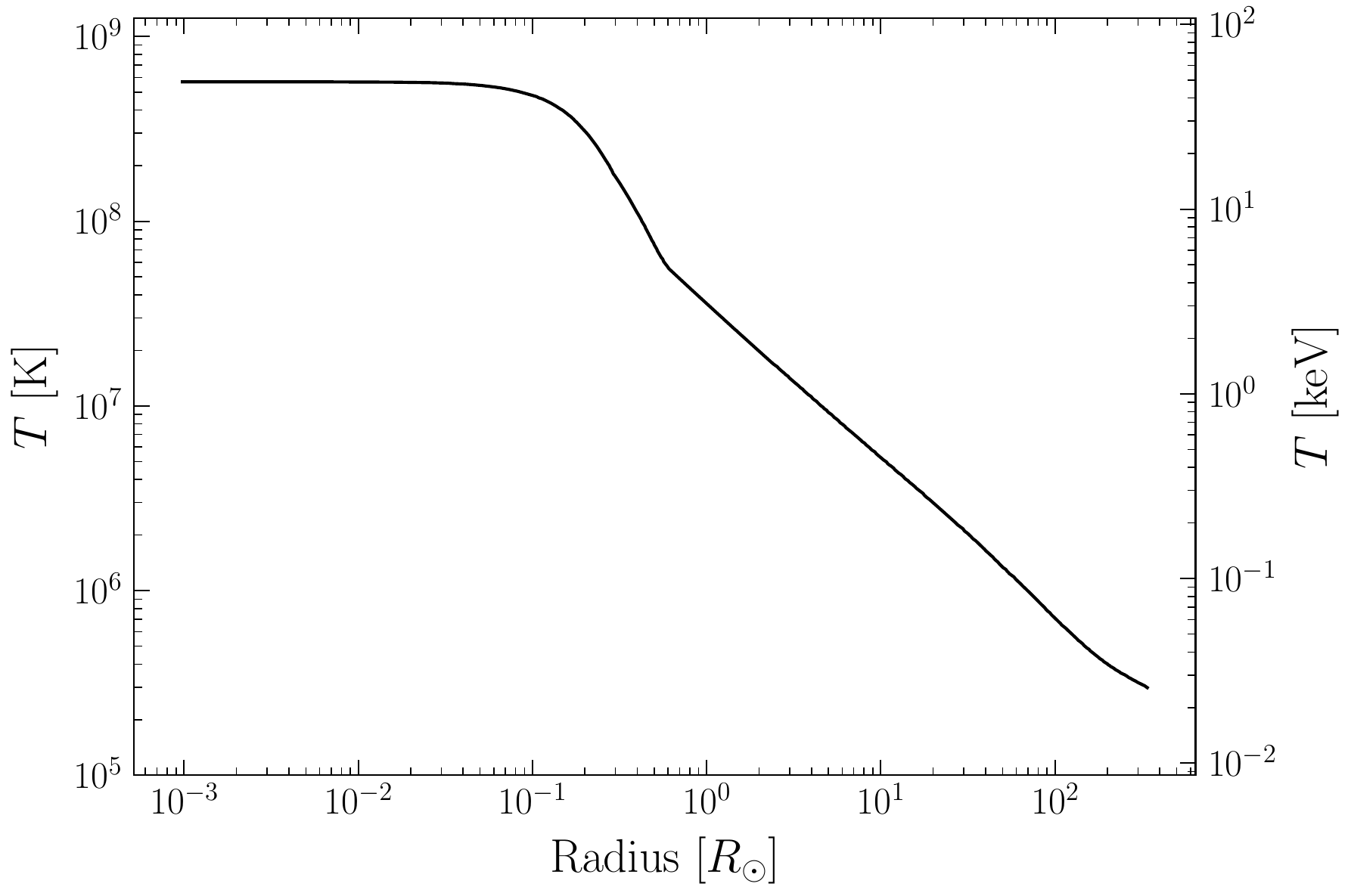}
\end{minipage}%
\begin{minipage}{.5\linewidth}
\centering
\includegraphics[scale=0.28]{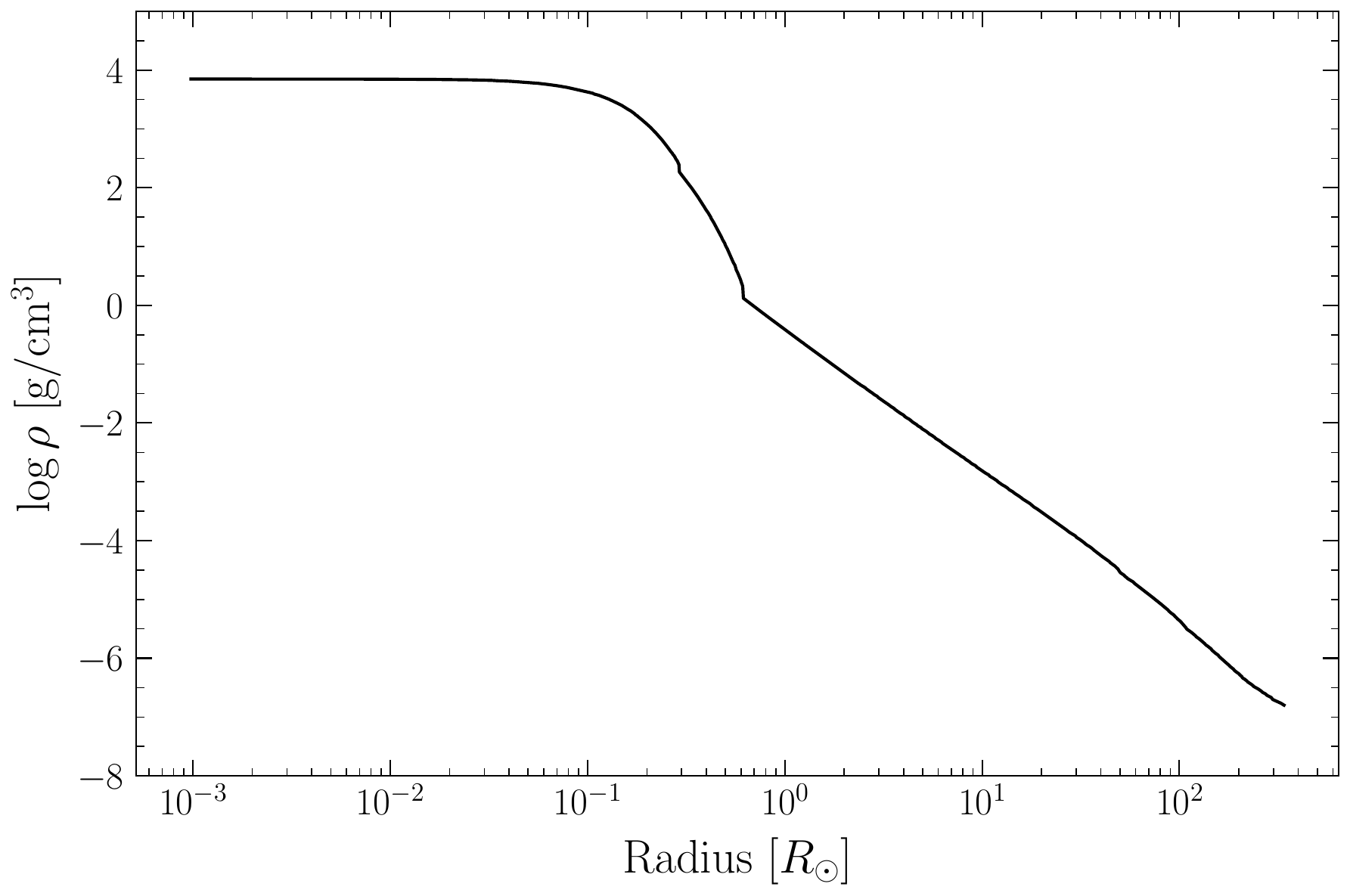}
\end{minipage}\par\medskip
\centering
\includegraphics[scale=0.28]{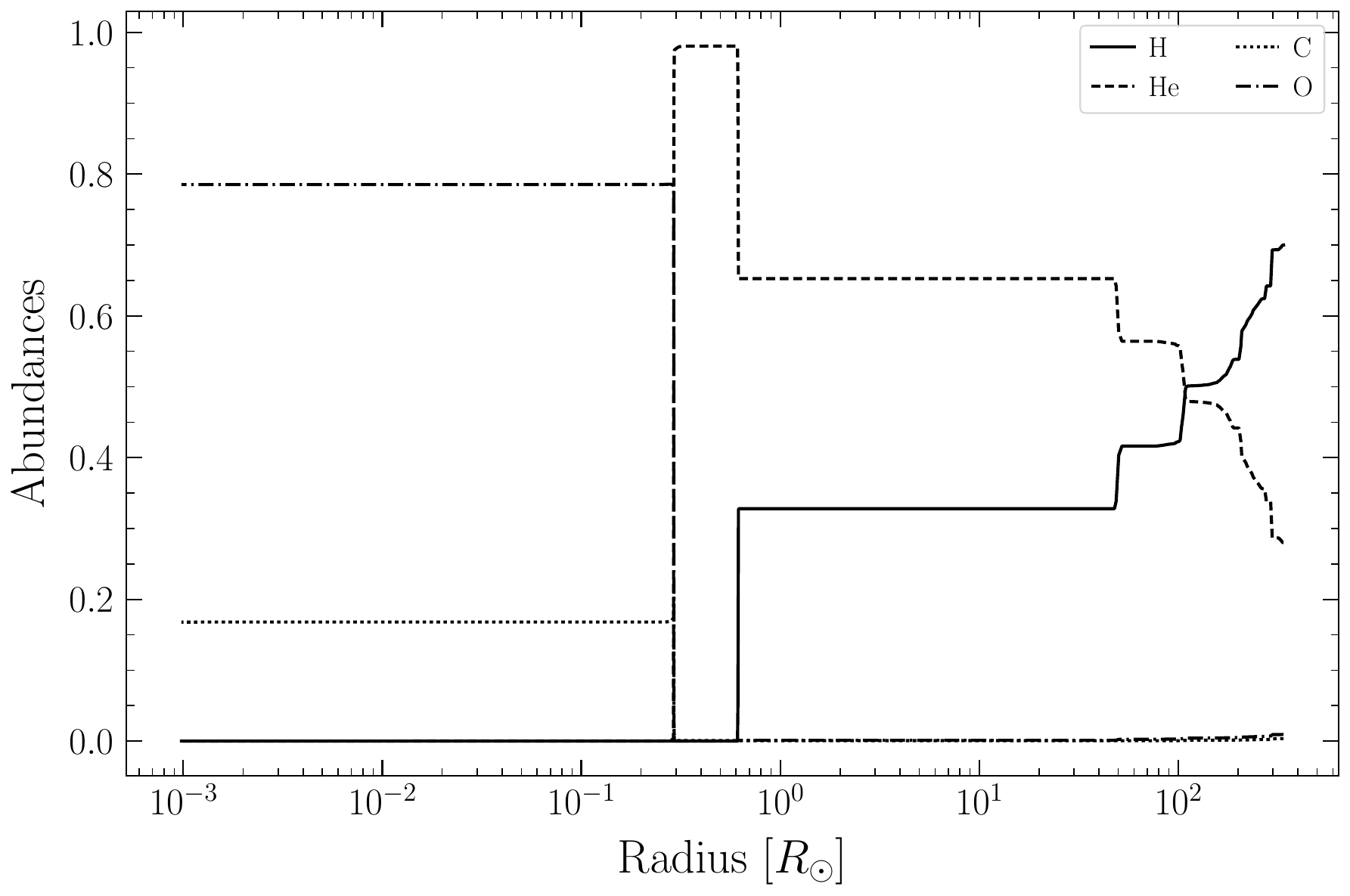}
\caption{Radial profiles over the interior of an initial $60$ $M_{\odot}$ star undergoing a BSG phase. The profiles are taken at an age of $\sim$$3.7$ Myr, which falls under the star's core carbon-burning phase. (Top Left) The temperature profile. (Top Right) The density profile. (Bottom) The abundance profile of hydrogen, helium, carbon, and oxygen. Compared to the RSG described in Fig.~\ref{fig:MESA_RSG}, also burning carbon in its core, we note our BSG shares essentially identical qualitative features. Generally, BSG stages of stars will have higher temperatures and densities than RSGs, while higher-mass stars in general will have increased mass loss from stellar winds, slightly affecting chemical abundances near the surface of the star (which can be seen here).}
\label{fig:MESA_BSG}
\end{figure}

\begin{figure}
\begin{minipage}{.5\linewidth}
\centering
\includegraphics[scale=0.28]{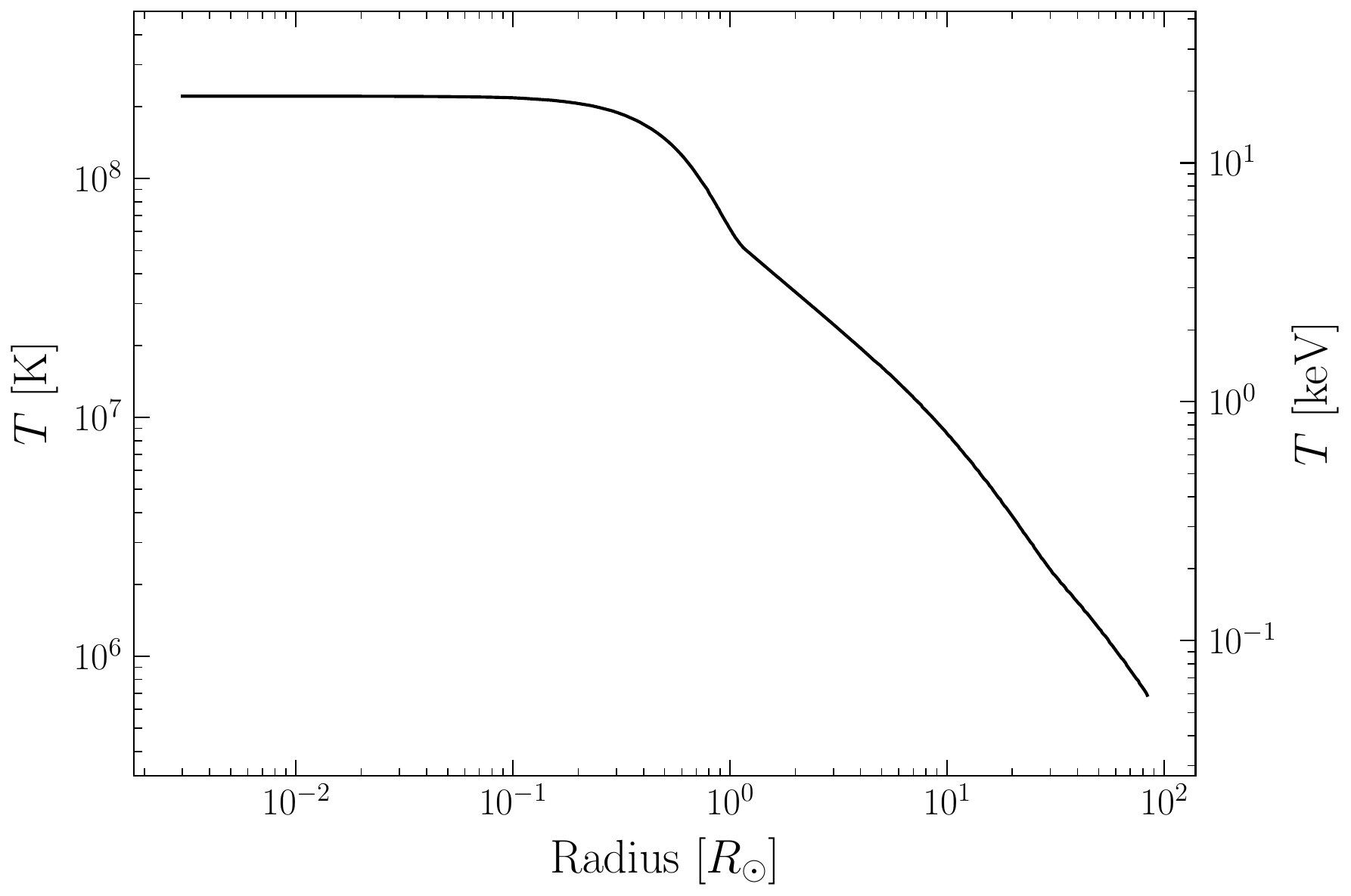}
\end{minipage}%
\begin{minipage}{.5\linewidth}
\centering
\includegraphics[scale=0.28]{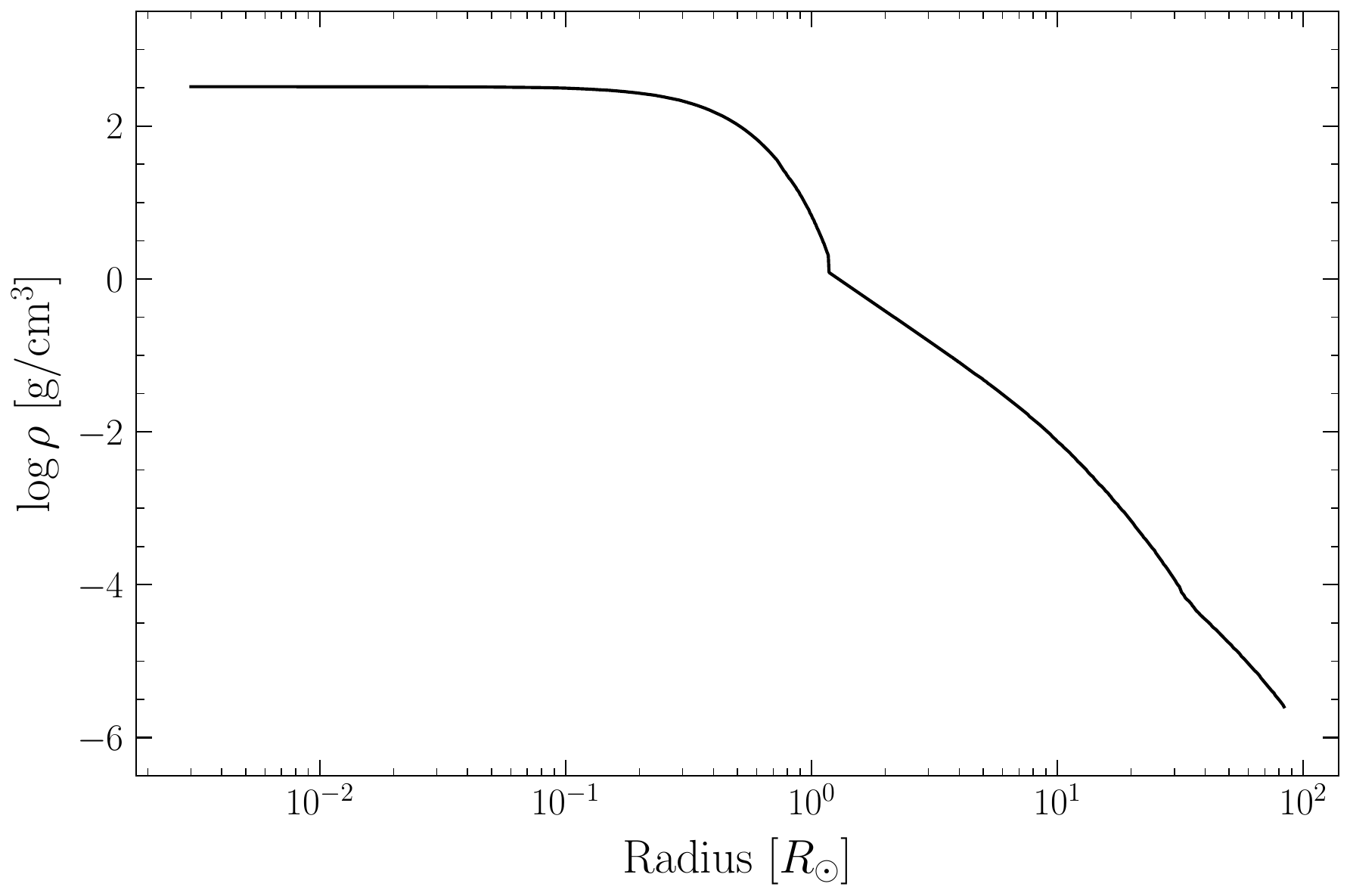}
\end{minipage}\par\medskip
\centering
\includegraphics[scale=0.28]{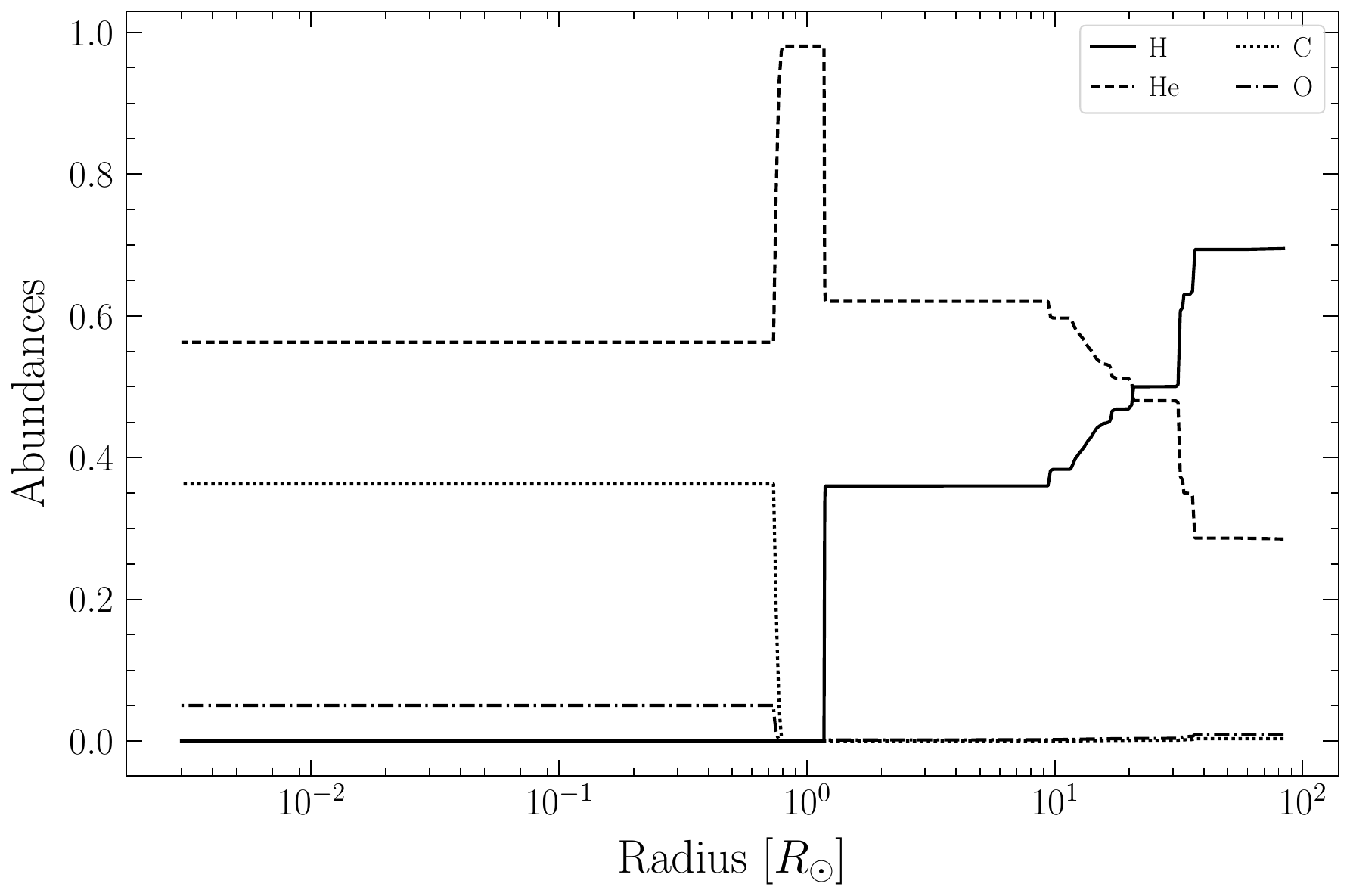}
\caption{Radial profiles over the interior of an initial $80$ $M_{\odot}$ O-type star after it has transitioned to its helium-burning phase (at an age of $\sim$$3.0$ Myr). (Top Left) The temperature profile. (Top Right) The density profile. (Bottom) The abundance profile of hydrogen, helium, carbon, and oxygen. This star will eventually develop into a supergiant, and already it possesses the same qualitative features as those stars in Figs.~\ref{fig:MESA_RSG} and \ref{fig:MESA_BSG}, with the only main difference being that the core helium is being burned in this stage, as opposed to carbon.}
\label{fig:MESA_O}
\end{figure}

\begin{figure}
\begin{minipage}{.5\linewidth}
\centering
\includegraphics[scale=0.28]{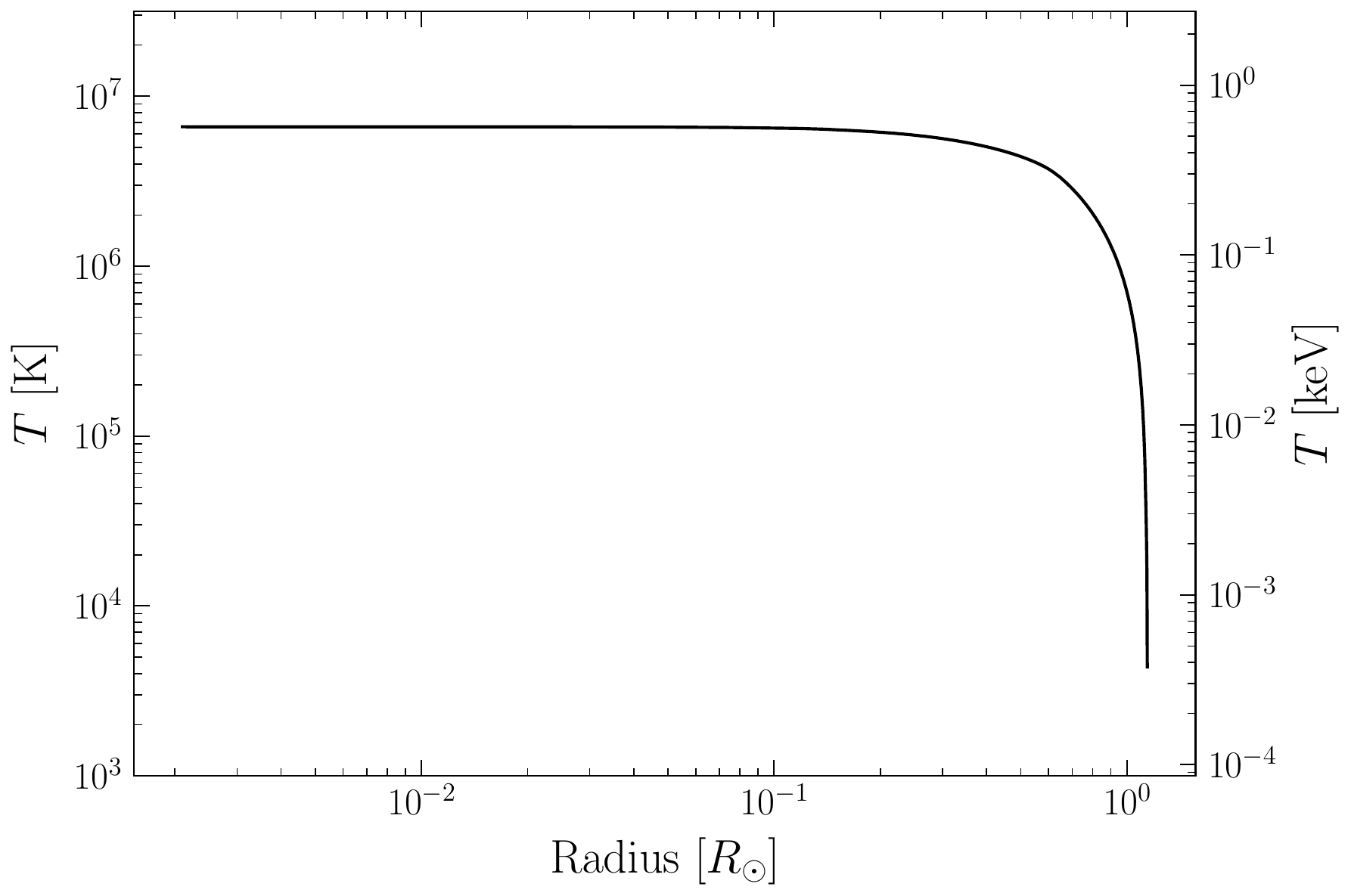}
\end{minipage}%
\begin{minipage}{.5\linewidth}
\centering
\includegraphics[scale=0.28]{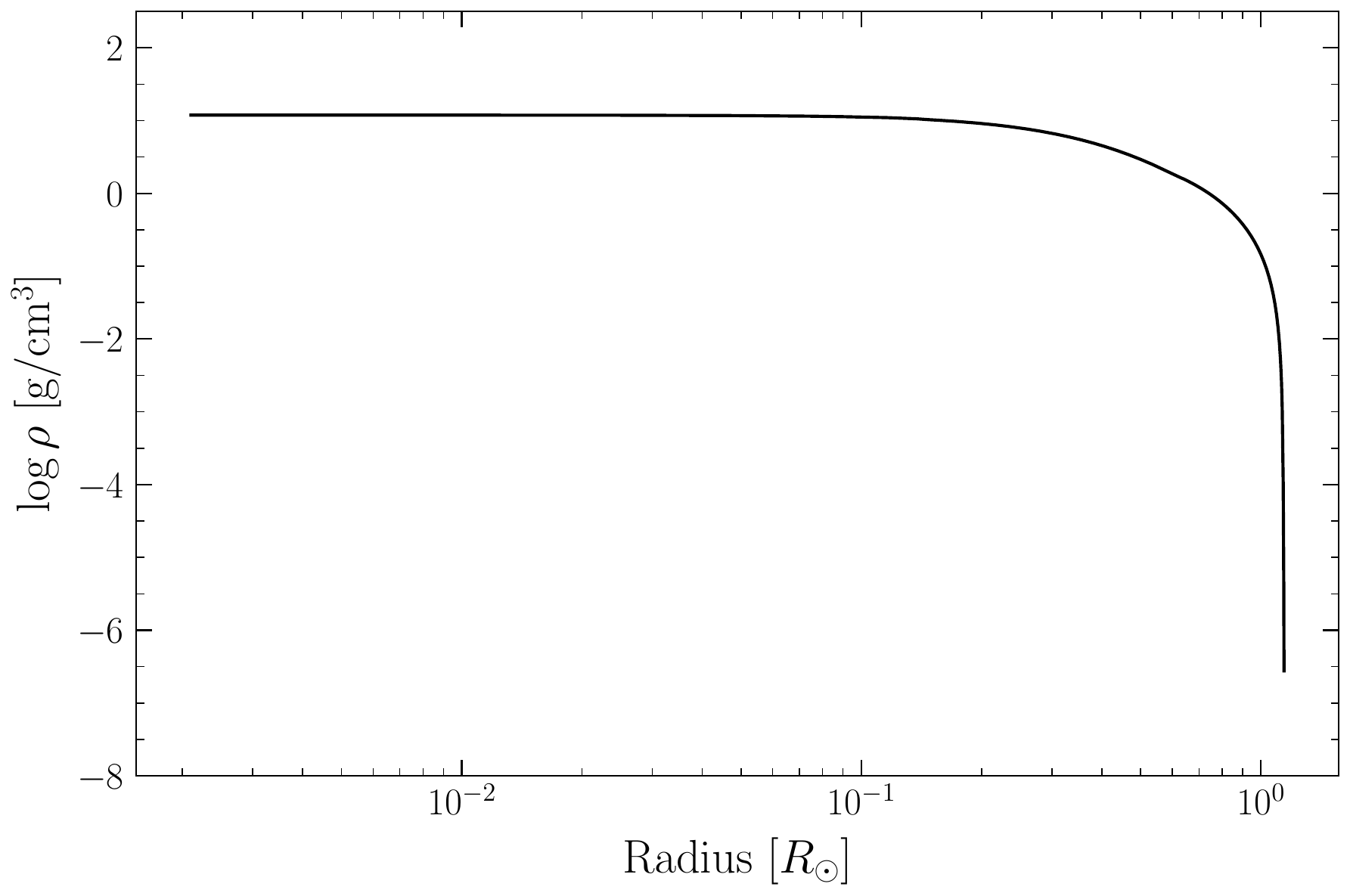}
\end{minipage}\par\medskip
\centering
\includegraphics[scale=0.28]{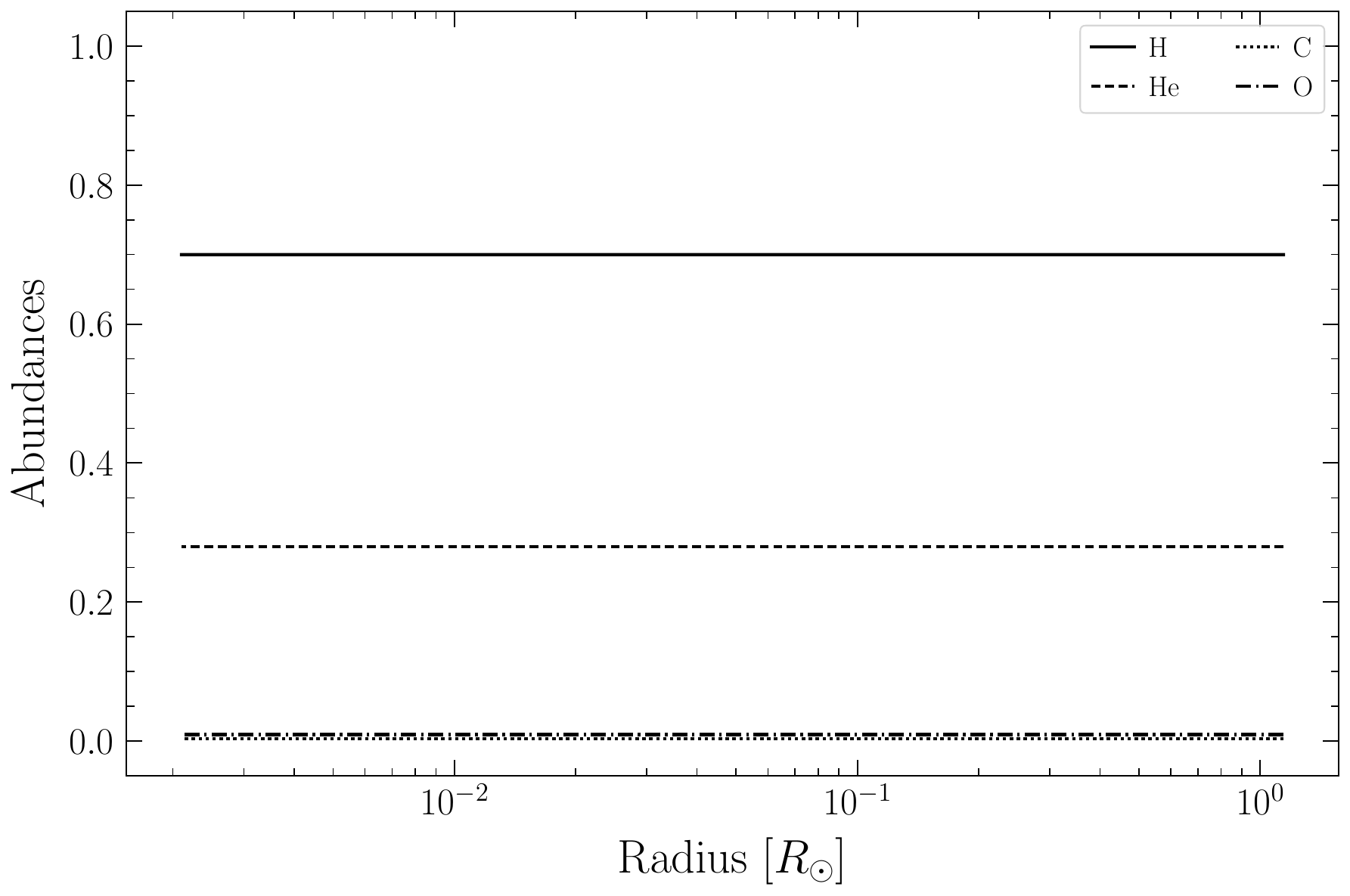}
\caption{Radial profiles over the interior of an initial $1$ $M_{\odot}$ star at an age of $\sim$$10$ Myr to represent a typical low-mass MS star in our simulations. (Top Left) The temperature profile. (Top Right) The density profile. (Bottom) The abundance profile of hydrogen, helium, carbon, and oxygen. Because this star sits squarely in its core hydrogen-burning MS phase and is low-mass, heavier elements than helium are not present, and the temperatures and densities are significantly lower than those of the supergiants.}
\label{fig:MESA_MS}
\end{figure}

\begin{figure}[!htb]
\centering
\includegraphics[width=0.45\textwidth]{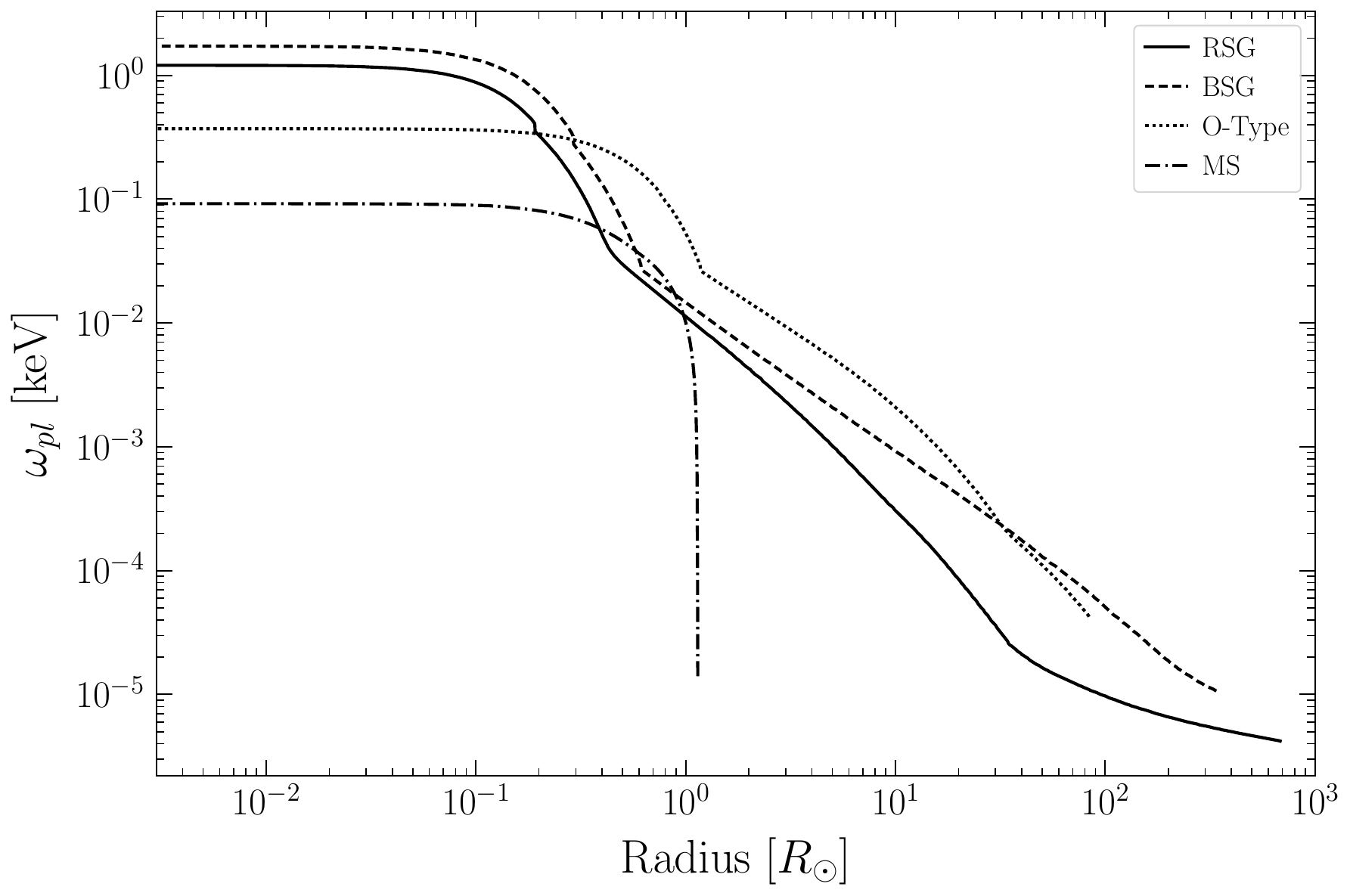}
\hspace{0.5cm}
\includegraphics[width=0.45\textwidth]{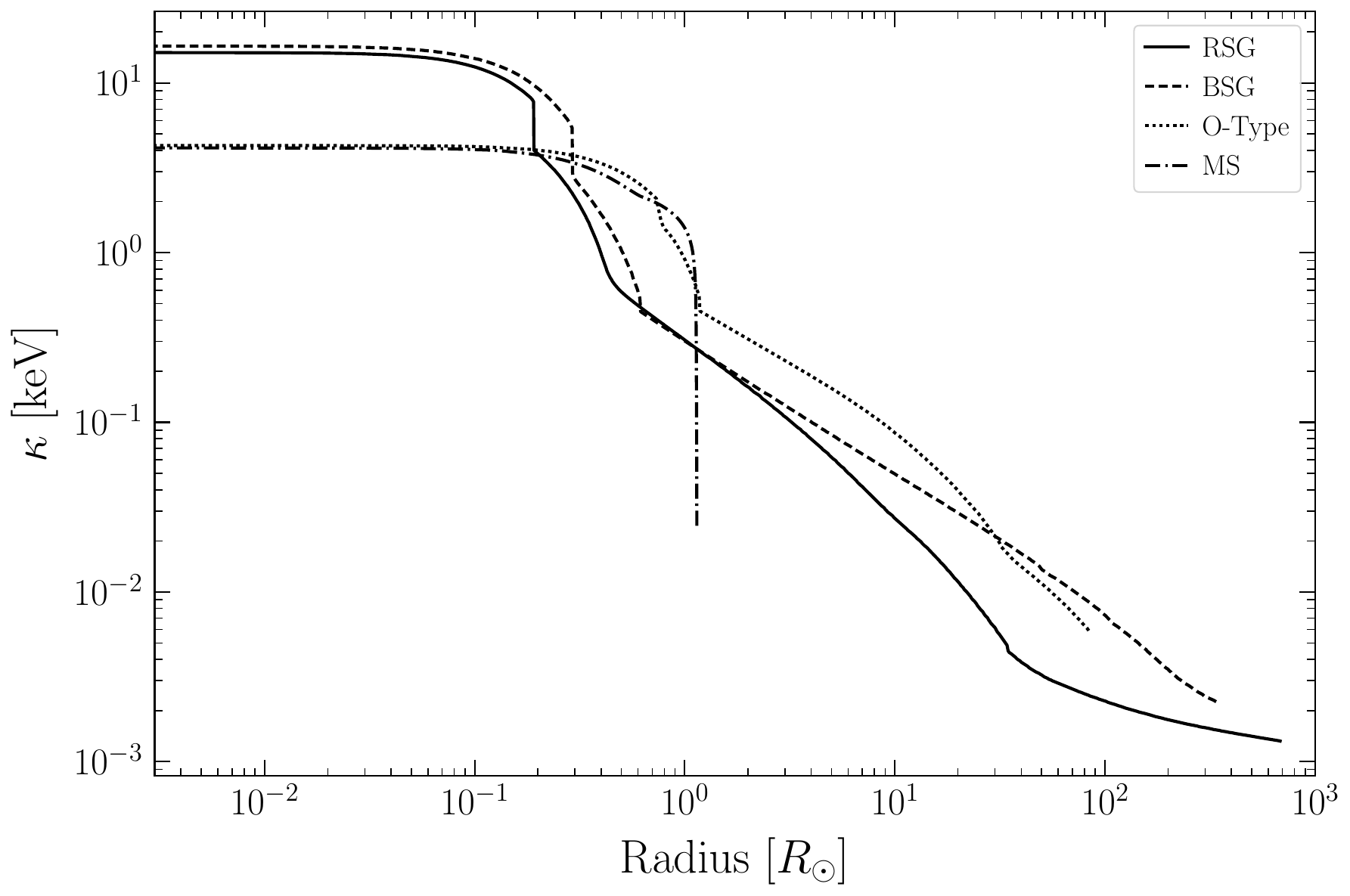}
\caption{Radial profiles over the interiors of the four stars discussed in Figs.~\ref{fig:MESA_RSG}-\ref{fig:MESA_MS} for $\omega_{pl}$, the plasma frequency (left), and $\kappa$, the Debye screening scale (right). Both of these are intermediate quantities involved in the calculation of the differential axion luminosity, given in~\eqref{eq:dLadE}.}
\label{fig:MESA_wpl_k}
\end{figure}

In Figs.~\ref{fig:MESA_RSG}-\ref{fig:MESA_wpl_k} we illustrate radial profiles at single time steps corresponding to examples of the three main stellar types that dominate the axion luminosity spectra for M82 and M87: RSGs, BSGs, and O-types. We additionally illustrate radial profiles for a typical low-mass MS star, as a point of comparison. We note that the examples are chosen at time steps which are away from important stellar evolution transition periods, such as the onset of helium or carbon ignition. The radial profiles are directly used in the computation of the axion spectrum. In Figs.~\ref{fig:MESA_RSG}-\ref{fig:MESA_MS} we show, for each example class, how the temperature, density, and chemical abundances vary over the star, and in Fig.~\ref{fig:MESA_wpl_k} we compare $\omega_{\rm pl}$, the plasma frequency, and $\kappa$, the Deybe screening scale, across each class; all of these quantities are crucial for the computation of the axion luminosity spectrum through~\eqref{eq:dLadE}. From these profiles one can see that generally the high-mass classes of stars  possess significantly higher temperatures and densities compared to our low-mass MS stars, illuminating the origin for why high-mass stars are so dominant in their contribution towards our final total axion luminosity spectra, Fig.~\ref{fig:signal_model_total}.

We make a final remark that future work can refine more detailed aspects of these simulations. For example, we do not consider detailed modeling of surface rotations in our simulated stars.  Very little is known about the surface rotation velocity distribution of stars in M82 and M87, and while rotation can become important in certain stars such as Wolf-Rayets (WR), their estimated axion emission~\cite{Dessert:2020lil} and relative rarity imply that they will likely make a small adjustment to the total axion luminosity. Likewise, we also do not consider detailed modeling of stars in binary systems, although this could potentially be an interesting avenue of further study for axion searches.

\section{Stellar Population Modeling}

Given the procedure for simulating a single star in MESA, we now turn to our stellar population synthesis method for all the estimated stars in M82 and M87. (See~\cite{Brocato:1997tu} for a related analysis that instead predicts the neutrino luminosity from stellar populations in galaxies instead of the axion luminosity.) We first give a brief overview of our method, and then later we go over each specific stellar population modeling ingredient and discuss associated uncertainties with each. 

As mentioned in the main Letter, we adopt specific IMFs and SFHs for both M82 and M87 and derive a total number of stars $N_{\rm tot}$ for each galaxy. From these choices, we draw from our distribution of MESA simulations $N_{\rm tot}$ times according to the probabilistic weights given by the IMF and SFH. For each MESA simulation draw, we compute the axion luminosity emitted by that simulated star at that age using the corresponding MESA stellar profiles. The total axion luminosity emitted from a galaxy, then, is the sum of all of these draws.

\begin{figure}[!htb]
\centering
\includegraphics[width=0.45\textwidth]{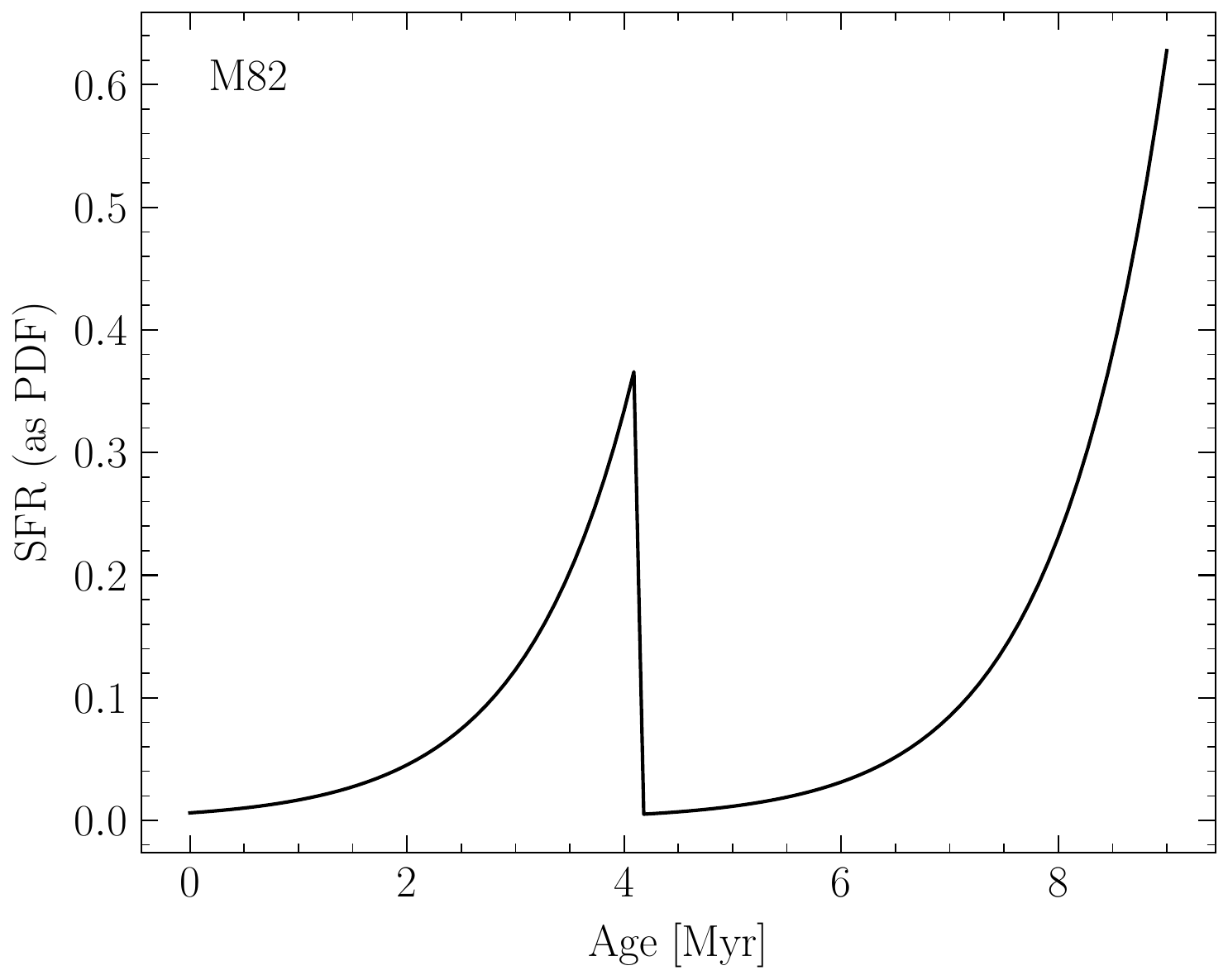}
\hspace{0.5cm}
\includegraphics[width=0.45\textwidth]{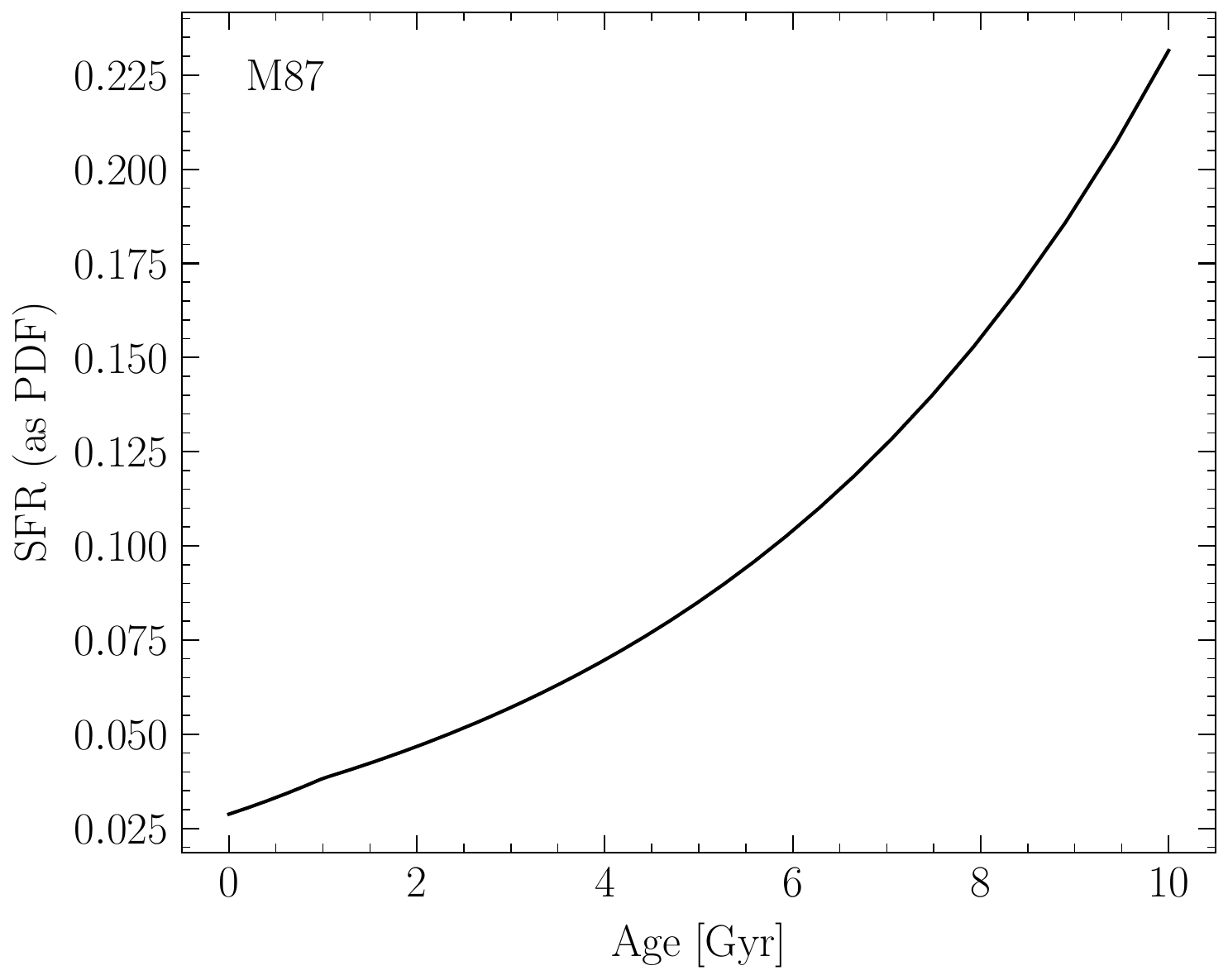}
\caption{(Left) Simple `two-burst' star formation history (SFH) model for M82. (Right) Simple double $\tau$ exponential SFH model for M87.}
\label{fig:sfh}
\end{figure}

{\bf SFH.---} The SFH dictates the distribution of stars in a galaxy according to their ages. The SFHs that we use for our fiducial models of both M82 and M87 are presented in Fig.~\ref{fig:sfh}. In the case of M82, we follow~\cite{ForsterSchreiber:2003ft} and describe the star formation history as primarily a `two-burst' model, with each burst modeled as (for $t < t_{\rm burst}$) $R_0 e^{-(t_{\rm burst} - t)/t_{\rm sc}}$ where $t_{\rm sc}$ is the characteristic decay time scale, given as $t_{\rm sc} = 1.0$ Myr. For the old burst, the complete constraints on $t_{\rm burst}$ and $R_0$ are given as $t_{\rm burst} = 9.0^{+0.2}_{-0.2}$ Myr and $R_0 = 31^{+7}_{-8}$ $M_\odot$/yr, while for the younger burst $t_{\rm burst} = 4.1^{+0.5}_{-0.7}$ and $R_0 = 18^{+9}_{-8}$ $M_\odot$/yr. As mentioned in the main Letter, for our fiducial model we use the quoted values of these parameters, and we also show in Fig.~\ref{fig:M82_systematics} that scanning over the uncertainties in all four of these parameters does not significantly affect our final limits. For M87, we follow~\cite{Cook:2020} and model the SFH as a combination of two exponential $\tau$ models: an exponential $\tau$ model with $\tau = 5$ Gyr for ages $t \geq 1$ Gyr, and another exponential $\tau$ model with $\tau = 3.5$ Gyr for ages $t < 1$ Gyr. Since the SFR below $1$ Gyr is relatively poorly constrained, we account for this uncertainty by examining reasonable ranges of $\tau$ in this age range based on~\cite{Cook:2020}, settling on the prescription where below $1$ Gyr we vary $2 \leq \tau \leq 5$ and show in Fig.~\ref{fig:M87_systematics} that our final limits are relatively unaffected by these bounds.

\begin{figure}[!htb]
\centering
\includegraphics[width=0.45\textwidth]{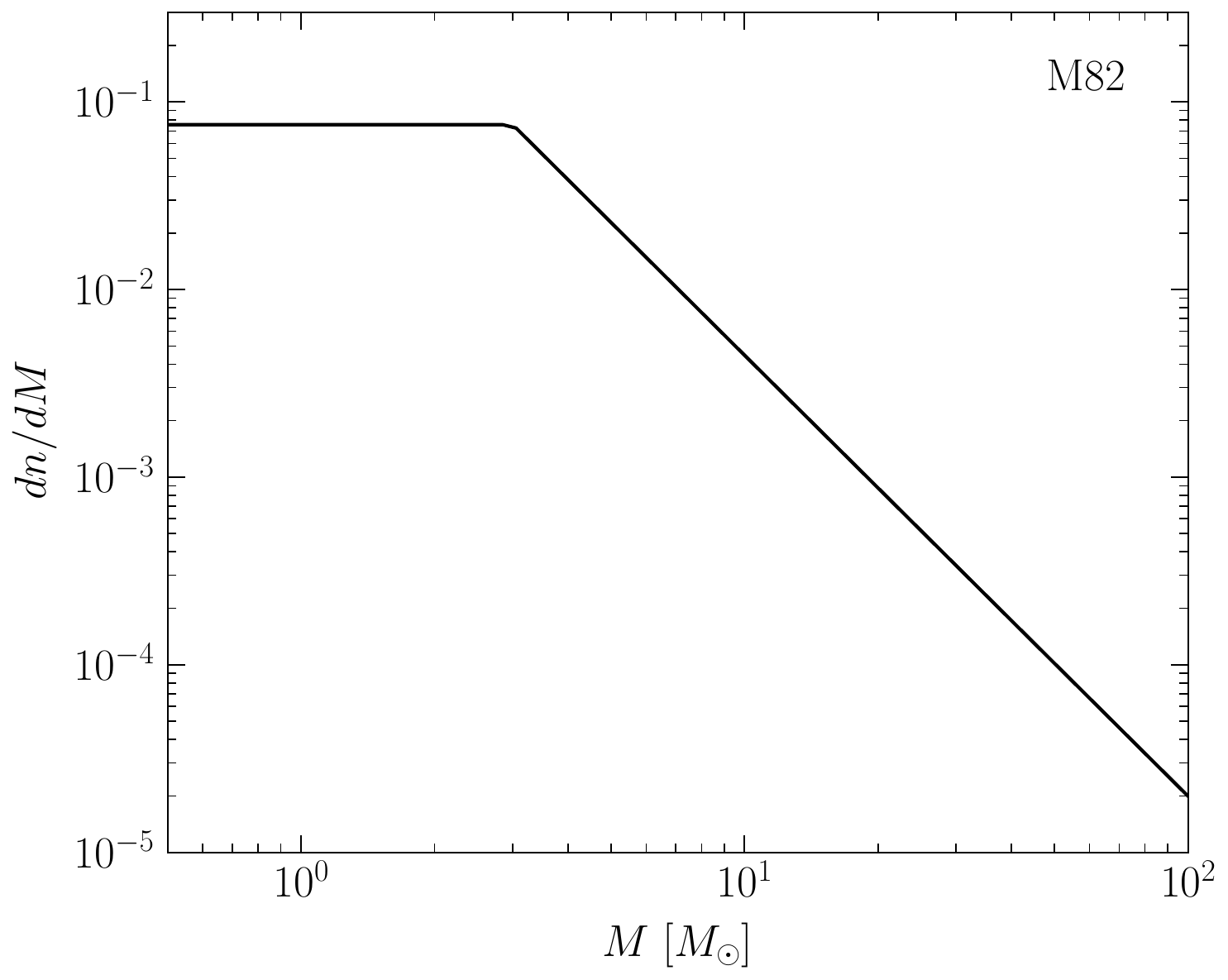}
\hspace{0.5cm}
\includegraphics[width=0.45\textwidth]{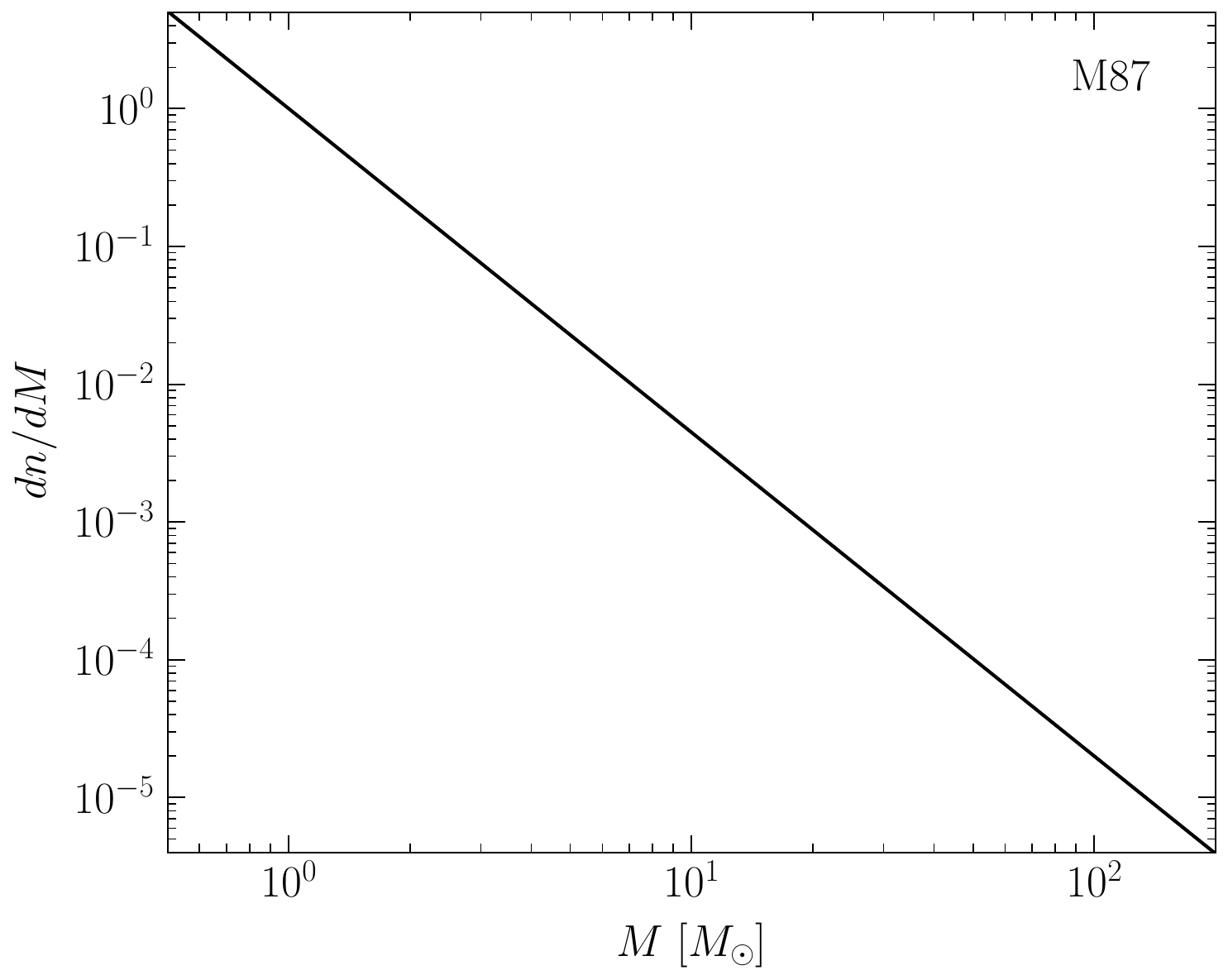}
\caption{(Left) Modified Salpeter IMF used for the M82 analysis. It is a Salpeter IMF with a slope of $\alpha = 2.35$, but flattened below $3 M_{\odot}$. (Right) The traditional Salpeter IMF, used for the M87 analysis.}
\label{fig:imf}
\end{figure}

{\bf IMF.---} The IMF dictates the distribution of stars in a galaxy according to their masses. The IMFs that we use for our fiducial models of both M82 and M87 are presented in Fig.~\ref{fig:imf}. In the case of M82, we follow~\cite{ForsterSchreiber:2003ft}, who compare population synthesis models with infrared spectroscopy data to determine that the IMF behaves as $dn/dM \propto 1/M^{2.35}$ at high masses and significantly `flattens out' below $\sim$3 $M_{\odot}$, with a high-mass cut-off of $100$ $M_{\odot}$. The stars most relevant to axion emission in this work are high-mass stars, and so we test the robustness of our final limits by showing that even with a high-mass cut-off of $200$ $M_{\odot}$, the resulting limit is still similar to our fiducial limit (see Fig.~\ref{fig:M82_systematics}). In the case of M87, as pointed out in the main Letter, the IMF is more difficult to constrain due to M87's complicated history of merger assembly~\cite{Montes:2014}. With that in mind, we follow the simple Salpeter IMF model ($dn/dM \propto 1/M^{2.35}$) proposed for M87 in~\cite{Cook:2020}, who compare photometric data of M87 to stellar evolution models using the pixel color-magnitude diagram (pCMD) framework. As was the case for M82, the final limits are not signficantly affected by variations on the high-mass cut-off part of the IMF (see Fig.~\ref{fig:M87_systematics}).

{\bf Number of Stars.---} As discussed in the main Letter, we roughly estimate the number of stars in M82 using luminosity observations and color-mass-to-light ratio relations for disk galaxies. We extract the color-mass-to-light ratio relations from~\cite{McGaugh:2014}, which provide robust and updated relations that ensures self-consistency of estimated stellar masses across different luminosity bands. We extract the $B - V$ color of M82 from the NASA/IPAC Extragalactic Database (NED)\footnote{\url{http://ned.ipac.caltech.edu/}} and use the relations in~\cite{McGaugh:2014} for $B- V$ color to extract a stellar mass-to-light ratio in the $I$ band. We use this ratio as well as the recorded $I$ band luminosity from NED to derive that the total stellar mass of M82 is given by approximately $1.49 \times 10^{10}$ $M_{\odot}$, with uncertainties ranging from $1.16 \times 10^{10}$ to $1.93 \times 10^{10}$ $M_{\odot}$ (and which agrees, within uncertainties, with~\cite{Oehm:2017}). Combining this with our discussion of the IMF above, this translates to roughly $1.8 \times 10^{10}$ stars in M82, with uncertainties ranging from around $1.4 \times 10^{10}$ to $2.2 \times 10^{10}$ stars. As mentioned in the main Letter, for our fiducial model we assume $1.8 \times 10^{10}$ stars, and we show in Fig.~\ref{fig:M82_systematics} that the derived range of uncertainties around this number does not significantly affect our final results. For M87, we deduce the number of stars using the stellar mass inferred in~\cite{DeLaurentis:2022} from photometry measurements, given by $(1.3 \pm 0.3) \times 10^{12}$ $M_{\odot}$. Combined with our discussion of the IMF above, this translates to roughly $2.1 \times 10^{12}$ stars in M87, with uncertainties ranging from around $1.6 \times 10^{12}$ to $2.5 \times 10^{12}$ stars. We use $2.1 \times 10^{12}$ in our fiducial model, and show in Fig.~\ref{fig:M87_systematics} that our final results are not significantly affected by the uncertainties in stellar number.

\begin{figure}[!htb]
\centering
\includegraphics[width=0.45\textwidth]{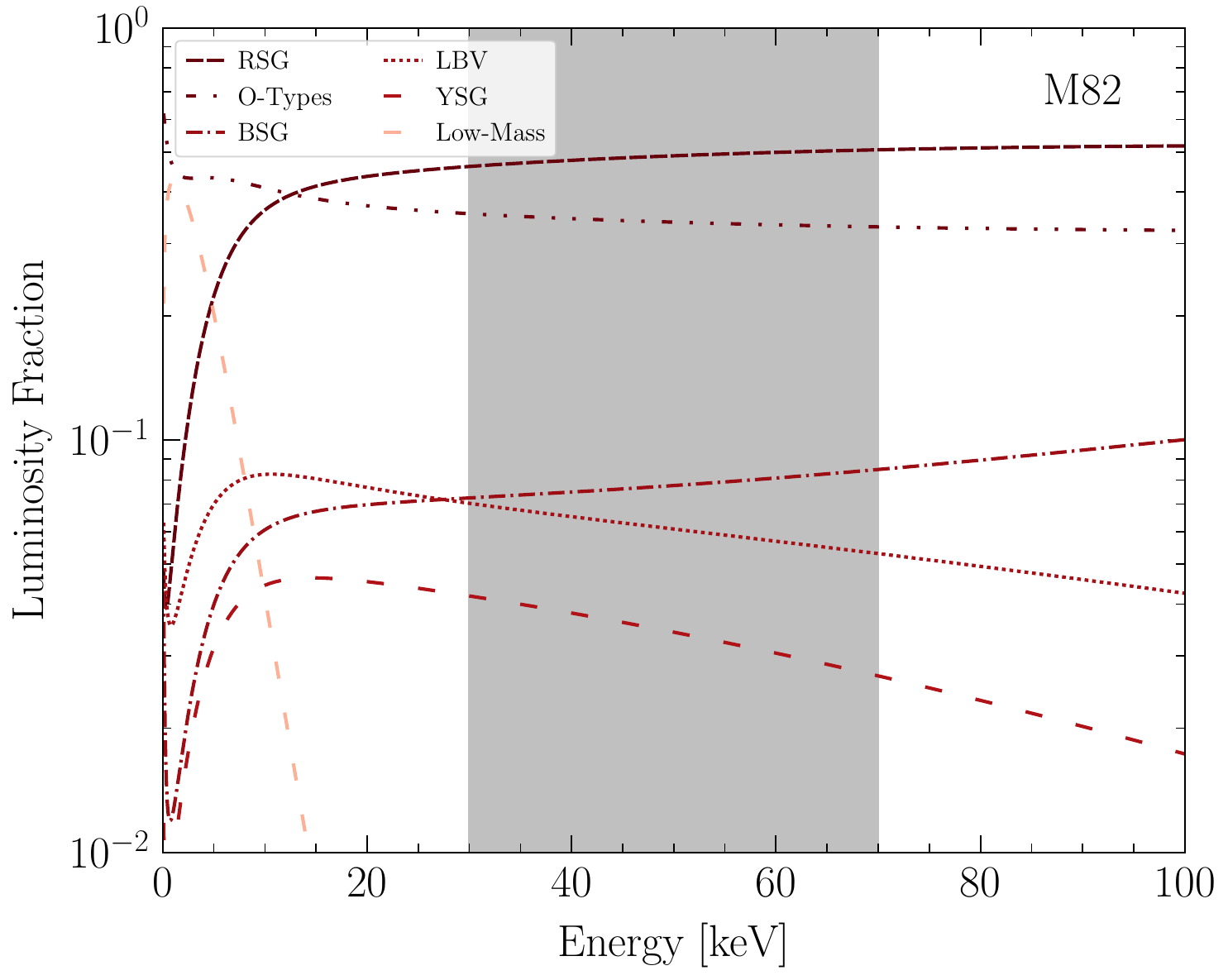}
\hspace{0.5cm}
\includegraphics[width=0.45\textwidth]{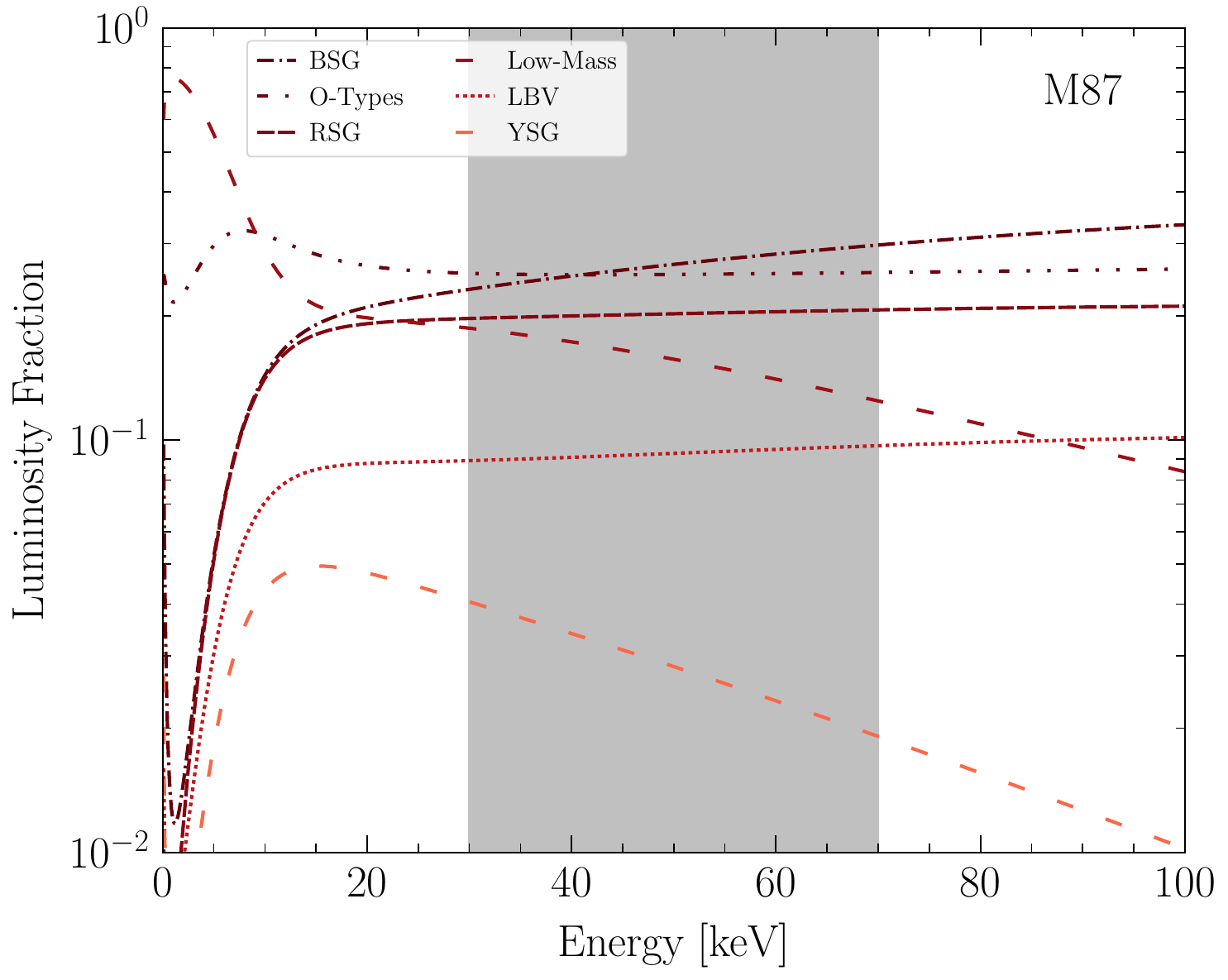}
\caption{(Left) The individual contributions of each stellar classification to the total M82 axion luminosity spectrum for a fixed $g_{a \gamma \gamma} = 1 \times 10^{-12}$ GeV$^{-1}$. We zoom in around our analysis region, in gray. (Right) The same but for M87.}
\label{fig:stellar_lum_frac}
\end{figure}

\section{Magnetic Field Model and Conversion Probabilities}

As mentioned in the main Letter, axions convert to photons in the strong magnetic fields permeating the M82 galaxy and, in the case of M87, the extended magnetic fields of the Virgo cluster.  Given a magnetic field profile ${\bf B}(s)$ and a free-electron density $n_e(s)$ along the line-of-sight parameterized by the distance $s$ extended from $s =0$ to $s = d$, one may compute the energy-dependent axion-to-photon conversion probability by~\cite{Raffelt:1987im,Dessert:2020lil,Safdi:2022xkm}
\begin{equation}
P_{a \to \gamma} \approx {g_{a\gamma\gamma}^2 \over 4} \sum_{j=1,2} \left| \int_0^d ds' B_j(s') e^{i \Delta_a s' - i \int_0^{s'} ds'' \Delta_{||}(s'')}\right|^2 \,, 
\label{eq:P_a_to_gamma}
\end{equation}
in the limit $P_{a\to\gamma} \ll 1$.
Above, $j$ is an index over the transverse directions, $\Delta_a \equiv - m_a^2 / (2 E)$, and $\Delta_{||}(s) = - \omega_{\rm pl}^2 / (2E)$, with $\omega_{\rm pl}$ determined by $n_e$.  The phase-factors in~\eqref{eq:P_a_to_gamma} account for the phase difference acquired between the axion wave and the electromagnetic wave after traversing a distance $d$, given their different dispersion relations in the presence of non-zero $m_a$ and $\omega_{\rm pl}$.  This axion-to-photon conversion probability is then convolved with the differential axion luminosity to predict the axion-induced X-ray flux incident on NuSTAR through~\eqref{eq:dFdE}. We note that we utilize the software code package \texttt{gammaALPs}~\cite{Meyer:2021} to help calculate these conversion probabilities.

To utilize~\eqref{eq:P_a_to_gamma} in deriving the conversion probabilities for axions to photons, we must use magnetic field and free-electron density models for our M82 and M87 galaxies. As discussed in the main Letter, to this end we utilize the simulation data releases from the IllustrisTNG TNG50 and TNG300 cosmological magnetohydrodynamical simulations for M82 and M87, respectively~\cite{Pillepich:2019bmb,Nelson:2019jkf}. 
We identify candidates in the TNG50 (TNG300) simulations for M82 (M87). We identify M82 analogue galaxies in the TNG50 simulation data by searching for candidates which have a stellar mass close to $\sim$$10^{10}$ $M_{\odot}$~\cite{Oehm:2017} and a star formation rate close to $\sim$$10$ $M_{\odot}$/yr~\cite{deGrijs:2001}. Ultimately, we identify three close candidates to this set of criteria; their masses fall between $\sim$$1 \times 10^{10}$ $M_{\odot}$ and $\sim$$4 \times 10^{10}$ $M_{\odot}$ and their star formation rates are at least $\sim$$7.5$ $M_{\odot}$/yr. These three candidates have subhalo indices 167408, 300907, and 544408 in the TNG50 database. For M87 we look for an analogue Virgo cluster in the TNG300 simulation data. Using~\cite{Kashibadze:2020}, we search for cluster candidates with masses falling in the range $(6.3 \pm 0.9) \times 10^{14}$ $M_{\odot}$, ultimately finding 7 candidates with subhalo indices 42631, 47315, 55060, 58081, 61682, 64929, and 70146 in the TNG300 database.  
To determine our fiducial orientation for both M82 and M87, we create an ensemble of all of our candidates, each with 10 equally spaced orientations, and, to be conservative, choose the candidate and orientation that ultimately reproduces the weakest upper limit on $g_{a \gamma \gamma}$ at $1\sigma$.

\begin{figure}[!htb]
\centering
\includegraphics[width=0.45\textwidth]{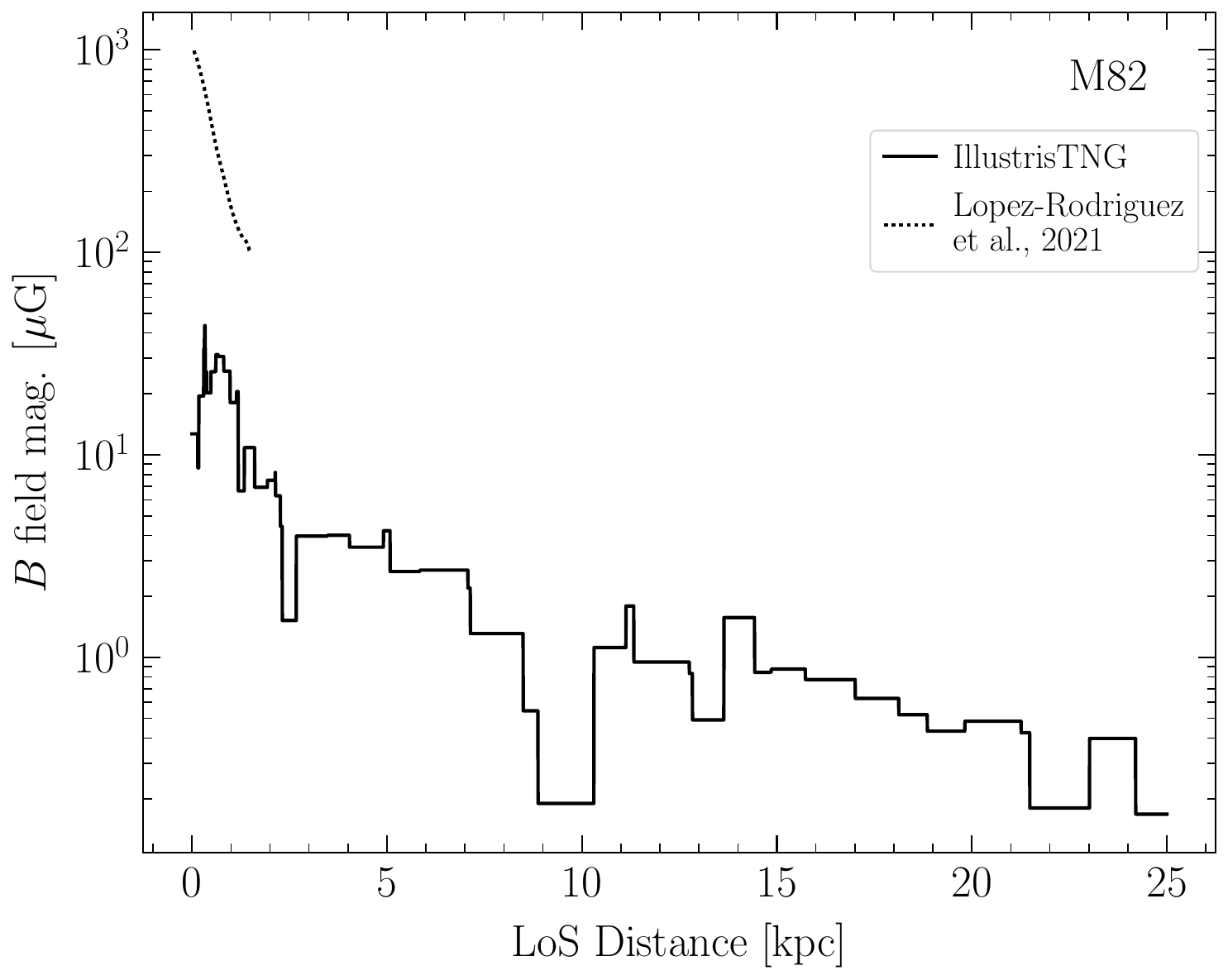}
\hspace{0.5cm}
\includegraphics[width=0.45\textwidth]{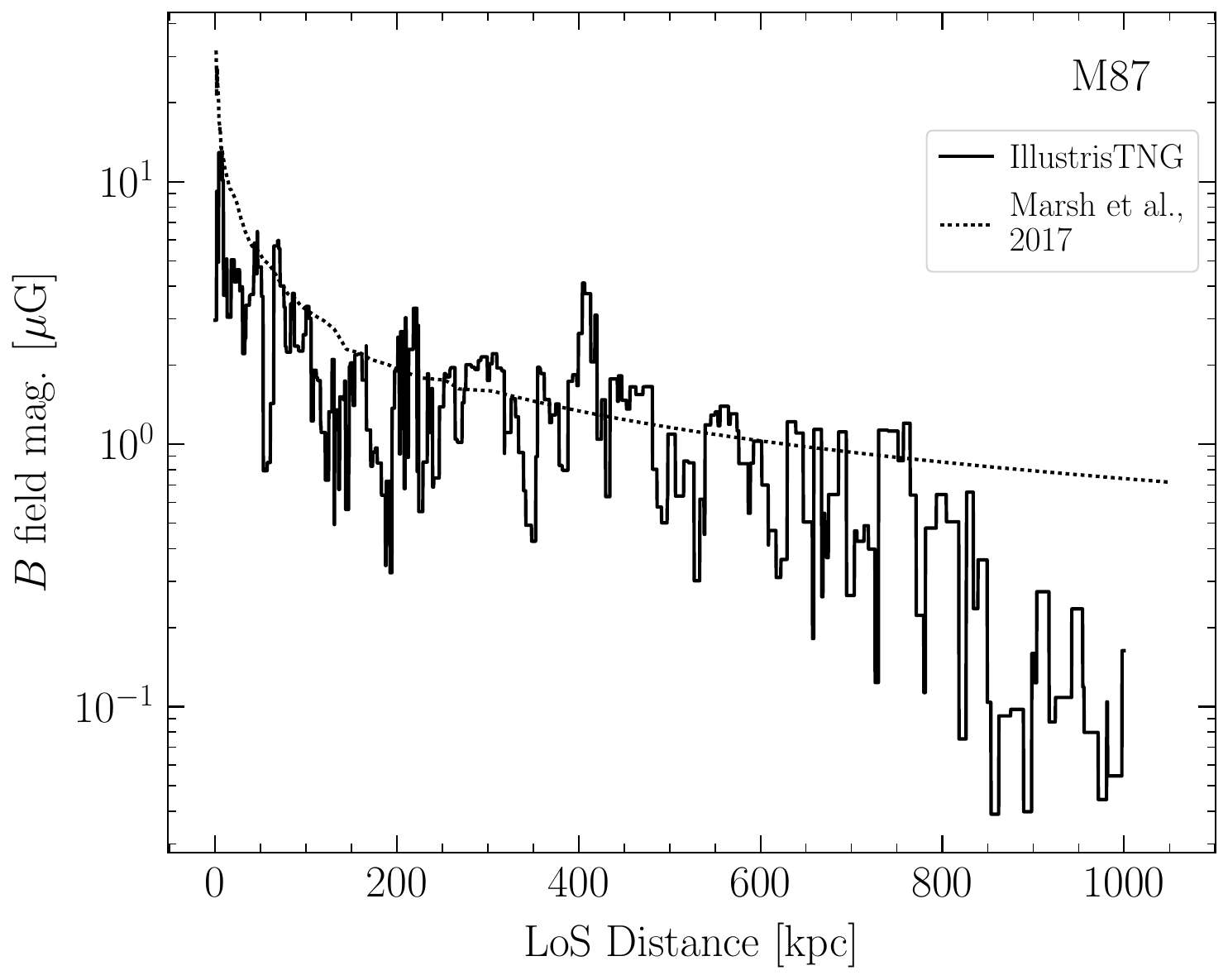}
\caption{(Left) An example profile of the magnetic field magnitude for M82 using our fiducial IllustrisTNG TNG50 subhalo and orientation. More exactly, we display the magnetic field magnitude from an example origin point near the center of the subhalo, and show its extent out to around 25 kpc in the direction of our fiducial orientation. We also illustrate the magnetic field profile deduced for the inner $\sim$2 kpc of M82 using thermal polarimetry observations from SOFIA/HAWC+ (black, dotted), as done in~\cite{2021ApJ...914...24L}. (Right) The same but for M87, using the fiducial IllustrisTNG TNG300 subhalo and orientation and illustrating the magnetic field extent out to $\sim$1 Mpc. We also illustrate the magnetic field profile adopted in~\cite{Marsh:2017yvc}, which comes from rotation-measure data. As mentioned in the text, our M82 model reaches high magnetic fields in the inner few kpc and slowly decreases further out, while our M87 model maintains reasonably sizable magnetic fields extended out to hundreds of kpc due to its position in the center of the Virgo cluster. The magnetic fields play a central role in the calculation of the conversion probabilities.}
\label{fig:B_prof}
\end{figure}

In Fig.~\ref{fig:B_prof} we show the typical magnetic field profile for M82 (M87) taken from our fiducial IllustrisTNG TNG50 (TNG300) subhalo and orientation; the line-of-sight (LoS) distances begin at an origin point near the center of the subhalo and extend in the direction of our fiducial orientation. In the case of M82, the magnetic field magnitudes tend to reach a sizable maximum of $\sim$30-40 $\mu$G in the inner few kpc, and eventually taper to around $\sim$1 $\mu$G in the range of $\sim$10-20 kpc. On the other hand, the magnetic fields pulled from our Virgo cluster model for M87 reach maxima of $\sim$10 $\mu$G in the inner tens of kpc, but, importantly, maintain a relatively consistent level of $\sim$1 $\mu$G out to hundreds of kpc. 

For M82 we also illustrate, in Fig.~\ref{fig:B_prof}, the magnetic fields deduced for the inner $\sim$2 kpc from far-infrared polarimetry observations of the High-resolution Airborne Wideband Camera-plus (HAWC+) on the Stratospheric Observatory for Infrared Astronomy (SOFIA), as done in~\cite{2021ApJ...914...24L}. We find that, in the same inner $\sim$2 kpc, the fields found in~\cite{2021ApJ...914...24L} are at least an order of magnitude greater than the ones found from the IllustrisTNG simulations; since the conversion probabilities directly depend on the strength of the magnetic fields, we can interpret our final probabilities using IllustrisTNG as likely more conservative than that which might have been obtained using the actual magnetic fields of M82. As a check, we perform an analysis implementing the ordered magnetic field component obtained in~\cite{2021ApJ...914...24L} for the inner 2 kpc and using the free-electron density obtained from our fiducial IllustrisTNG simulation within the same region. We orient the magnetic field perpendicular to the line of sight, and restrict our analysis to the inner 2 kpc, only counting the fraction of stars inside this inner region. We obtain a strong upper limit of $|g_{a\gamma \gamma}|\lesssim 3.8 \times 10^{-13}$ GeV$^{-1}$ under these assumptions, illustrated in Fig.~\ref{fig:M82_systematics}, and hence show that our fiducial analysis is indeed more conservative.  This also suggests our limit may be improved with a more robust and accurate magnetic field description of M82.

For the case of M87 we also illustrate, in Fig.~\ref{fig:B_prof}, the magnetic field model adopted in~\cite{Marsh:2017yvc}, which is deduced from rotation-measure data that additionally involves the free-electron density determined via X-ray observations. While there is some overlap between $\sim$200-500 kpc, we find that this magnetic field model is generally higher in strength than our simulation results (especially in the inner few kpc and beyond $\sim$800 kpc), and so our final conversion probabilities using IllustrisTNG will likely be somewhat more conservative. As an example, we perform an analysis using the magnetic field model and free-electron densities adopted in~\cite{Marsh:2017yvc}. The magnetic field models in~\cite{Marsh:2017yvc} are stochastically generated with random domain lengths partitioning the line of sight, as well as random, isotropic B-field pointings within each domain. We generate $N=1000$ of these realizations to compare to our fiducial IllustrisTNG results for M87, and find that the models in~\cite{Marsh:2017yvc} generally have only slightly greater conversion probabilities when comparing within our fiducial analysis energy range, resulting in the upper limits shown in Fig.~\ref{fig:M87_systematics}, with uncertainties illustrated at $1\sigma$. Our fiducial limit using IllustrisTNG falls within these uncertainties.

\begin{figure}[!htb]
\centering
\includegraphics[width=0.45\textwidth]{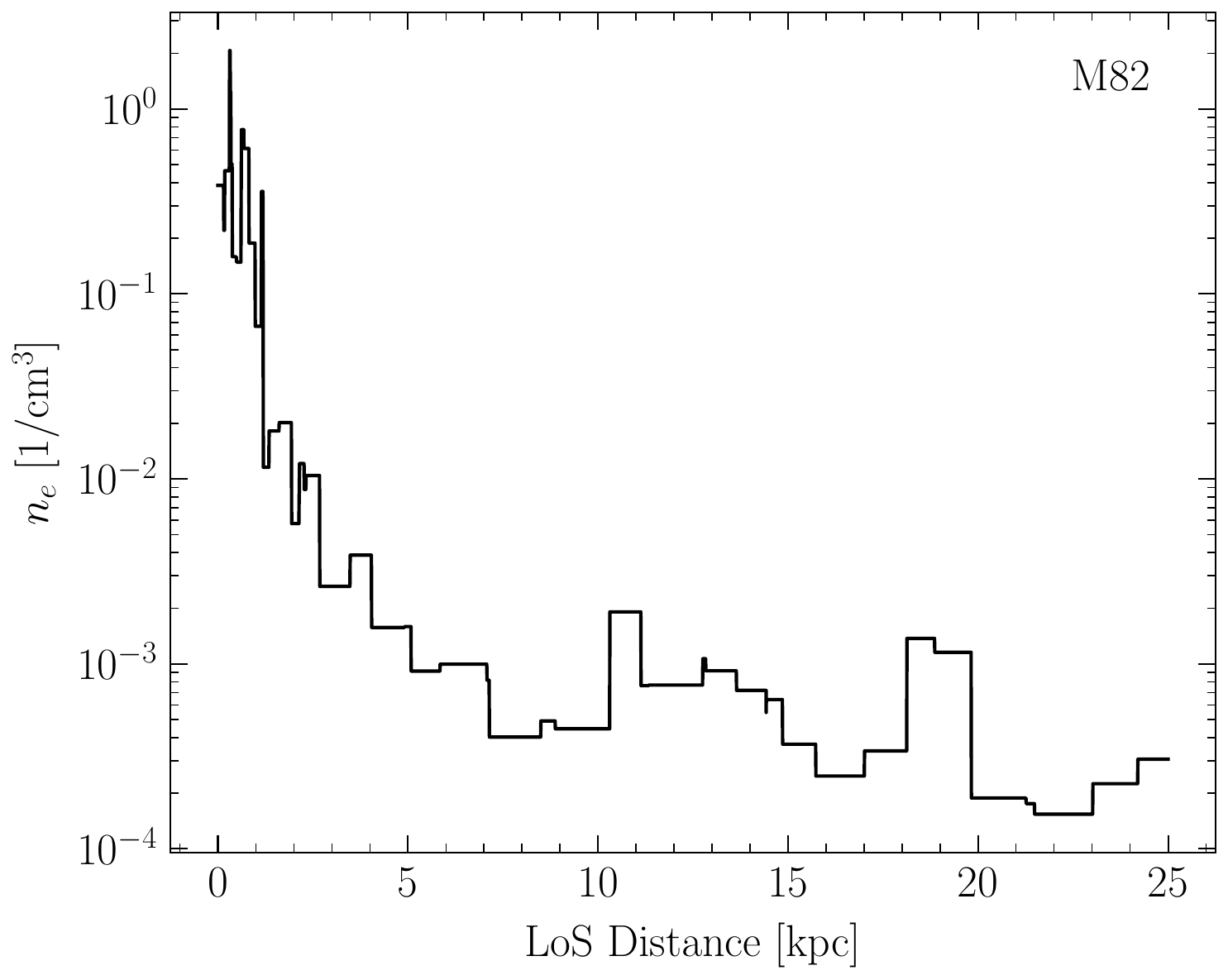}
\hspace{0.5cm}
\includegraphics[width=0.45\textwidth]{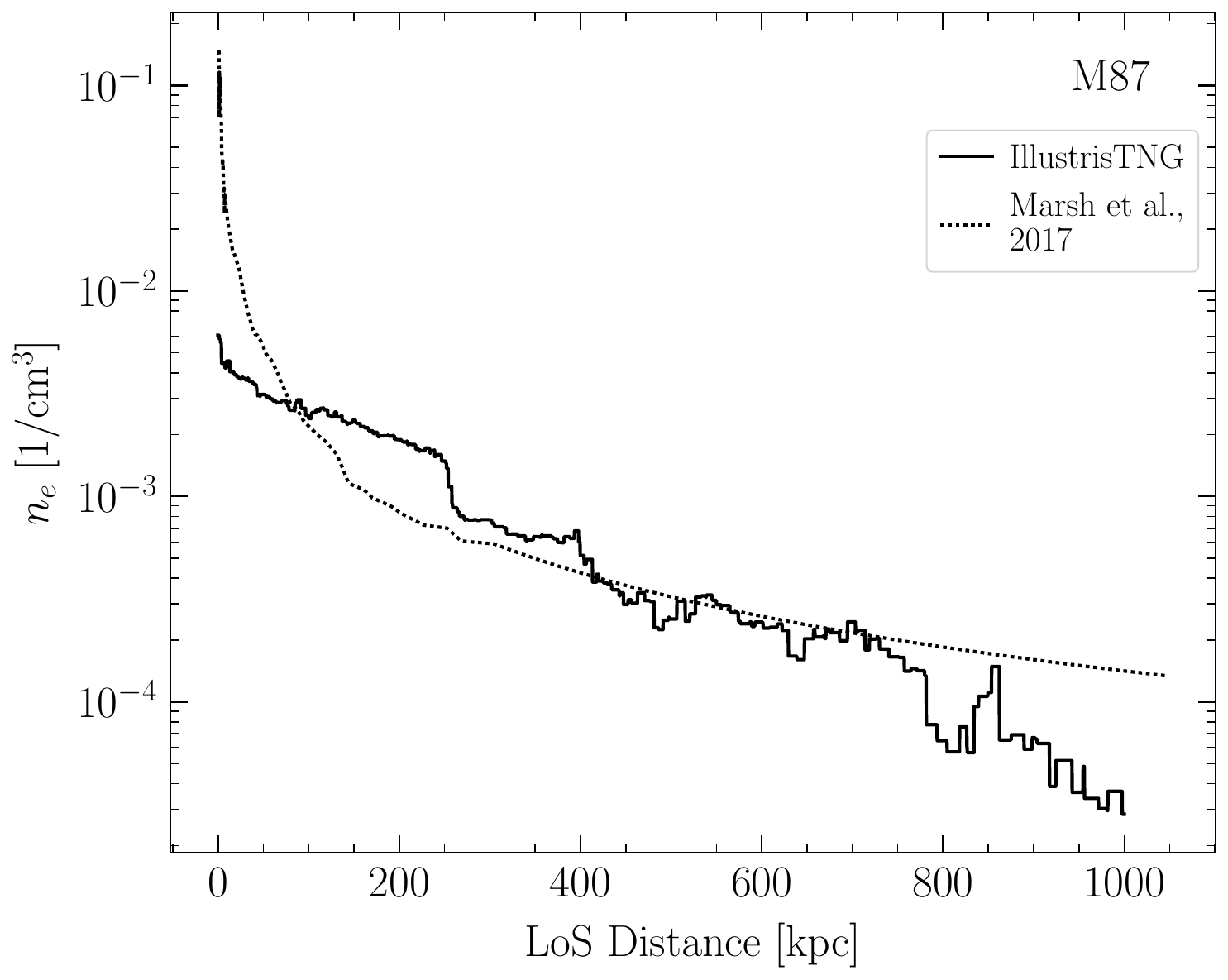}
\caption{(Left) The profile of the free-electron density $n_e$ for M82 using our fiducial IllustrisTNG TNG50 subhalo and orientation, as in Fig.~\ref{fig:B_prof}. (Right) The same but for M87, using the fiducial IllustrisTNG TNG300 subhalo and orientation. For M87 we also illustrate the free-electron density profile used in~\cite{Marsh:2017yvc}, which comes from X-ray observations. The free-electron density gives the photon an effective mass and thus affects the axion-photon conversion probability.}
\label{fig:ne_prof}
\end{figure}

\begin{figure}[!htb]
\centering
\includegraphics[width=0.45\textwidth]{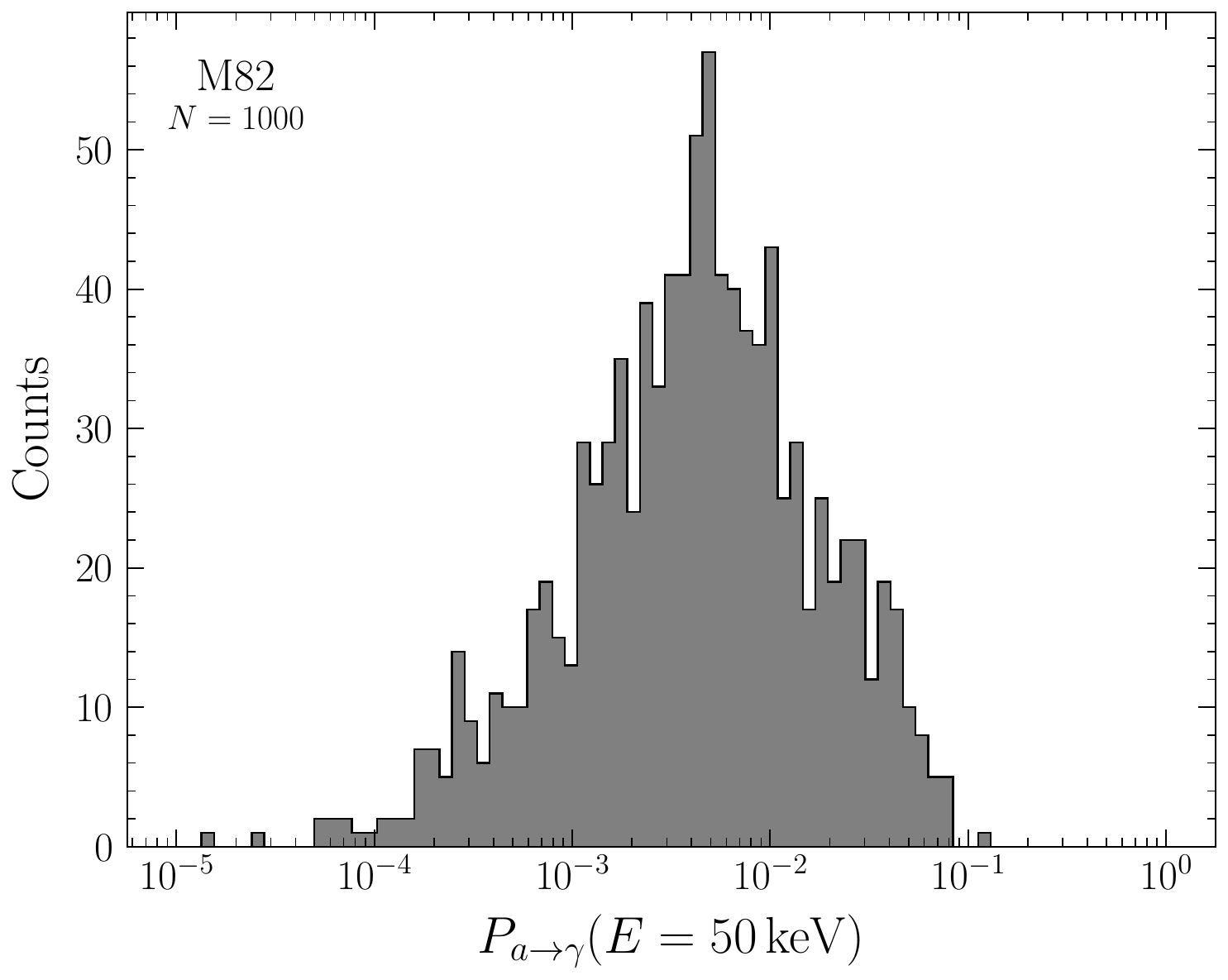}
\hspace{0.5cm}
\includegraphics[width=0.45\textwidth]{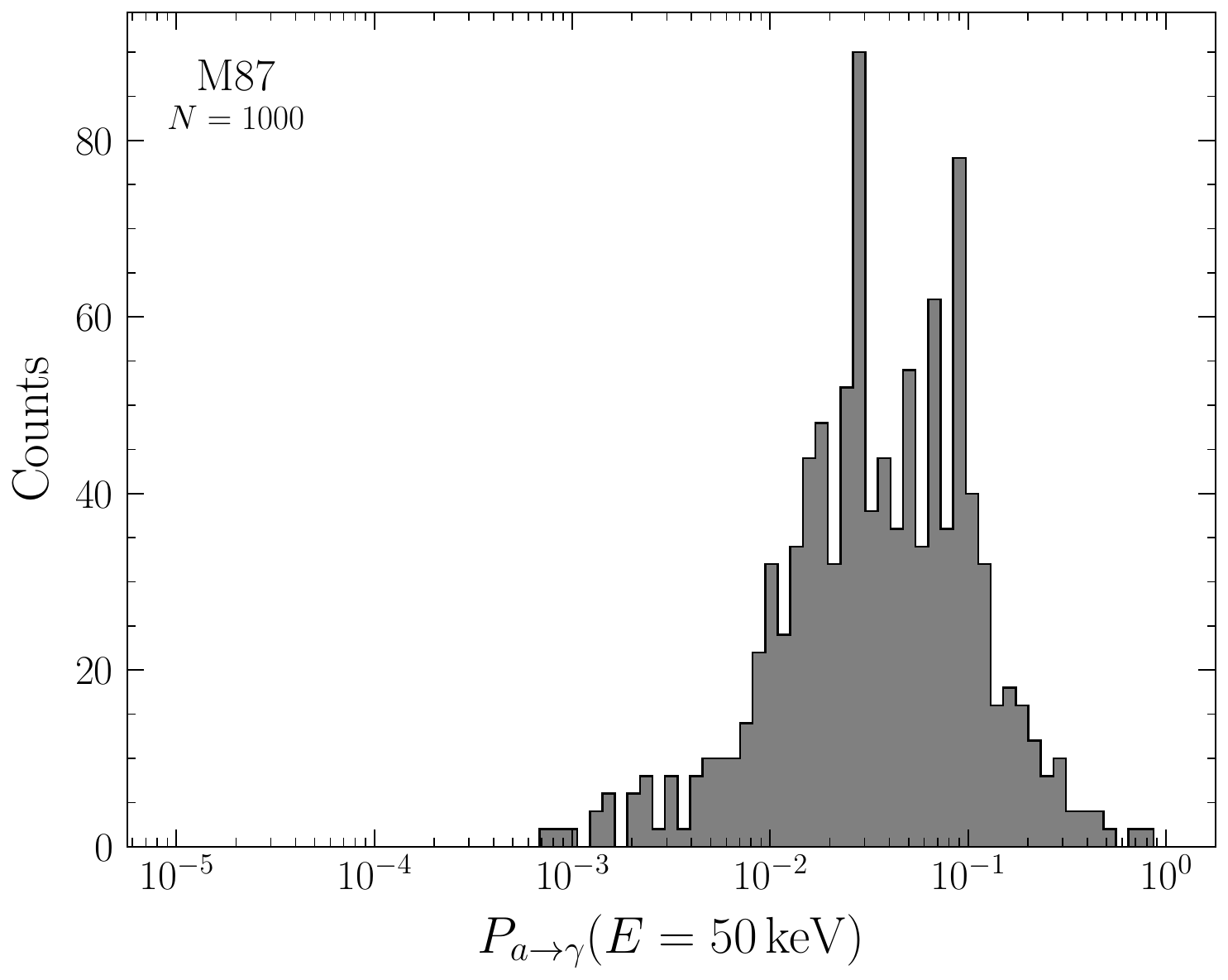}
\caption{(Left) Sample distribution of the conversion probabilities for M82 using our fiducial IllustrisTNG TNG50 subhalo and orientation. Samples are drawn according to the baryon probability density in the simulation output. Results are shown for $N=1000$ draws in the limit $m_a = 0$ at an energy of $50$ keV, which is at the center of our fiducial energy analysis range. (Right) As in the left panel but for M87, using the fiducial IllustrisTNG TNG300 subhalo.  Note that for M87 all stars within the galaxy have comparable conversion probabilities, since the conversion predominantly takes place in the cluster magnetic field. To obtain the distribution shown here we thus do not sample the stars from different locations within the galaxy but rather show the sampling of different lines of sight out of the central galaxy M87 within the (simulated) Virgo cluster.}
\label{fig:conv_prob}
\end{figure}

In Fig.~\ref{fig:ne_prof} we show the typical free-electron density profiles for our fiducial IllustrisTNG subhalo and orientation for both M82 and M87. We combine both the magnetic field models in Fig.~\ref{fig:B_prof} and the free-electron models in Fig.~\ref{fig:ne_prof} to compute the axion-photon conversion probabilities through~\eqref{eq:P_a_to_gamma}, for a given axion energy $E$. In Fig.~\ref{fig:conv_prob} we show, for M82, the distribution of conversion probabilities (across all stellar locations) in the limit $m_a = 0$ for an example energy of 50 keV, which lies at the center of our fiducial energy analysis range.  In that figure we also show, in the right panel, the distribution of conversion probabilities for lines of sight starting at the central galaxy M87 and extending out in different directions. (All stars within M87 have comparable conversion probabilities, for a fixed line of sight to Earth, since the conversion to photons predominantly takes place far away from the galaxy in the surrounding cluster.) Relative to that for M82, the M87 model generally has stronger conversion probabilities due to the extended magnetic fields present in the surrounding cluster. In Figs.~\ref{fig:M82_systematics} and ~\ref{fig:M87_systematics} we show how the extent of our uncertainties associated with the magnetic field (over all candidates and orientations) affects our final upper limits on $g_{a \gamma \gamma}$. For both M82 and M87, these magnetic field uncertainties serve as the dominant source of uncertainty in our overall analysis. 

Note that, in both cases, we also add in the Milky Way conversion probability, computed using the JF and NE2001 models, though this conversion probability is subdominant compared to the M82 and M87 conversion probabilities ($P_{a \to \gamma} \approx 2 \times 10^{-6}$ and $P_{a \to \gamma} \approx 5 \times 10^{-5}$ for M82 and M87, respectively, at $g_{a\gamma\gamma} = 10^{-12}$ GeV$^{-1}$).

\section{Comparing Upper Limit Systematics}
\begin{figure}[!htb]
\centering
\includegraphics[width=0.75\textwidth]{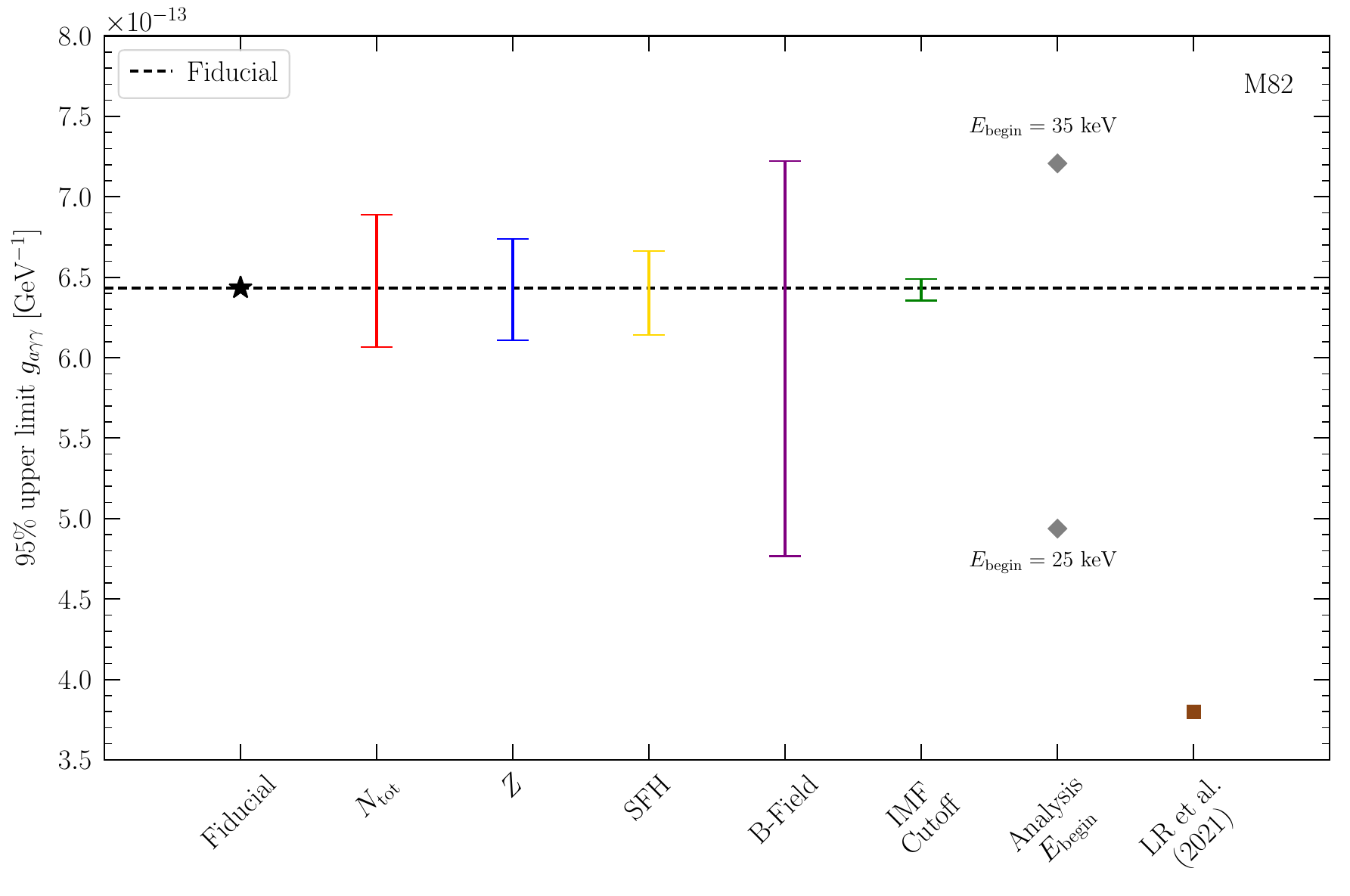}
\caption{The range of 95\% upper limits on $g_{a\gamma \gamma}$ at $m_a = 0$ from M82 as a result of systematic analyses across all major sources of uncertainties in our work for M82. As discussed in the text, we account for uncertainties in the number of stars ($N_{\rm tot}$), the metallicity ($Z$), the star formation history (SFH), the magnetic field models provided by IllustrisTNG (B-field), and the upper cutoffs on the IMF (IMF Cutoff). Additionally, we examine the resulting upper limits if we had begun our analysis one bin above and below our fiducial starting energy range value of 30 keV (Analysis $E_{\rm begin}$). The upper limits from our fiducial model is given by the black star and the dotted black line. We also illustrate the upper limit obtained as a result of implementing the ordered magnetic field obtained for the inner $\sim$2 kpc in~\cite{2021ApJ...914...24L}. As we mention in the text, the dominant source of uncertainty in our analysis comes from the magnetic field models from IllustrisTNG.}
\label{fig:M82_systematics}
\end{figure}

\begin{figure}[!htb]
\centering
\includegraphics[width=0.75\textwidth]{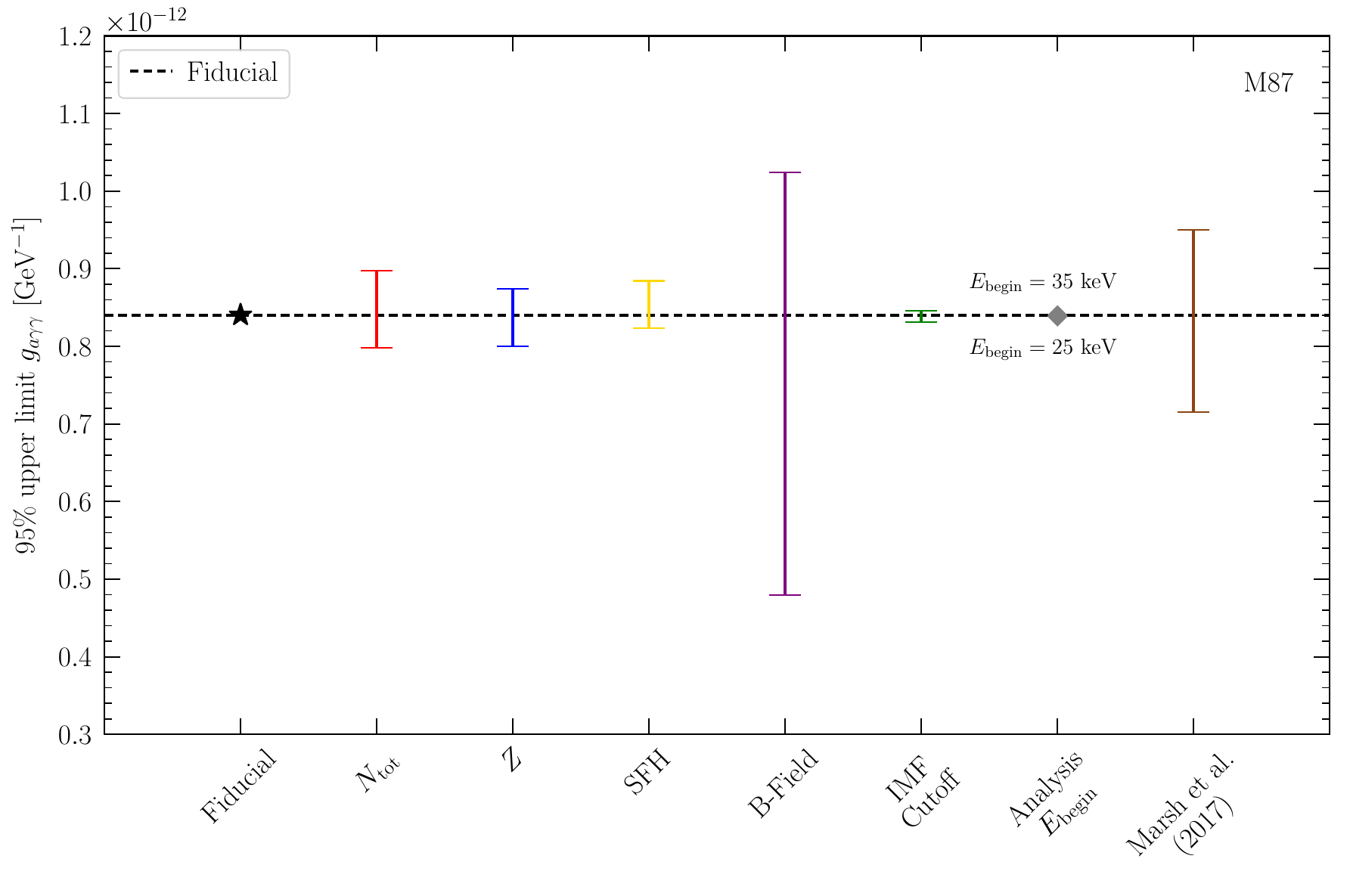}
\caption{The same as Fig.~\ref{fig:M82_systematics}, but for M87. In addition, we illustrate the upper limits and uncertainties obtained as a result of implementing the magnetic field and free-electron density models in~\cite{Marsh:2017yvc}. As was the case for M82, for M87 the dominant source of uncertainty in our analysis comes from the magnetic field models from IllustrisTNG.}\label{fig:M87_systematics}
\end{figure}

In this section we present Figs.~\ref{fig:M82_systematics} and ~\ref{fig:M87_systematics} which compare the 95\% upper limits on $g_{a\gamma \gamma}$ obtained from our fiducial model against the range of 95\% upper limits on $g_{a\gamma \gamma}$ obtained from our systematic analyses across all major sources of uncertainties in our work for both M82 and M87. These include the number of stars, metallicity, star formation history, magnetic field models provided by IllustrisTNG, upper cutoffs on the IMF, and the starting energy value used in our fiducial analysis. Also illustrated are the upper limits obtained as a result of implementing the magnetic fields in~\cite{2021ApJ...914...24L} and~\cite{Marsh:2017yvc} for M82 and M87, respectively. As can be seen for both M82 and M87, the dominant source of uncertainty comes from the magnetic field models.

\end{document}